\documentclass[aps, prd, 10pt, twocolumn, superscriptaddress,noshowpacs, preprintnumbers, longbibliography, nofootinbib,bibnotes,hyperref,floatfix]{revtex4-1}
\usepackage[colorlinks=true,breaklinks=true]{hyperref}
\usepackage{color}	    
\hypersetup{allcolors=[rgb]{0.0 0.0 1.0},linkcolor=[rgb]{0.75 0.05 0.05}}
\usepackage[dvipsnames]{xcolor}

\usepackage{listings}
\usepackage{mathtools}
\usepackage{amsmath}
\usepackage{amsfonts}
\usepackage{amssymb}
\usepackage{bbold}
\usepackage{epsfig}
\usepackage{graphicx}
\usepackage{bm}
\usepackage{array}
\usepackage{hyperref}
\usepackage{listings}
\usepackage{color}
\usepackage{float}
\usepackage[normalem]{ulem}
\hypersetup{
    urlcolor=magenta,
    }

\newenvironment{system}%
{\left\lbrace\begin{array}{@{}l@{}}}%
{\end{array}\right.}

\usepackage{tikz,xcolor,hyperref}

\definecolor{lime}{HTML}{A6CE39}
\DeclareRobustCommand{\orcidicon}{\hspace{-1mm}
	\begin{tikzpicture}
	\draw[lime, fill=lime] (0,0) 
	circle [radius=0.16] 
	node[white] {{\fontfamily{qag}\selectfont \tiny \,ID}};
	\draw[white, fill=white] (-0.0525,0.095) 
	circle [radius=0.007];
	\end{tikzpicture}
	\hspace{-3mm}
}

\foreach \x in {A, ..., Z}{\expandafter\xdef\csname orcid\x\endcsname{\noexpand\href{https://orcid.org/\csname orcidauthor\x\endcsname}
			{\noexpand\orcidicon}}
}

\begin{document}

\title{State-of-the-Art Collapsar Jet Simulations Imply\\ Undetectable Subphotospheric Neutrinos}

\author{Ersilia Guarini\orcidA{}}
\author{Irene Tamborra\orcidB{}}%
\affiliation{Niels Bohr International Academy \& DARK, Niels Bohr Institute, University of Copenhagen, Blegdamsvej 17, 2100, Copenhagen, Denmark}%
\author{Ore Gottlieb\orcidC{}}
\affiliation{
Center for Interdisciplinary Exploration \& Research in Astrophysics (CIERA), Physics \& Astronomy, Northwestern University, Evanston, IL 60202, USA}%

\date{\today}

\begin{abstract}
Mounting evidence suggests that the launching of collapsar jets is magnetically driven. Recent general relativistic magneto-hydrodynamic simulations of collapsars reveal that the jet is continuously loaded with baryons, owing to strong mixing with the cocoon. This results in a high photosphere at $ \gtrsim 10^{12}$~cm. Consequently,  collisionless internal shocks below the photosphere are disfavored, and   neutrino production in the deepest jet regions is prevented, in contrast to what has been assumed in the literature. We find that subphotospheric neutrino production could take place in the presence of collisionless sub-shocks or magnetic reconnection. Efficient particle acceleration is not possible in the cocoon, at the cocoon-counter cocoon shock interface, or at the shock driven by the cocoon in the event of a jet halted in an extended envelope. These subphotospheric neutrinos have energy $E_\nu \lesssim 10^5$~GeV for  initial jet magnetizations $\sigma_0= 15$--$2000$.  {More than one neutrino  event is expected to be observed in  Hyper-Kamiokande and IceCube DeepCore  for bursts occurring at $z \lesssim \mathcal{O}(0.1)$}.  Because of their energy, these neutrinos  {cannot} contribute to the diffuse  flux detected by the IceCube Neutrino Observatory. Our findings have implications on neutrino searches ranging from  gamma-ray bursts to luminous fast blue optical transients. 
\end{abstract}

\maketitle

\section{\label{sec:intro} Introduction}
Relativistic jets are known to play a crucial role in a wide range of astrophysical transients, however many aspects of the jet physics remain poorly understood. For example, the mechanism powering short- and long-duration gamma-ray bursts (GRBs) is still puzzling~\citep{Klebesadel:1973iq, Kumar:2014upa, Piran:2004ba}, with  hydrodynamic~\citep{Piran:1993jm} or magnetized jets~\citep{1977MNRAS.179..433B} being proposed. Relativistic jets have also been invoked to model the explosion of some core-collapse supernovae as well as common envelope jet supernovae~\citep{Piran:2017owm, Soker:2022vdg,Kuroda:2020bdq}. 
Intriguingly, also the emerging class of luminous fast blue optical transients (LFBOTs) may harbor relativistic jets, likely choked~\citep{Gottlieb:2022old}, as suggested by the asymmetry of the outflow responsible for the radiation observed in the ultraviolet, optical, infrared, radio, and X-ray bands~\cite{Margutti:2018rri,Coppejans:2020nxp}.

Independently on the source, the central engine of collapsar jets is expected to be a compact object (CO), which can  either be a hyper-accreting black hole or a rapidly spinning magnetar~\citep{Petropoulou:2020jmp,Liu:2017kga,Gottlieb:2022old,Metzger:2022xep,Berger:2013jza}.  The outflow is powered over a limited time interval, during which energy is extracted electromagnetically---by tapping into the rotational energy of the CO or the harbored magnetic field~\citep{1977MNRAS.179..433B, Meszaros:1996ww}---or thermodynamically, through neutrino annihilation~\citep{Chen:2006rra, Eichler:1989ve, Popham:1998ab}.  After its launch, the jet propagates through the stellar envelope and may break out or be choked, e.g.~if it is too weak or the stellar envelope is too dense~\cite{Mazzali:2008tb, Margutti:2014gha, Nakar:2015tma}. Independently of its fate,  the jet inflates the cocoon, while piercing through the stellar mantle, and the cocoon inevitably breaks out from the stellar envelope~\cite{2011ApJ...740..100B, MacFadyen:1999mk, Ramirez-Ruiz:2002szz, Zhang:2002yk, Lazzati:2005xv}.

Multi dimensional simulations of hydrodynamic jets contributed to shed light on the jet properties and evolution~\citep{2013ApJ...767...19L, Lopez-Camara:2016zqe, Ito:2015jfm, Ito:2018cxu, 2018MNRAS.477.2128H, Gottlieb:2019aae, Gottlieb:2020ifs, Gottlieb:2020raq}, though it is currently understood that, while energy deposition through neutrino annihilation can accelerate outflows with large Lorentz factors,  if the baryon loading is low along the polar funnel,  jet launching is more efficient if magnetically driven~\cite{2013ApJ...766...31K,Leng:2014dfa,Just:2015dba}.
The first simulations of magnetized jets, e.g.~Refs.~\cite{Burrows:2007yx, 2010MNRAS.402....7M, 2013MNRAS.429.2482P, Bromberg:2015wra, 2016MNRAS.462.2970S}, could not successfully follow  the jet upon its breakout from the star, and the jet was  artificially launched at the  boundary of the  simulation grid.
More recently, Ref.~\cite{Gottlieb:2022tkb} carried out the first 3D general relativistic magneto-hydrodynamic (GRMHD) simulation of a highly magnetized relativistic jet that breaks out from a star, expanding  on the findings of Ref.~\cite{Gottlieb:2021srg} and illustrating the need for strong  magnetic fields to allow for successful jet breakout with relativistic Lorentz factors.  

Relativistic jets are deemed to be factories of ultra-high-energy cosmic rays and  neutrinos up to $\mathcal{O}(10^{10})$~GeV~\citep{Waxman:1997ti, Meszaros:2017fcs, Guetta:2003wi,Dermer:2000yd}. Neutrinos could be produced in jets through photo-hadronic ($p \gamma$)~\citep{Waxman:1997ti, Guetta:2003wi, Wang:2018xkp} or hadronic ($pp$ and $pn$) interactions (the latter are expected to be more efficient in the innermost regions of the outflow where the baryon density is large~\cite{Razzaque:2003uv, 2011MNRAS.415.2495M, Heinze:2020zqb}), as pointed out through a number of analytical models~\cite{Waxman:1997ti, Pitik:2021xhb, Waxman:1999ai,  Guarini:2021gwh,Guarini:2022uyp, Senno:2015tsn,Gottlieb:2021pzr,Tamborra:2015fzv,Tamborra:2015qza,Denton:2017jwk,Lunardini:2016xwi,Winter:2020ptf,Dai:2016gtz}.  
But before breakout, the jet is subject to strong mixing with the cocoon, which results in heavy baryon loading~\cite{Gottlieb:2022tkb}; this reduces the Lorentz factor of the outflow and substantially increases its opaqueness, preventing the formation of collisionless shocks and potentially disfavoring neutrino production~\citep{Matsumoto:2020lsw, Gottlieb:2020mmk, Gottlieb:2021avb, Gottlieb:2022tkb}. In addition, the mixing between the highly magnetized jet with the weakly magnetized stellar material leads to reduction of the jet magnetic energy, which also impacts neutrino production~\citep{Pitik:2021xhb,Zhang:2010jt}.

Neutrinos with $\gtrsim\mathcal{O}(10)$~TeV energy could  be produced  in optically thick regions of relativistic jets~\citep{Murase:2013ffa, He:2018lwb, Kimura:2018vvz,Tamborra:2015fzv,Senno:2015tsn, Wang:2008zm, Murase:2013hh, Guarini:2022uyp,Razzaque:2004yv,Fasano:2021bwq,Grichener:2021xeg,Ando:2005xi}. Internal shocks occurring at large densities in the outflow, or in an extended envelope surrounding the star,  {have been} deemed to lead to efficient neutrino production. Most of the aforementioned work relies on the criterion for the formation of collisionless shocks~\cite{Murase:2013ffa}; the latter is fulfilled by jets with low luminosity and reaching high Lorentz factors before  {undergoing collimation}.  {However, such properties} may not be common to all  jet-powered transients, nor supported by numerical simulations  {of collapsar jets}.  Subphotospheric neutrino production  has been explored in Ref.~\cite{Gottlieb:2021pzr}, in the context of short GRBs;  it was found that the  production of high energy neutrinos in the optically thick part of the outflow is highly suppressed, due to the large baryon density and magnetic field that  limit the maximum energy up to which protons can be accelerated.  Unsuccessful jets, dark in gamma-rays and producing neutrinos while still inside the stellar progenitor, have also been suggested as major contributors to the diffuse flux detected by the IceCube Neutrino Observatory~\citep{Senno:2015tsn, Murase:2015xka,Tamborra:2015fzv,Denton:2018tdj}.

The non-thermal production of neutrinos could take place in the subphotospheric region through other poorly explored processes. Even though collisionless shocks  {are disfavored} within the optically thick region of the outflow, collisionless sub-shocks may emerge in the outflow in the presence of  mild magnetization~\citep{Beloborodov:2016jmz}. Furthermore,  Ref.~\citep{Gottlieb:2022tkb} reveals that  magnetic energy may be dissipated in the jet, while the latter is still embedded in the stellar envelope. Hence, magnetic reconnection may be another viable mechanism for particle acceleration in the optically thick regions~\citep{1977MNRAS.179..433B, Drenkhahn:2002ug, Drenkhahn:2001ue,Beniamini:2017fqh, Gill:2020oon}.  

In this paper, for the first time,  we  carry out a realistic modeling  of subphotospheric neutrino production by post-processing the outputs of the  3D GRMHD simulations presented in Ref.~\citep{Gottlieb:2022tkb}. We find that neutrino production can occur in mildly-magnetized collisionless sub-shocks~\citep{Beloborodov:2016jmz} and because of magnetic reconnection~\citep{1977MNRAS.179..433B, Drenkhahn:2001ue, Drenkhahn:2002ug}  in the innermost regions of the outflow. 
We also investigate possible neutrino production in the cocoon and at the interface between the cocoon and the counter-cocoon, showing that particle acceleration is hindered.  {To date, the simulations presented in Ref.~\citep{Gottlieb:2022tkb} are among the most advanced ones of collapsar jets, yet affected by some limitations. A larger and more advanced simulation set would be   needed to comprehensively assess  subphotospheric neutrino production in collapsar jets.}

Our work is organized as follows. In Sec.~\ref{sec:simulation}, we present our benchmark models of collapsar jets. In Sec.~\ref{sec:particleDist}, we introduce the energy distributions of photons and protons produced at the acceleration sites, as well as neutrinos.  In Sec.~\ref{sec:acceleration}, we discuss  viable acceleration mechanisms below the photosphere, namely sub-shocks and magnetic reconnection. 
In Sec.~\ref{sec:fatejets}, we investigate subphotospheric production of neutrinos in the outer regions of the jet and show under which conditions the jet is halted in the stellar envelope or an extended outer envelope. 
The expected neutrino production from  subphotospheric acceleration sites is summarized in Sec.~\ref{sec:fluence}, while the detection prospects are presented in Sec.~\ref{sec:results}. 
Finally, in Sec.~\ref{sec:conclusions} we draw conclusions on our findings.  A discussion on the thermalization of the photon spectrum is reported in Appendix~\ref{app:A}, while Appendix~\ref{app:B} summarizes the main proton and meson cooling times.  {Appendix~ \ref{app:C} explores possible acceleration sites linked to the cocoon, while we  outline the production of neutrinos in successful jets  in Appendix~\ref{app:D} for reference.}


\section{Jet model}\label{sec:simulation} 
We rely on the  3D GRMHD simulations  presented  Ref.~\citep{Gottlieb:2022tkb}. The simulations have been carried out through the code \textsc{h-amr}~\citep{Liska:2019uqw} (we refer the interested reader to Refs.~\citep{Gottlieb:2021srg,Gottlieb:2022tkb} for details on the numerical implementation). The initial magnetic field configuration allows for a self-consistent jet launching and production of a long-lived jet, which breaks out from the stellar progenitor.

\begin{figure*}[t]
\includegraphics[width=0.8\textwidth]{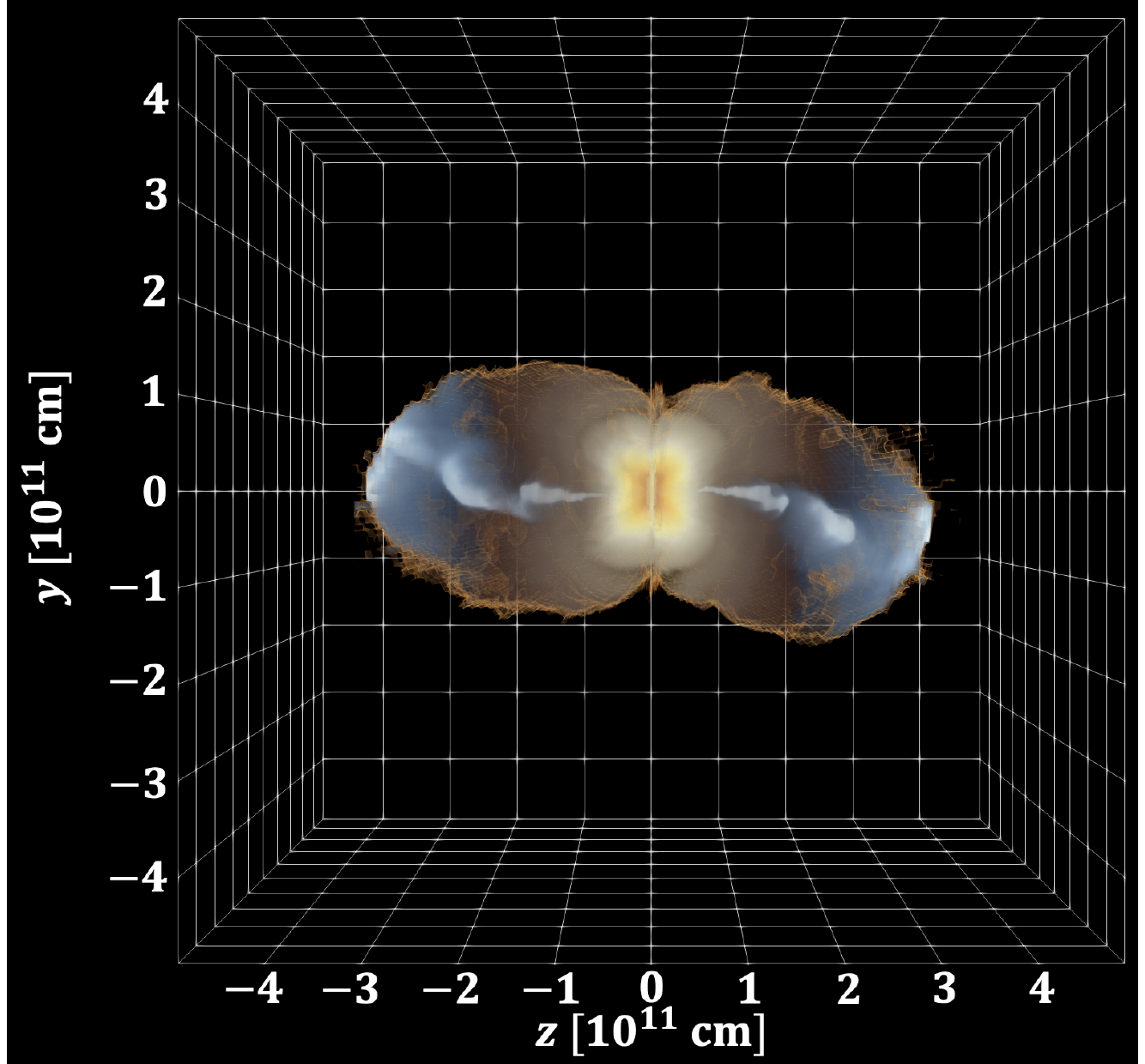}
\caption{Isocontour of the matter density of the star (yellow) and the cocoon (white-brown) combined with the asymptotic proper velocity of the jet (grey/blue) for the simulation with $\sigma_0=15$ extracted when the jet head is at $R \simeq 10 R_\star =4 \times 10^{11}$. The jet is collimated by the cocoon, which breaks out from the star. A shock develops at the interface between the cocoon and the counter-cocoon (same colors as the cocoon, but on the opposite axis).} 
\label{fig:simulation}
\end{figure*}

The CO powering the jet is a Kerr BH with mass $M_{\rm{BH}, 0}= 4~M_\odot$  and dimensionless spin $s_0=0.8$. The BH is embedded in a Wolf-Rayet star of mass $M_\star = 14~M_\odot$, extended up to  $R_\star = 4 \times 10^{10}$~cm. 
The initial magnetic field is uniform and vertical inside the magnetic core, which extends up to $\simeq 10^8$~cm; outside the core, the magnetic field profile decreases as $R^{- 1.5}$, being $R$ the distance from the CO. 
The simulation tracks the collapse of the stellar envelope onto the CO and subsequent formation of an accretion disk. A bipolar jet is launched a few milliseconds after the collapse, as shown from the snapshot in Fig.~\ref{fig:simulation}. The CO powering the jet exhibits an intrinsic variability on a timescale $10$~ms~$\lesssim t_v \lesssim$~$100$~ms.
The simulation runs for $18$~s after the launching of the jet.

 A jet with opening angle $\theta_j \simeq 0.1$~rad  and time-varying luminosity $\tilde{L}_j$ forms. The jet  is powered over the time interval $\tilde{t}_j$, so that the total energy the CO injects in it is $\tilde{E}_j = \int_0^{\tilde{t}_j} d\tilde{t} \tilde{L}_j(\tilde{t})$~\footnote{We adopt three different reference frames throughout this paper: the CO frame, the observer frame and the jet comoving frame. Quantities in each of these frames are denoted as: $\tilde{X}$, $X$, and $X^\prime$, respectively.}.
The simulation reveals that the disk-jet system develops misalignment relative to the CO axis. This results in the jet wobbling with an angle $\theta_{w} \simeq 0.2$~rad throughout its propagation. The effective opening angle of the jet is $\simeq \theta_j + \theta_w =0.3$~rad. 
It is useful to define the total isotropic-equivalent luminosity of the jet $\tilde{L}_{\rm{iso}}= \tilde{L}_j/\left( \theta_j^2/2 \right)$, since it is directly related to the observed quantities on Earth~\citep{Piran:2004ba}. The post-breakout jet isotropic luminosity is $\tilde{L}_{\rm{iso}} \simeq 10^{54}$~erg s$^{-1}$, although it might seem that this luminosity lies in the tail of the luminosity distribution of long duration GRBs~\citep{Liang:2007rn},   $\tilde{L}_{\rm{iso}}$ effectively observed would be smaller because of the jet wobbling and therefore within average or just above the peak of the luminosity distribution of long GRBs~\citep{Liang:2007rn}; see Ref.~\citep{Gottlieb:2022qow} for a  detailed discussion. 
Our benchmark simulation does not constrain the jet lifetime. Hence, we assume  $t_j=10$~s, which is  representative of long GRBs~\citep{Paciesas:2012vs}. Note that other sources of interest---such as LFBOTs or low luminosity GRBs---have typical luminosity smaller than the ones of long GRBs, see e.g.~Refs.~\citep{Ho:2018emo, Ho:2020hwf,Coppejans:2020nxp}. 

The magnetic field of the CO plays a crucial role in the launching of the jet. A fundamental quantity entering the dynamics of the outflow is its magnetization,
\begin{equation}
\sigma= \frac{B^{\prime 2}}{4 \pi \rho^\prime c^2 } \ ,
\label{eq:magnetization}
\end{equation}
where $B^\prime$ is the comoving magnetic field strength and $\rho^\prime$ is the comoving matter density in the jet.  Simulations are performed for two  initial magnetizations: $\sigma_0=15$ and $\sigma_0=200$. 
The initial magnetization of the jet corresponds to the maximum asymptotic velocity that each fluid element in the outflow can reach, if no mixing takes place. 

\begin{figure*}[t]
\centering
\includegraphics[width=0.4\textwidth]{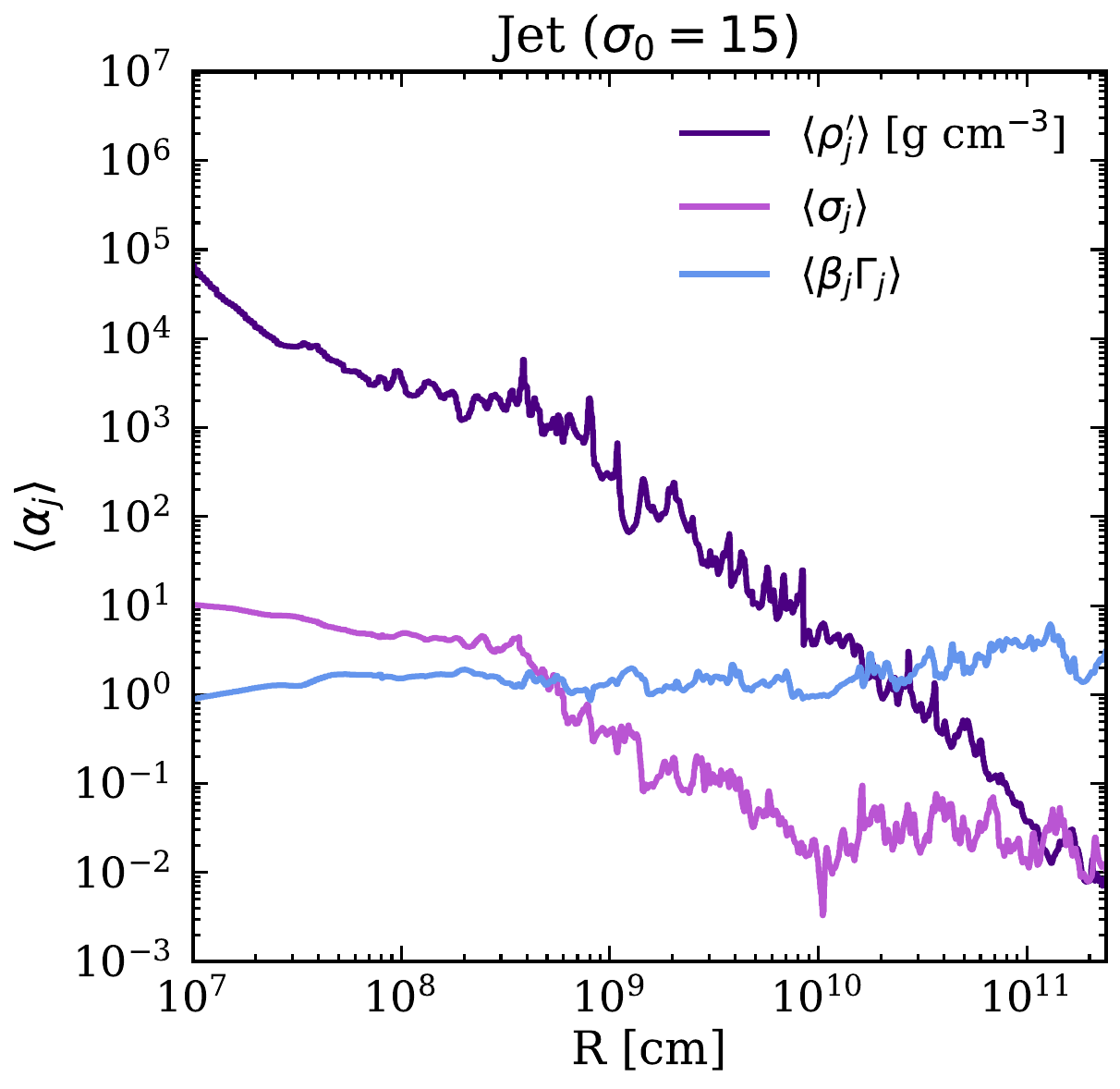}
\includegraphics[width=0.4\textwidth]{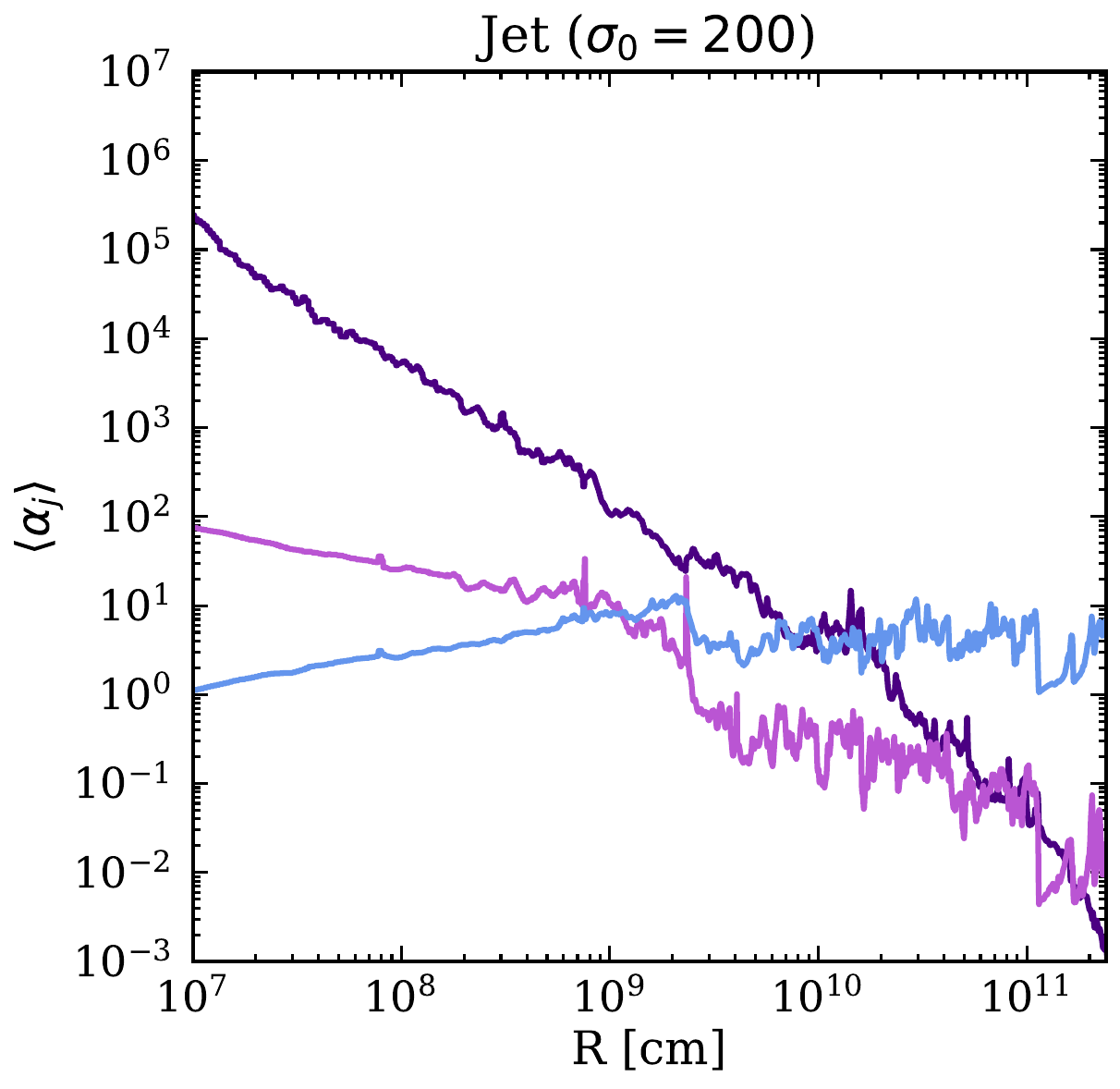}
\includegraphics[width=0.4\textwidth]{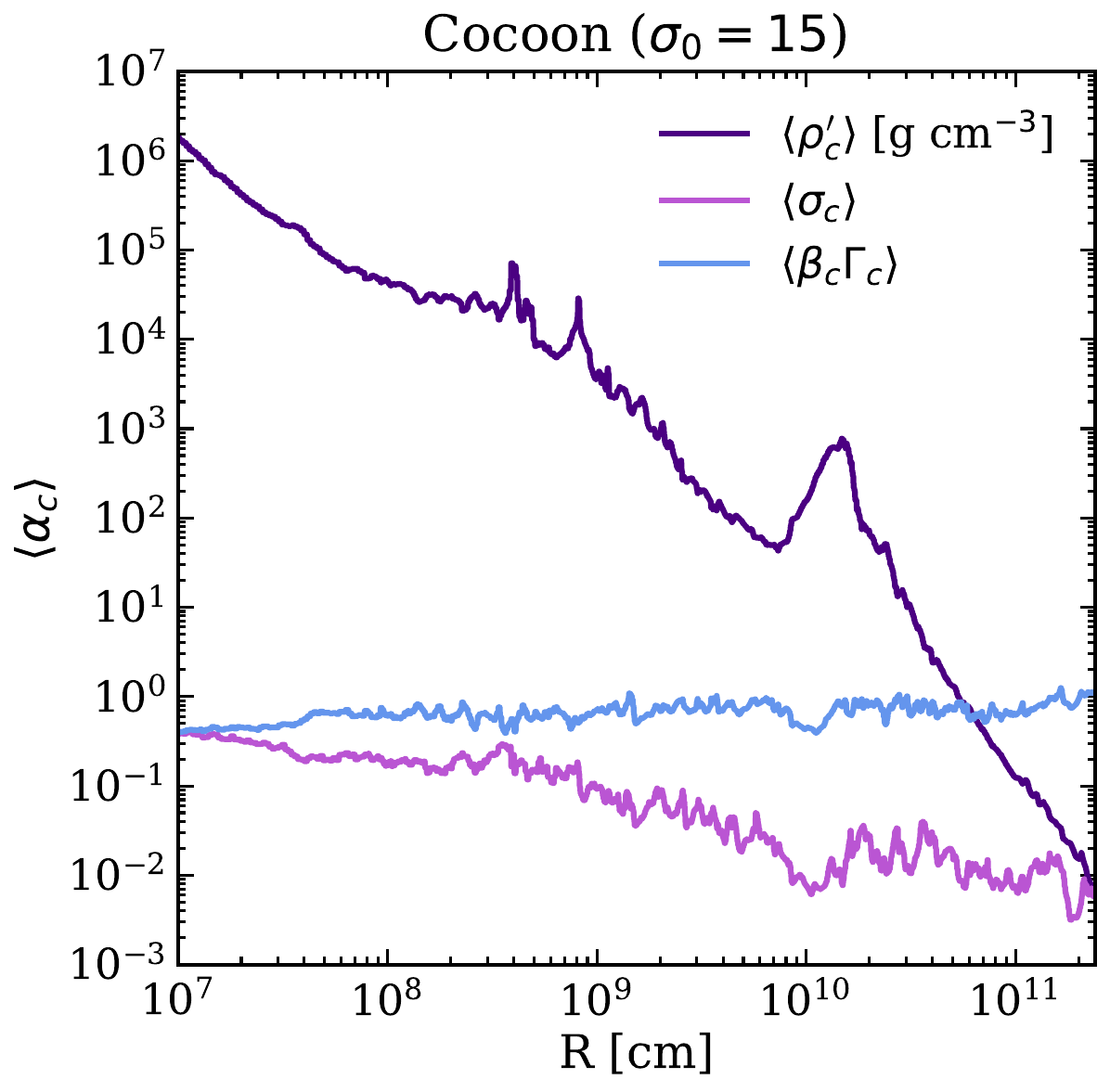}
\includegraphics[width=0.4\textwidth]{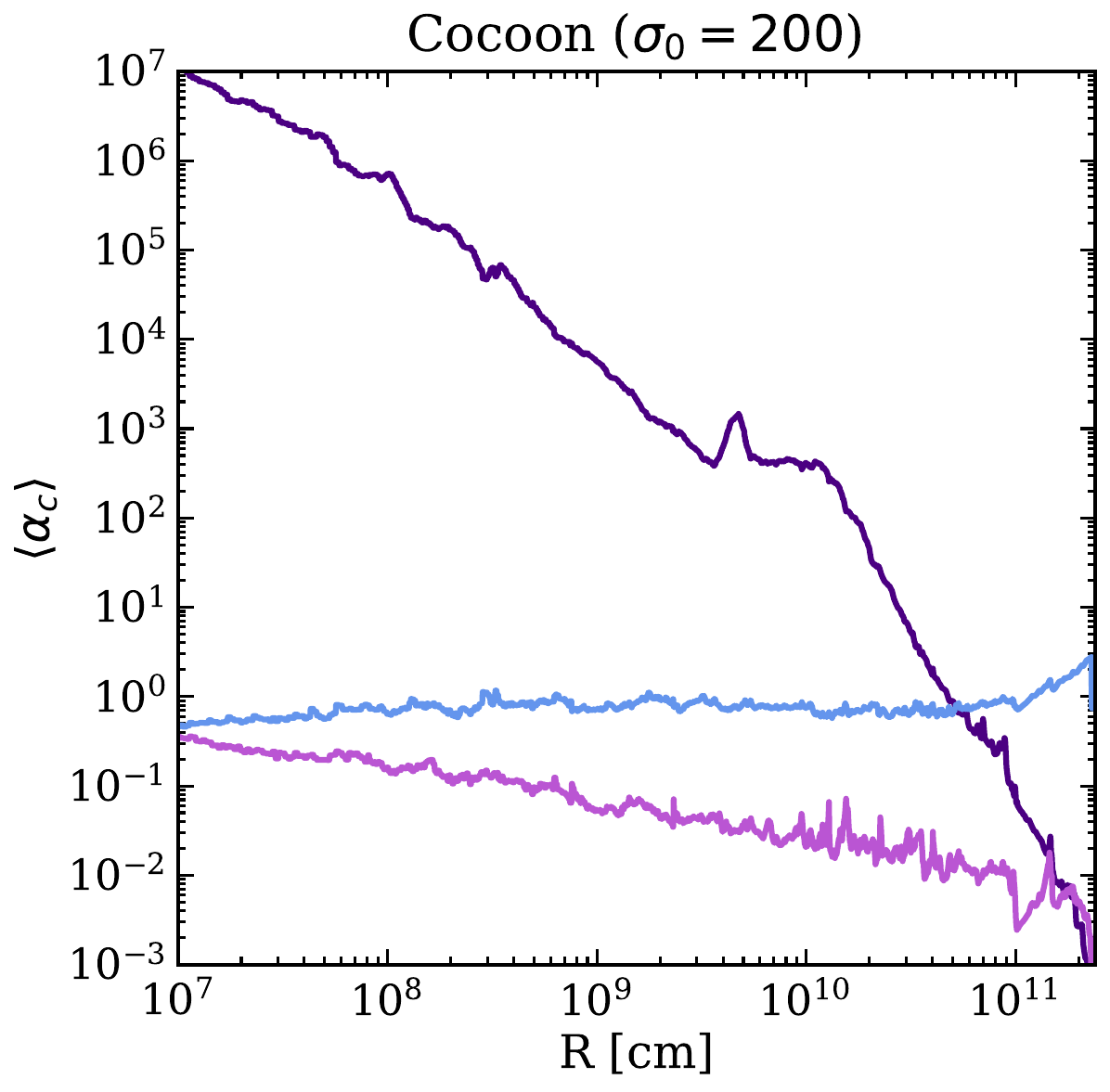}
\caption{The top (bottom) panels show the radial profiles of the angle averaged proper velocity (blue line), magnetization (pink line) and matter density (purple line) in the jet (cocoon) for initial magnetization $\sigma_0=15$ (left panels) and $\sigma_0=200$ (right panels). These quantities have been extracted  when the jet head is at $R \simeq 10 R_\star$~cm. The magnetization in the jet decreases, while its proper velocity increases as a function of the radius. This hints  that magnetic energy is efficiently converted into kinetic energy of the jet up to $R \simeq 3 \times 10^8$~cm ($R \simeq 2 \times 10^9$~cm) for $\sigma_0=15$ ($\sigma_0=200$), where both the magnetization and the Lorentz factor start showing an erratic behavior. The cocoon has roughly constant magnetization   and  proper velocity $ \langle \beta_c \Gamma_c \rangle \lesssim 1 $ throughout the whole evolution.}
\label{fig:profiles}
\end{figure*}
\begin{figure*}[t]
\centering
\includegraphics[width=0.4\textwidth]{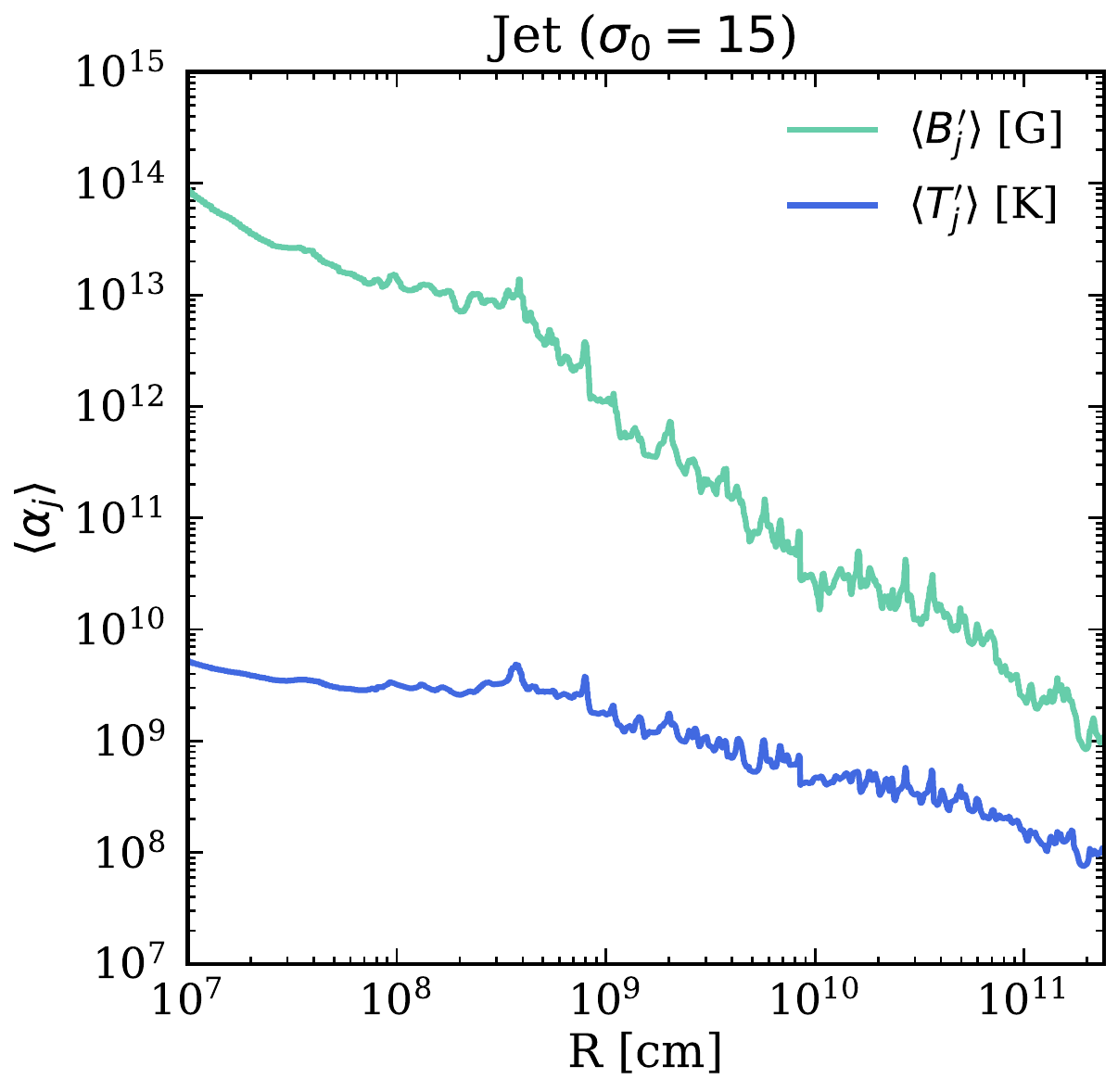}
\includegraphics[width=0.4\textwidth]{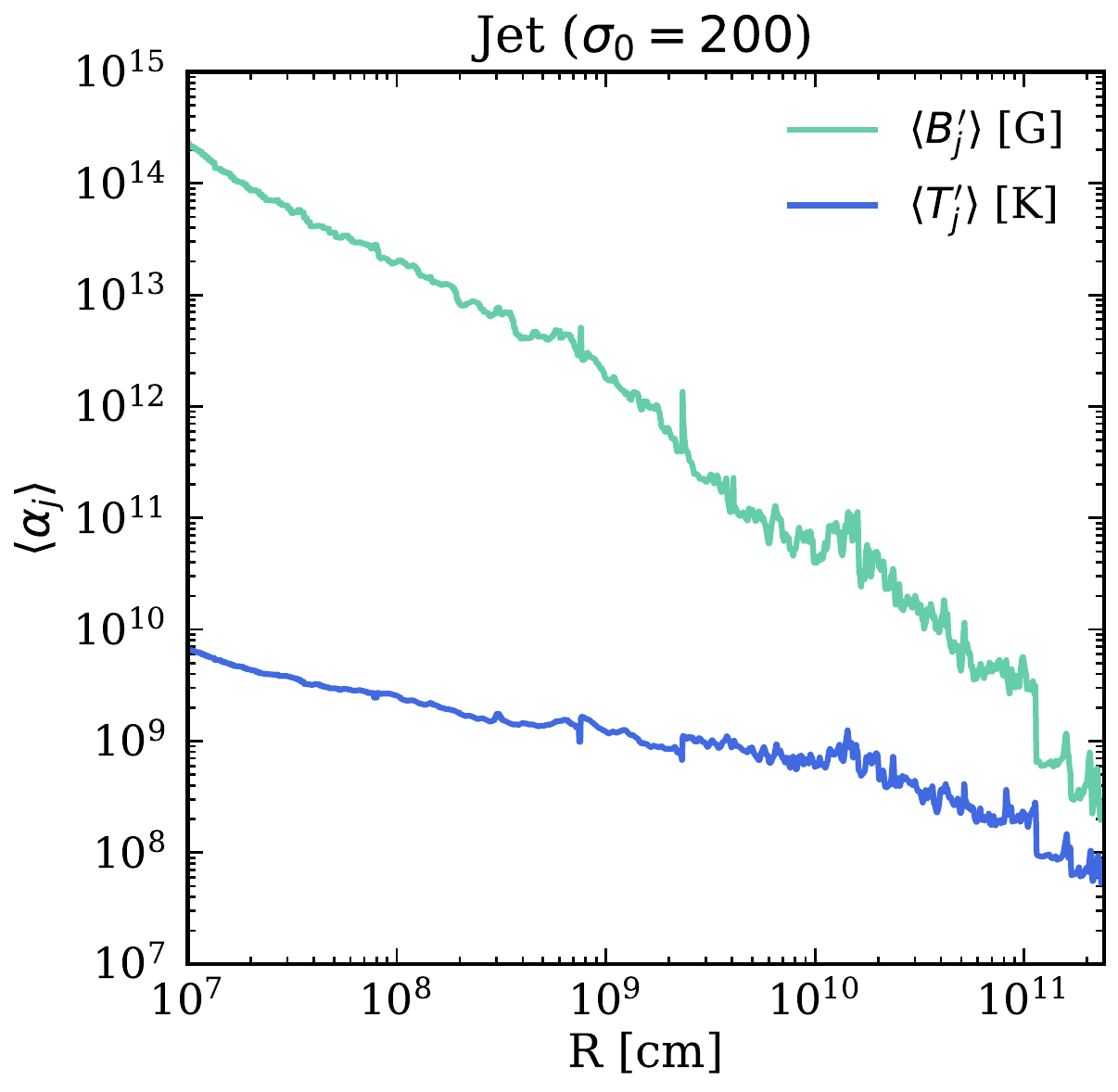}
\caption{Angle averaged radial profile of the comoving temperature $\langle T^\prime_j \rangle$ (blue line) and magnetic field magnitude $\langle B^\prime_j \rangle$ (green line) for $\sigma_0=15$ (left panel) and $\sigma_0=200$ (right panel). These quantities have been extracted when the jet head is at $R \simeq 10 R_\star$.}
\label{fig:TB}
\end{figure*}

Because the jet wobbles, it is convenient to describe the jet dynamics in terms of angle averaged quantities, namely the energy-flux weighted quantities. The top panels of Fig.~\ref{fig:profiles} show the jet proper velocity $\langle \beta_j \Gamma_j \rangle $,  magnetization $\langle \sigma_j \rangle$, and comoving matter density $\langle \rho^\prime_j \rangle $, where the symbol $\langle ... \rangle$ denotes angle averaged quantities. Here, $\beta_j$ and $\Gamma_j$ are the dimensionless velocity and the Lorentz factor of the jet, respectively. The left (right) panel has been obtained for   $\sigma_0=15$ ($\sigma_0=200$), and all  quantities have been extracted  when the jet head is at $R \simeq 10~R_\star$. The magnetization of the jet $\langle \sigma_j \rangle $ decreases with the radius, a fraction  of which is dissipated, while some is invested in accelerating the bulk motion, hence the increase in $\langle \beta_j \Gamma_j \rangle $. This hints towards efficient conversion of magnetic energy into kinetic energy, up to $R \simeq 3 \times 10^8$~cm ($R \simeq 2 \times 10^9$~cm) for $\sigma_0=15$ ($\sigma_0=200$). At this distance from the CO, both  $\langle \sigma_j \rangle $ and $\langle \beta_j \Gamma_j \rangle $ start showing an erratic behavior, induced by the entrainment of stellar material from the cocoon in the jet.  In Fig.~\ref{fig:TB} we show the comoving angle averaged temperature $\langle T_j^\prime \rangle $ and magnetic field $\langle B^\prime_j \rangle$ along the jet, when the jet head reaches $R = 6 R_\star$, as in Fig.~\ref{fig:profiles}. The temperature and the magnetic field profiles are similar for both  initial configurations with $\sigma_0=15$ and $\sigma_0=200$.

While it propagates through the star, the jet inflates a high pressure region, the cocoon, which plays a fundamental role in the collimation of the jet~\cite{2011ApJ...740..100B, MacFadyen:1999mk, Ramirez-Ruiz:2002szz, Zhang:2002yk, Lazzati:2005xv}.  The cocoon, see also Fig.~\ref{fig:simulation}, is characterized  by the average proper velocity $\langle \beta_c \Gamma_c \rangle$, magnetization $\langle \sigma_c \rangle$, and comoving matter density $\langle \rho^\prime_c \rangle$, whose radial profiles are shown in the bottom panels of Fig.~\ref{fig:profiles}. 
The cocoon magnetization is $\langle \sigma_c \rangle \lesssim 0.1$ throughout its whole evolution. The cocoon propagates at non-relativistic to mildly relativistic velocities, with $ \langle \beta_c \Gamma_c \rangle \lesssim  1$.   The isocontour in Fig.~\ref{fig:simulation} shows the existence of the counter-cocoon (white/brown region), which collides with the cocoon outside the star at the distance $R \simeq 2 R_\star$.

\begin{figure*}[t]
\centering
\includegraphics[width=0.7\textwidth]{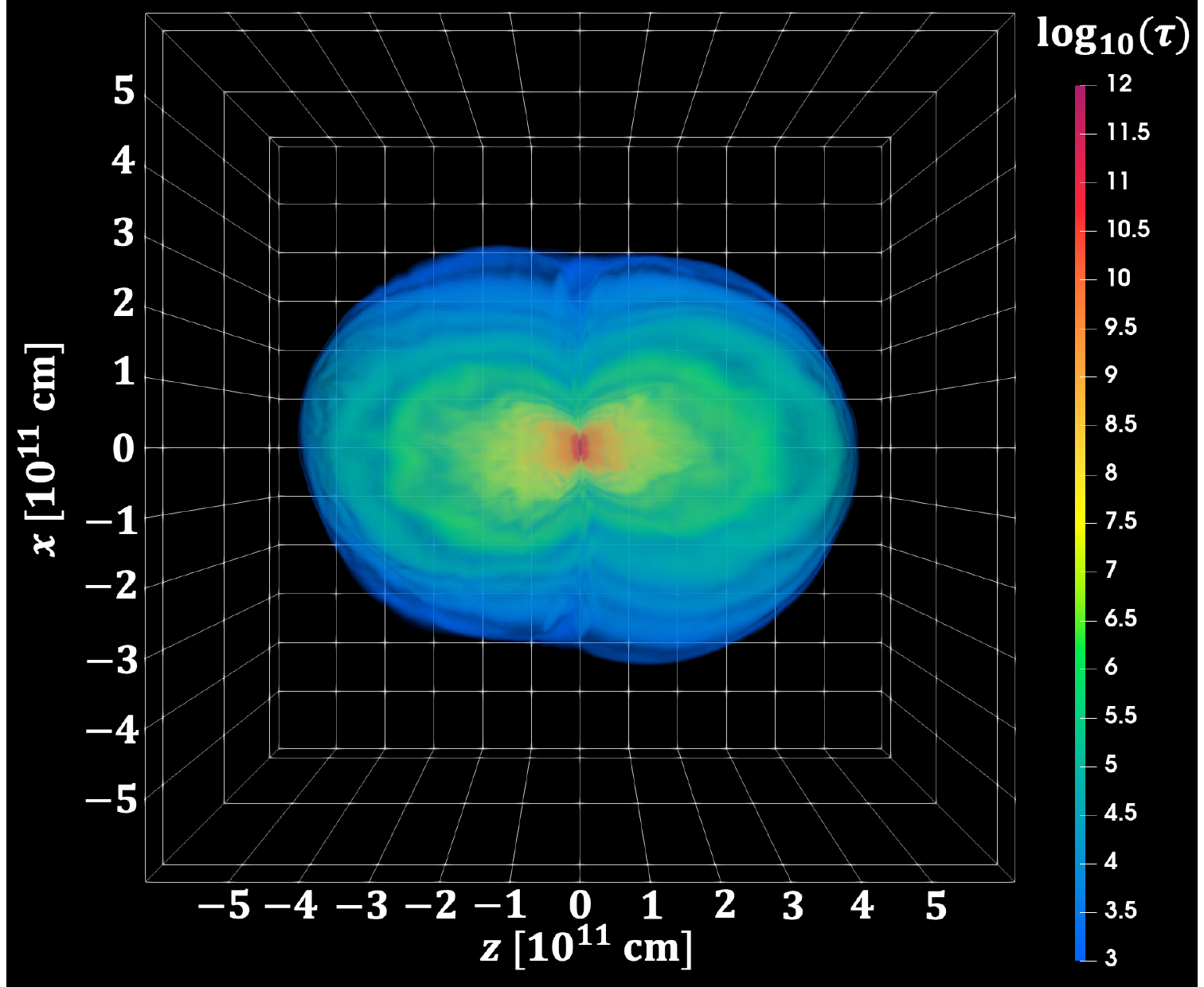}
\caption{Isocontour of the outflow optical depth  extracted when the jet head is at $R \simeq 10 R_\star$. 
The jet is highly optically thick at all times, with $\tau \gtrsim 10^3$. From the simulation it can be extrapolated that the photosphere is located at $R_{\rm{PH}} \gtrsim 10^{12}$~cm, independently on the initial magnetization.}
\label{fig:opticalDepth}
\end{figure*}

The jet-cocoon mixing observed in Fig.~\ref{fig:profiles} plays a crucial role in the definition of the outflow optical depth, since it increases the jet baryon density and it reduces the jet Lorentz factor.
Hence,  we show a contour plot of the Thompson optical depth $\tau$ of the outflow in Fig.~\ref{fig:opticalDepth}. The latter is highly optically thick throughout the simulation duration, while we find that the jet becomes optically thin ($\tau \simeq 1$) at the photospheric radius $R_{\rm{PH}} \gtrsim 10^{12}$~cm,  independently on the initial magnetization of the jet (see Ref.~\citep{Gottlieb:2022old} for a discussion).  
The role of jet-cocoon  mixing has been overlooked in the literature; this led to underestimate the optical depth of relativistic outflows, with consequent optimistic conclusions on particle acceleration efficiency~\citep{Murase:2013ffa}.

 {Lower baryon densities may be possible if the jet achieves Lorentz factors of $\mathcal{O}(100)$ early on. In this scenario, optically thin regions  may form deeply embedded in the star. However, state-of-the-art numerical simulations suggest that the jet is likely  loaded with baryons as soon as collimation starts, both for low- and high-luminosity collapsar jets~\citep{Matsumoto:2020lsw, Gottlieb:2020mmk,Gottlieb:2020ifs, Gottlieb:2020raq}. Therefore, acceleration to ultra-relativistic Lorentz factors at small radii seems  unlikely in collapsar jets. Further  work is needed to shed light on  possible exceptions.}

\section{Energy distributions of photons, protons, and neutrinos}\label{sec:particleDist}
The main goal of this paper is to investigate neutrino production below the photosphere in collapsar jets. We do so, by relying on the jet model outlined in Sec.~\ref{sec:simulation}. 
Neutrinos can be copiously produced through photo-hadronic ($p \gamma$) and hadronic ($p p$) interactions. The former take place when accelerated protons interact with a photon target, while the latter involve the collision of relativistic protons on   proton targets in the outflow.

The process responsible for particle acceleration is still subject of active research. To date, the most commonly  invoked mechanisms are diffusive shock acceleration~\citep{Piran:1999kx, Piran:2004ba, Meszaros:2006rc,Kumar:2014upa} and magnetic reconnection~\citep{Spruit:2000zm, Giannios:2007yj, Kagan:2014hea}. The outcome of both processes are non-thermal distributions of particles, which we introduce in this section.

\subsection{Photon energy distribution}
Both in diffusive shock acceleration and magnetic reconnection processes, accelerated electrons are expected to cool by emitting synchrotron radiation~\citep{2009ApJ...707L..92S,Beniamini:2017fqh,Gill:2020oon}. Since dissipation of energy occurs in a highly optically thick region (see Fig.~\ref{fig:opticalDepth}), synchrotron photons quickly thermalize to a black-body distribution~\citep{Beniamini:2017fqh}. The timescale over which the synchrotron spectrum thermalizes is much faster than any other relevant timescale for photon interactions, see Appendix~\ref{app:A}. Hence, the photon energy distribution in the region of interest is given by [in units of GeV$^{-1}$ cm$^{-3}$]:
\begin{equation}
n^{\prime}_\gamma(E^{\prime}_\gamma) = A^{\prime}_\gamma \frac{E^{\prime 2}_\gamma}{e^{E^{\prime}_\gamma/ k_B \langle T_j^{\prime} \rangle}-1}  \ ,
\label{eq:photons} 
\end{equation}
where $k_B$ is the  Boltzmann constant and $A^\prime_{\gamma}= a \langle T^{\prime_4}_j \rangle \left[ \int_0^\infty d E^\prime_\gamma E^\prime_\gamma n^\prime_\gamma(E^\prime_\gamma) \right]^{-1}$, with $a$ being the radiation constant. The radial profile of the comoving temperature $\langle T^\prime_j \rangle $ is extracted from our benchmark jet simulations and it is shown in Fig.~\ref{fig:TB}. The photon distribution in Eq.~\ref{eq:photons} is evaluated at each radius $R$ where photons are produced. 

Note that   synchrotron photons might not be abundant enough to ensure complete thermalization. In this case, photons would resemble a Wien distribution rather than a black-body one~\citep{Begue:2014kxa, Chhotray:2015lva}. 
The photon spectrum may  adjust to a Wien distribution also because of pair production, which maintains the photon bath at a comoving temperature $T^\prime_{W} \simeq 50 $~keV~\citep{Gottlieb:2019aae}. This result holds for hydrodynamic jets and it is yet to be proven for magnetized outflows~\citep{Gottlieb:2022tkb}. However,  since the optical depth in the region of interest is extremely large, as shown in Fig.~\ref{fig:opticalDepth}, we assume that deviations from  the black-body distribution  (Eq.~\ref{eq:photons})  are negligible   in the region of interest. This approximation is also justified as we have tested that the neutrino distribution  is  not sensitive to  differences between the black-body and Wien distributions (results not shown here; see also Sec.~\ref{sec:neutrino}).

\subsection{Proton energy distribution}
The  non-thermal proton energy distribution is [in units of GeV$^{-1}$ cm$^{-3}$]:
\begin{equation}
n^{\prime}_p(E^\prime_p) = A^\prime_p E^{\prime - k_p}_p\Theta \left( E^{\prime}_p - E^\prime_{p, \min} \right) \exp \left[  - \left( E^\prime_p / E^\prime_{p, \max} \right)^{\alpha_p} \right] \ , 
\label{eq:proton}
\end{equation}
where $k_p$ is the proton spectral index, $\alpha_p=2$ takes care of  the exponential cutoff~\citep{Hummer:2010vx}, and $\Theta$ is the Heaviside function. 
$E^\prime_{p, \rm{min}}$ is the minimum energy of accelerated protons and $E^\prime_{p, \rm{max}}$ is the maximum energy at which protons can be accelerated. 
The latter is fixed by requiring that the proton acceleration rate $t^{\prime -1}_{p, \rm{acc}}$ is smaller than the total cooling rate $t^{\prime -1}_{p, \rm{cool}}$, with the proton cooling rates being outlined in  Appendix~\ref{app:B}.

The normalization constant $A^\prime_p = \varepsilon_p \varepsilon_{d} e^\prime_{k} \left[ \int_{E^\prime_{p, \rm{min}}}^{E^\prime_{p, \rm{max}} } dE^\prime_p E^\prime_p n^\prime_p(E^\prime_p) \right]^{-1}$, with  $\varepsilon_d$ being the dissipation efficiency  and $\varepsilon_p$  the fraction of the dissipated kinetic energy stored in accelerated protons. 
Finally, $e^\prime_{k} = \langle \rho_j^\prime \rangle c^2 (\langle \Gamma_j \rangle -1)$  is the kinetic energy density of the outflow. 
The specific values for the parameters entering in Eq.~\ref{eq:proton} depend on the mechanism responsible for particle acceleration; we  introduce their values in Sec.~\ref{sec:acceleration}. 

\subsection{Neutrino energy distribution}\label{sec:neutrino}
Neutrinos can be produced through $p \gamma$ or $pp$ interactions. In the following, we introduce these interaction channels and the resultant neutrino distributions. 

\subsubsection{Neutrino production through $p \gamma$ interactions} \label{pg}
When accelerated protons interact with the  photons thermalized in the jet, efficient $p \gamma$ interactions take place (see Appendix~\ref{app:B}). Hereafter, unless otherwise specified, we do not distinguish between neutrinos and antineutrinos and we refer to their sum $\nu_\alpha \equiv \nu_\alpha + \bar{\nu}_\alpha$, where $\alpha = e, \mu, \tau$ is the neutrino flavor. 
The main channels for $p \gamma$ interactions are
\begin{eqnarray}
   p+ \gamma & \rightarrow & \Delta \rightarrow n + \pi^+, p + \pi^0  \\ 
p + \gamma & \rightarrow & K^+ + \Lambda/\Sigma \ .
    \label{reaction_channel}
\end{eqnarray} 
Subsequently, neutral pions decay into gamma rays: $\pi^0 \rightarrow  2 \gamma$. Neutrinos are produced through the charged pion (kaon) decay chain $\pi^+ (K^{+}) \rightarrow \mu^ + + \nu_\mu$, followed by the muon decay $\mu^+ \rightarrow \bar{\nu}_\mu + \nu_e + e^+$, and $n \rightarrow p + e^- + \bar{\nu}_e$, and the related antiparticle decay channels.

In order to compute the neutrino distribution, we rely on the photo-hadronic model of Ref.~\cite{Hummer:2010vx}.  For the given injected energy distribution of protons $n^\prime_p(E^\prime_p)$ and distribution of target photons  $n^\prime_\gamma(E^\prime_\gamma)$, secondary mesons $l$ (with $l= \pi^{\pm}, \pi^0, K^{\pm}$) are produced in the comoving frame at a rate [in units of GeV$^{-1}$~cm$^{-3}$~s$^{-1}$]:
\begin{equation}
Q^\prime_l(E^\prime_l) = c \int_{E^\prime_l}^\infty \frac{d E^\prime_p}{E^\prime_p} n^\prime(E^\prime_p) \int_{E_{\rm th}/2 \gamma^\prime_p}^\infty dE^\prime_\gamma n^\prime_\gamma(E^\prime_\gamma) R(x, y)\ ,
\label{eq:rate_secondaries}
\end{equation}
where $x = E^\prime_l/E^\prime_p$ is the fraction of the proton energy which goes in the secondaries, $y = \gamma^\prime_p E^\prime_l$, and $R(x, y)$ takes into account the interaction physics. 
The photon and proton distributions are given by Eqs.~\ref{eq:photons} and \ref{eq:proton}.

Charged mesons $l$ undergo energy losses, quantified by the cooling time $t^{\prime -1}_{l, \rm{cool}}$. Energy losses of secondaries are particularly important when  the magnetic field and  the baryon density are very large, as shown in Appendix~\ref{app:B}.  The spectrum of mesons at  decay is:
\begin{equation}
Q^{\prime \rm{dec}}_l(E^\prime_l) = Q^\prime_{l}(E^\prime_{l}) \biggr[1 - \exp\biggl(- \frac{t^\prime_{l, \rm{cool}} m_l}{E^\prime_l \tau^\prime_l}\biggr)\biggr]\  ,
\label{eq:decayed_spectrum}
\end{equation}
with $\tau^\prime_l$ being the lifetime of the meson $l$. The comoving neutrino production rate from decayed mesons is [in units of GeV$^{-1}$~cm$^{-3}$~s$^{-1}$]:
\begin{equation}
Q^\prime_{\nu_\alpha}(E^\prime_{\nu}) = \int_{E^\prime_{\nu}}^{\infty} \frac{dE^\prime_l}{E^\prime_l} Q^{^\prime \rm{dec}}_{l}(E^\prime_l) F_{l \rightarrow \nu_\alpha} \biggl(\frac{E^\prime_{\nu}}{E^\prime_l} \biggr)\ ,
\label{eq:rate_neutrini}
\end{equation}
where $\alpha$ is the neutrino flavor at production and $F_{l \rightarrow \nu_\alpha}$ is  provided in Ref.~\cite{Lipari:2007su}. 
The cooling of secondaries affects the resulting neutrino spectral energy distribution~\citep{Lipari:2007su}. In particular, when kaons cool before decaying they  contribute significantly to the neutrino spectrum at high energies~\citep{He:2012tq, Asano:2006zzb, Petropoulou:2014lja, Tamborra:2015qza}. 

\subsubsection{Neutrino production through $pp$ interactions}
Because of the large proton densities in the innermost regions of the outflow, $pp$ interactions copiously contribute to sub-photospheric neutrino production.  Accelerated protons interact with the static proton target in the jet, producing  charged and neutral pions in equal numbers. 

At each radius along the jet, the proton number density is given by 
\begin{equation}
    n^{\prime}_{p, j}= \frac{\langle \rho^\prime_j  \rangle}{2 m_p}\ , 
    \label{eq:protonStatic}
\end{equation}
where we  assume an equal amount of baryons and leptons in the jet. The radial profile of the angle averaged matter density $\langle \rho^\prime_j \rangle $ is shown in Fig.~\ref{fig:profiles}. 

As for the modeling of $pp$ interactions, we rely on Ref.~\citep{Kelner:2006tc} and, in particular, focus on  $E^\prime_p < 0.1$~TeV. This is justified, since the contribution of $pp$ interactions dominates over $p \gamma$ interactions for $E^\prime_p \lesssim 10^2$~GeV, as shown in Appendix~\ref{app:B}. 

The comoving pion production rate [in units of GeV$^{-1}$ cm$^{-3}$ s$^{-1}$] is given by
\begin{equation}
    Q^{\prime}_\pi(E^{\prime}_\pi)= \tilde{n} \frac{c n^\prime_{p, j}}{K_\pi} \sigma_{pp}\left( m_p + \frac{E^\prime_\pi}{K_\pi} \right) n^\prime_p(E^\prime_p)  \ ,
    \label{eq:pionpp}
\end{equation}
where $n^\prime_{p, j}$ is defined in Eq.~\ref{eq:protonStatic} and $n^{\prime}_p(E^\prime_p)$ is the energy distribution of accelerated protons in Eq.~\ref{eq:proton}. The free parameters are assumed to be: $\tilde{n} \simeq 1$ and $K_\pi \simeq 0.17$; the former is a valid approximation~\citep{Kelner:2006tc}, the latter is the pion multiplicity for $E^{\prime}_p \leq 0.1$~TeV~\citep{ParticleDataGroup:2020ssz}. Finally, $\sigma_{pp}$ is the energy-dependent 
cross-section for $pp$ interactions, which is provided in Ref.~\citep{ParticleDataGroup:2020ssz}. 

Since secondaries are affected by strong energy losses in the optically thick region, the cooling of pions must be taken into account. The pion spectrum at decay can be approximated as in Eq.~\ref{eq:decayed_spectrum}, using the initial rate in Eq.~\ref{eq:pionpp}. 
The production rate of muon neutrinos from  pion decay reads [in units of GeV$^{-1}$ s$^{-1}$ cm$^{-3}$]: 
\begin{equation}
    Q^\prime_{\pi \rightarrow \nu_\mu} (E^\prime_\nu) = \int_{E^\prime_{\rm{min}}}^{\infty} \frac{dE^\prime_\pi }{\sqrt{E^\prime_\pi -m_\pi^2 c^4}}  Q^{\prime \rm{dec}}_\pi (E^\prime_\pi) f_{\nu_\mu}^{(1)}\left(\frac{E^\prime_\nu}{E^{\prime}_\pi} \right) \ ,
\end{equation}
where $E^{\prime}_{\rm{min}}= E^\prime_\nu + m_\pi^2/4 E^\prime_\nu$ is the minimum energy of pions and $f_{\nu_\mu}^{(1)}$ is a function given in Ref.~\citep{Kelner:2006tc}.

As for muons from  pion decay, the treatment in Ref.~\citep{Kelner:2006tc} does not include their cooling before decaying and producing neutrinos. We therefore follow Ref.~\citep{Tamborra:2015qza} and assume that the cooling of muons results in an additional term in the neutrino spectrum approximated by $\left[1-\exp \left(-t^{\prime}_{\mu, \rm{cool}} m_\mu / E^\prime_\mu \tau^\prime_\mu \right) \right]$, with $E^{\prime}_\nu \approx E^\prime_\mu/3$~\footnote{Muons are produced by the cooled population of pions and then they undergo further energy losses. As a consequence, the spectrum of neutrinos from  muon decay is highly suppressed compared to the one produced in the direct decay of pions. Hence, the approach adopted  in Ref.~\citep{Tamborra:2015qza}  is a good approximation to our purposes, since we do not expect muons to contribute significantly to the neutrino signal, see also Sec.~\ref{sec:fluence}.}.
The neutrino production rate from muon decay is [in units of GeV$^{-1}$ s$^{-1}$ cm$^{-3}$]: 
\begin{eqnarray}
    &Q^\prime_{\mu \rightarrow \nu_\mu} (E^\prime_\nu)& = 2 \left[1-\exp \left(-\frac{t^{\prime}_{\mu, \rm{cool}} m_\mu}{E^\prime_\mu \tau^\prime_\mu} \right)\right] \times  \\  \nonumber &  & \; \int_{E^\prime_{\rm{min}}}^\infty \frac{dE^\prime_\pi }{\sqrt{E^\prime_\pi -m_\pi^2 c^4}}  Q^{\prime \rm{dec}}_\pi (E^\prime_\pi) f_{\nu_\mu}^{(2)}\left(\frac{E^\prime_\nu}{E^{\prime}_\pi}\right) \ ,  \\ 
    &Q^\prime_{\mu \rightarrow \nu_e } (E^\prime_\nu) &= 2 \left[1-\exp \left(-\frac{t^{\prime}_{\mu, \rm{cool}} m_\mu}{E^\prime_\mu \tau^\prime_\mu} \right)\right] \times \\  \nonumber &   & \; \int_{E^\prime_{\rm{min}}}^{\infty} \frac{dE^\prime_\pi }{\sqrt{E^\prime_\pi -m_\pi^2 c^4}}   Q^{\prime \rm{dec}}_\pi (E^\prime_\pi) f_{\nu_e}\left(\frac{E^\prime_\nu}{E^{\prime}_\pi}\right) \ ,
\end{eqnarray}
where the functions $f_{\nu_\mu}^{(2)}$ and $f_{\nu_e}$ are given in Ref.~\citep{Kelner:2006tc}. 
The total production rates of muon and electron neutrinos are 
\begin{eqnarray}
    Q^\prime_{\nu_\mu}(E^\prime_\nu) &=& Q^{\prime}_{\pi \rightarrow \nu_\mu}(E^\prime_\nu) + Q^\prime_{\mu \rightarrow \nu_\mu} \label{eq:rate_mupp}\ , \\
    Q^{\prime}_{\nu_e}(E^{\prime}_{\nu}) & \equiv &  Q^\prime_{\mu \rightarrow \nu_e } \label{eq:rate_epp} \ .
\end{eqnarray}

\subsection{Neutrino flux at Earth}
Neutrinos undergo flavor oscillation on their way to Earth~\cite{Farzan:2008eg,Anchordoqui:2013dnh}. Hence, the resulting observed fluence for the flavor $\alpha$ is [in units of GeV$^{-1}$~cm$^{-2}$]:
\begin{eqnarray}
\Phi_{\nu_\alpha}(E_{\nu}, z) &=& V^{\prime} t_j \frac{(1+z)^2}{4 \pi d_L^2(z)} \sum_\beta P_{\nu_\beta \rightarrow \nu_\alpha}(E_{\nu})  \times \\ & & \mathcal{Q}^{\prime}_{\nu_\beta}\left(\frac{E_{\nu} (1+z) }{\langle \Gamma_i \rangle} \right)\ , \nonumber
\label{eq:neutrino_flux}
\end{eqnarray}
where $z$ is the redshift of the source harboring the jet, $i=j, c$ depending on the neutrino production site (i.e.~the jet or the cocoon), $\mathcal{Q}^{\prime}_{\nu_\beta}\left(E_{\nu} (1+z) / \langle \Gamma_i \rangle \right)$ is the comoving neutrino production rate for $p\gamma$ or $pp$ interactions, given by Eq.~\ref{eq:rate_neutrini} and Eqs.~\ref{eq:rate_mupp}-\ref{eq:rate_epp}, respectively.
The  comoving volume of the interaction region is $V^\prime \simeq 2 \theta_i^2 \pi R_{\rm{int}}^3 /(2 \langle \Gamma_i \rangle)$~\citep{Baerwald:2011ee}, where $R_{\rm{int}}$ is the distance from the CO where the interaction takes place. The outflow lifetime measured on Earth is $t_j = \tilde{t_j} (1+z)$. The neutrino oscillation probabilities, $P_{\nu_\beta \rightarrow \nu_\alpha}= P_{\bar{\nu}_\beta \rightarrow \bar{\nu}_\alpha}$, are given by~\citep{Anchordoqui:2013dnh}: 
\begin{eqnarray}
P_{\nu_e \rightarrow \nu_\mu} &=& P_{\nu_\mu \rightarrow \nu_e} = P_{\nu_e \rightarrow \nu_\tau} = \frac{1}{4} \sin^2 2\theta_{12}\ , \\
P_{\nu_\mu \rightarrow \nu_\mu} &=& P_{\nu_\mu \rightarrow \nu_\tau  }= \frac{1}{8}(4-\sin^2 \theta_{12})\ ,\\
P_{\nu_e \rightarrow \nu_e} &=& 1- \frac{1}{2} \sin^2 2\theta_{12}\ ,
\end{eqnarray}
where $\theta_{12} \simeq 33.5^\circ$~\citep{ParticleDataGroup:2020ssz, Esteban:2020cvm}.

In a standard flat $\Lambda \rm{CDM}$ cosmology, the luminosity distance is
\begin{equation}
d_L(z) = (1+z) \frac{c}{H_0} \int_0^z \frac{dz^\prime}{\sqrt{\Omega_\Lambda + \Omega_M(1+z^\prime)^3}}\ ,
\end{equation}
where we adopt $H_0 =  67.4$~km~s$^{-1}$~Mpc$^{-1}$, $\Omega_M = 0.315$, and $\Omega_\Lambda = 0.685$~\citep{ParticleDataGroup:2020ssz}. In the following, unless otherwise specified, we assume that the source harboring the collapsar jet is located at $z=2$, namely at the peak of the redshift distribution of long GRBs~\citep{2012ApJ...752...62J}.

\section{Inner subphotospheric particle acceleration sites} 
\label{sec:acceleration} 
\begin{figure*}[t]
\includegraphics[scale=0.35]{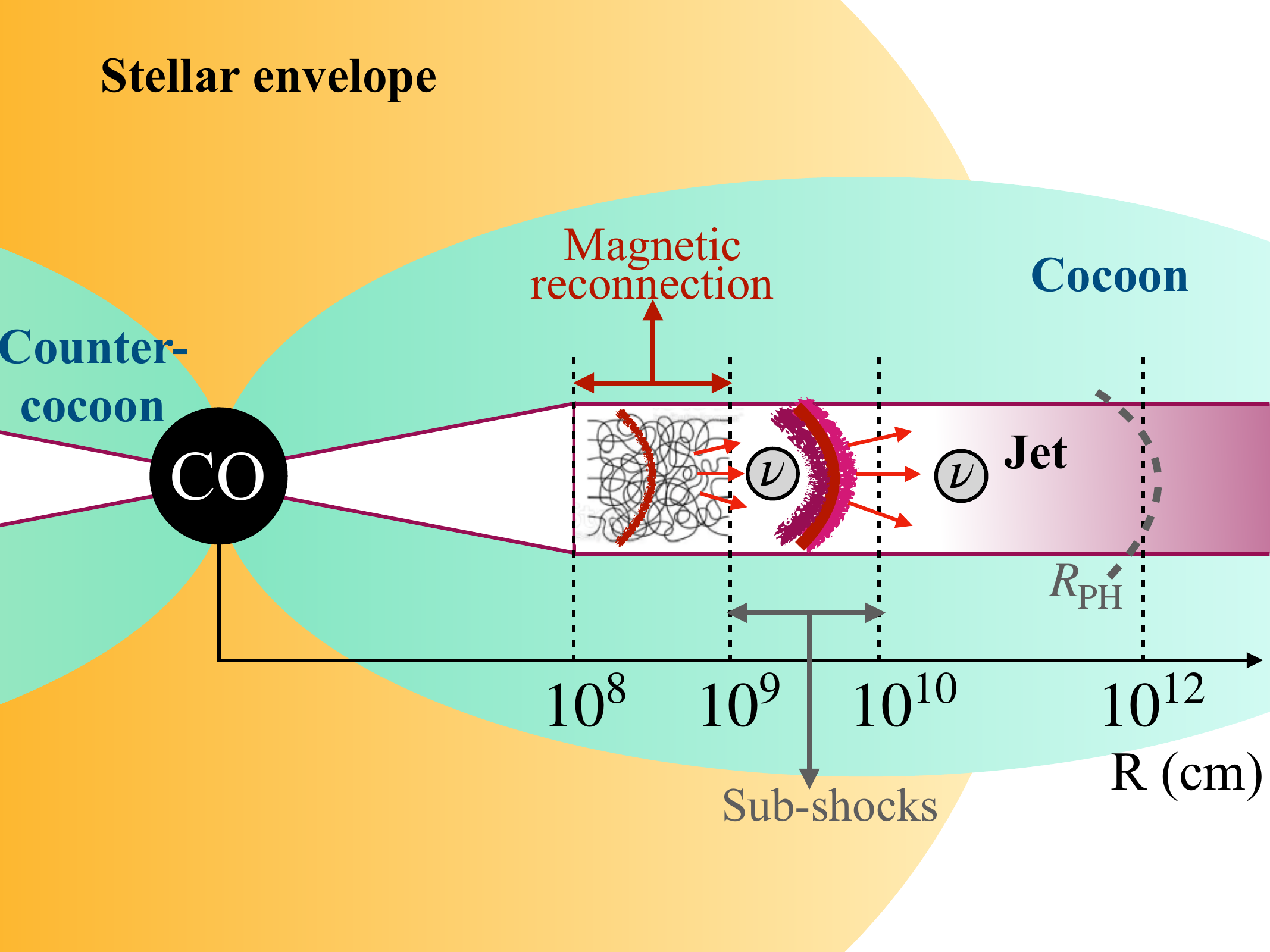}
\caption{Sketch of the particle acceleration sites in the optically thick regions of the jet shown Fig.~\ref{fig:simulation} (not in scale). The jet  (white-purple) is launched by the CO (black) and it inflates the cocoon (aqua region) and the counter-cocoon while propagating in through the star (yellow); the photosphere radius is marked for orientation.  Magnetic reconnection is efficient deep in the outflow ($R \simeq 10^8$--$10^9$~cm), where the magnetic energy is dissipated and converted into kinetic energy of the jet. At larger radii ($R \simeq 10^9$--$10^{10}$~cm) particles can be accelerated at the collisionless sub-shocks where the outflow is mildly magnetized. Both magnetic reconnection and sub-shocks lead to neutrino production. } 
\label{fig:sketch}
\end{figure*}

As discussed in Sec.~\ref{sec:particleDist}, efficient neutrino production occurs where particles can be  accelerated efficiently. 
In this section, we outline two possible mechanisms for particle acceleration in the optically thick region of collapsar jets: magnetic reconnection and collisionless mildly magnetized sub-shocks emerging within radiation mediated shocks, and  present the corresponding neutrino fluence. {We stress that our results are based on the physics of our benchmark jet model~\citep{Gottlieb:2022tkb}. Nevertheless, ours is a first step towards a more realistic modeling of  particle acceleration in collapsar jets.}
A schematic summary of the particle acceleration regions is displayed in Fig.~\ref{fig:sketch}.
We rely on the angle averaged profiles shown in Figs.~\ref{fig:profiles} and \ref{fig:TB}. 

 \subsection{Magnetic reconnection}
 \label{sec:magnetic}
 When the central engine hosts a highly variable magnetic field, particle acceleration  can take place through magnetic reconnection~\citep{Guo:2014via, Nalewajko:2015gma, Petropoulou:2018bvv, Kilian:2020yyw}. In the standard picture, magnetic energy is gradually dissipated along the jet, starting below the photosphere and extending over a wide range of radii~\citep{Drenkhahn:2002ug, Drenkhahn:2001ue}. 

The central engine powering the outflow  changes polarity on a length scale $\lambda$. When magnetic lines of inverse polarity reconnect,  magnetic energy is dissipated. Half of the dissipated energy is converted into kinetic energy of the jet. The remaining half is believed to go into particle acceleration. Magnetic reconnection is no longer efficient when the magnetization of the outflow drops to $\sigma_j \simeq 1$, where the jet stops accelerating. For a review on the analytical modelling of magnetic reconnection see, e.g., Refs.~\cite{Beniamini:2017fqh, Gill:2020oon}. 

Our two benchmark jet simulations show  polarity inversion of the magnetic field lines over a typical length scale $\lambda \simeq 2 \times 10^8$~cm, both for $\sigma_0=15$ and $\sigma_0=200$.
This length scale is in very good agreement with the one usually adopted in the literature (i.e.~$\lambda \simeq 10^8$--$10^9$~cm)~\citep{Beniamini:2017fqh}. Magnetic energy is efficiently converted in kinetic energy along the outflow, as shown in Fig.~\ref{fig:profiles} and discussed in Sec.~\ref{sec:simulation}. 

\begin{figure*}[t]
 \centering
 \includegraphics[width=0.47\textwidth]{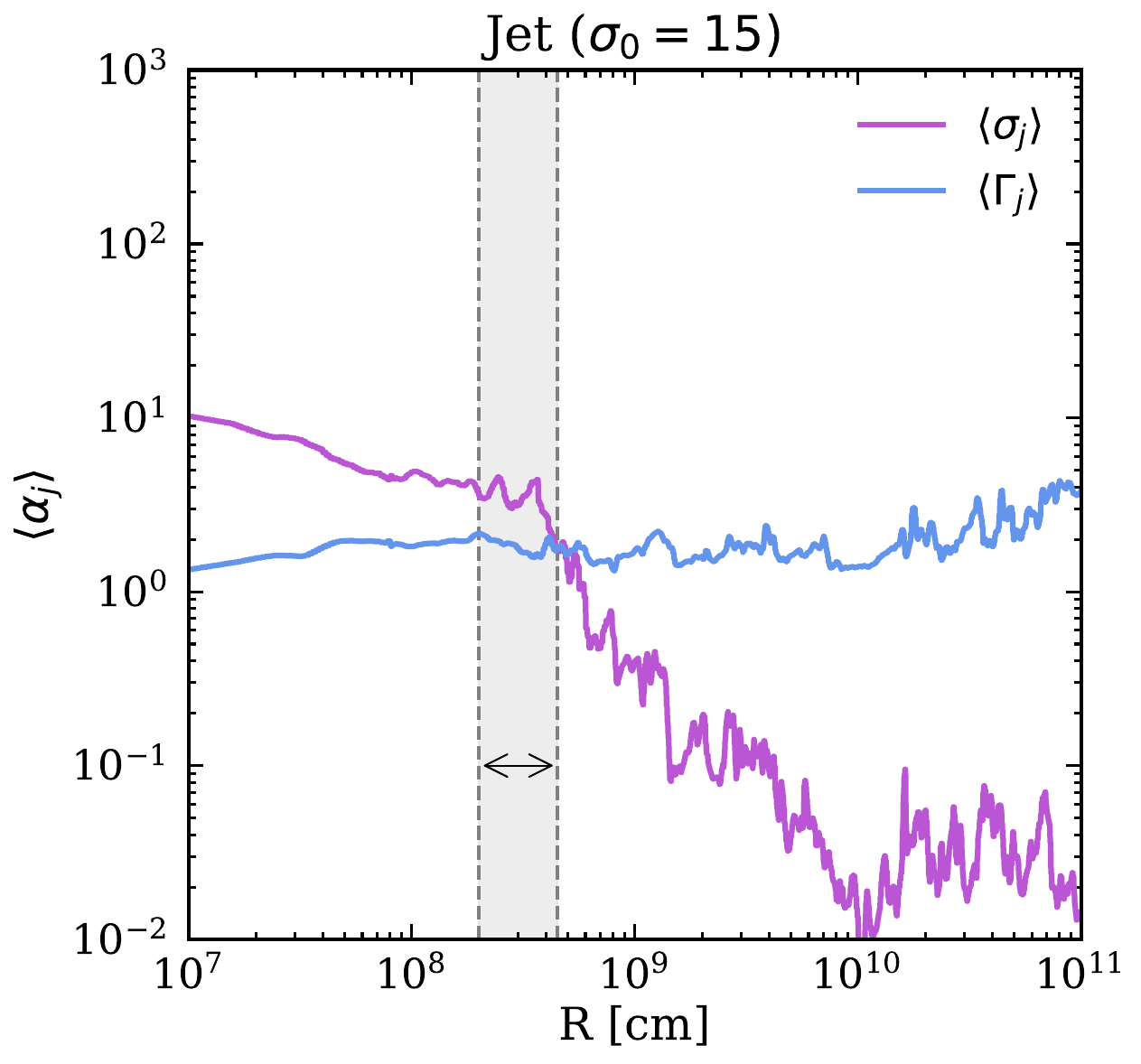}
 \includegraphics[width=0.47\textwidth]{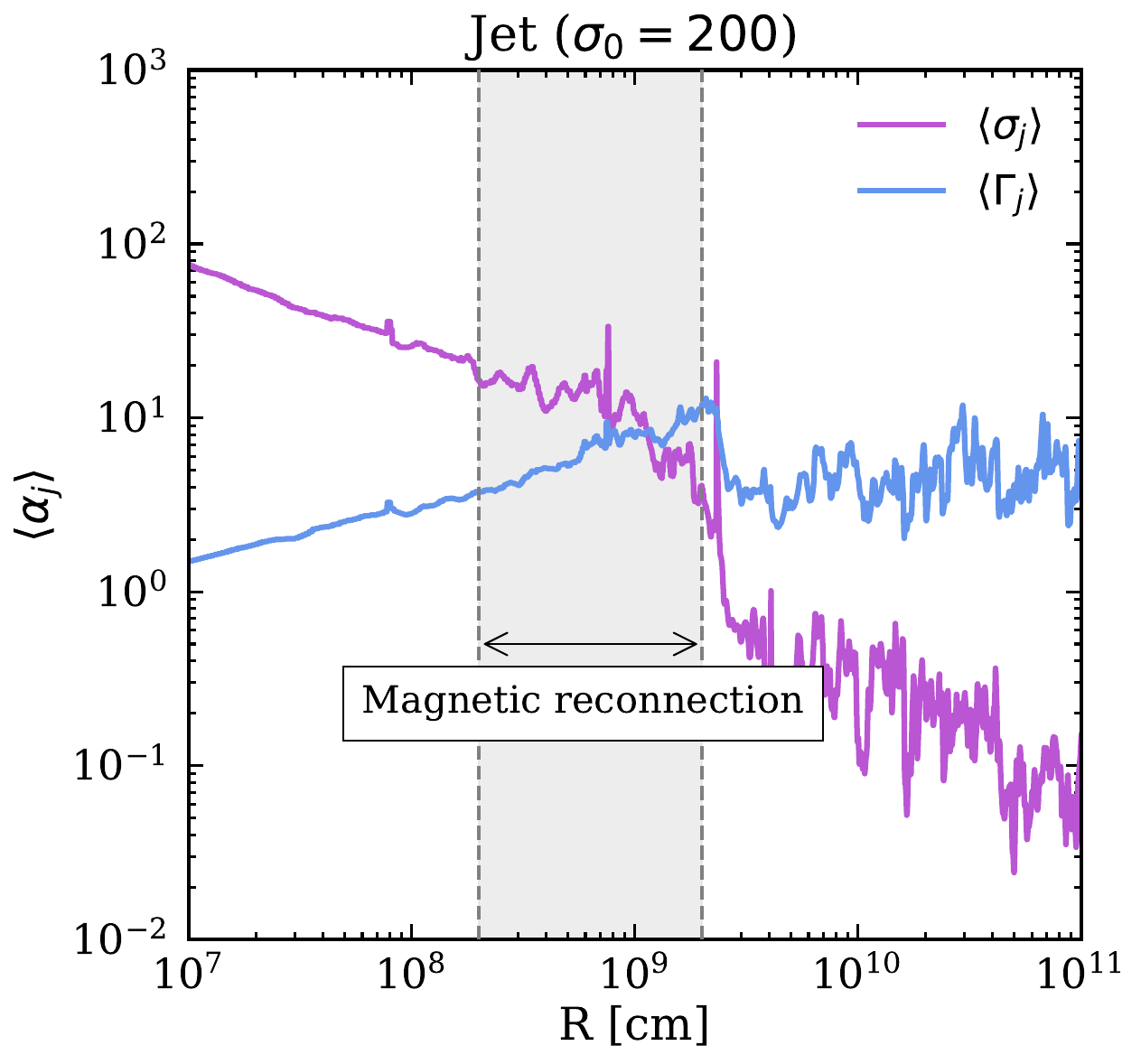}
  \caption{Radial profiles of the angle averaged magnetization $\langle \sigma_j \rangle $ and Lorentz factor $\langle  \Gamma_j \rangle $ in the jet, for $\sigma_0=15$ (left panel) and $\sigma_0=200$ (right panel), same as the top panels of Fig.~\ref{fig:profiles}. Magnetic reconnection  can efficiently occur  from $R \equiv \lambda \simeq 2 \times 10^8$~cm (namely the length scale over which the magnetic field inverts polarity) through  the radius such that $\langle \sigma_j \rangle \simeq 1$.
 This corresponds to the following radial ranges: $R \in [2 \times 10^8, 3.5 \times 10^8]$~cm for $\sigma_0=15$ (gray shaded region in the left panel) and $R \in  [2 \times 10^8, 2 \times 10^9]$~cm for $\sigma_0=200$ (gray shaded region in the right panel). Magnetic reconnection is not efficient for the jet with $\sigma_0=15$.} 
 \label{fig:magrec}
 \end{figure*}

The left panel of Fig.~\ref{fig:magrec}  shows the radial evolution of   the jet Lorentz factor and the magnetization for $\sigma_0=15$. One can see that  particle acceleration through magnetic reconnection can only occur over a very  narrow  radial range, since the jet magnetization drops to unity at $R \simeq 3.5 \times 10^8$~cm. We conclude that magnetic reconnection is therefore inefficient for $\sigma_0=15$.
On the other hand, the right panel of Fig.~\ref{fig:magrec} shows that  the jet Lorentz factor increases up to $R\simeq 2 \times 10^9$~cm for $\sigma_0=200$, where its magnetization approaches $\langle \sigma_j \rangle \simeq 1$.  This  hints that  magnetic energy is efficiently dissipated up to this radius, where $\langle \Gamma_j \rangle $ starts displaying an erratic behavior and the jet becomes mildly magnetized. Hence, magnetic reconnection can take place over the range $R \simeq 2 \times 10^8$--$2 \times 10^9$~cm, outlined with a gray shaded  band in Fig.~\ref{fig:magrec}. We warn the reader that the range of radii highlighted in Fig.~\ref{fig:magrec}  is sensitive to the initial magnetization of the jet: a larger $\sigma_0$ may stretch the region over which magnetic reconnection occurs, since the jet would reach $\langle \sigma_j \rangle \simeq 1$ at $R \gg 10^9$~cm. On the other hand, the erratic behavior of $\langle \Gamma_j \rangle$ could inhibit magnetic reconnection before the jet magnetization drops below unity. 

When protons are accelerated through magnetic reconnection, the proton energy distribution (Eq.~\ref{eq:proton}) depends on the outflow magnetization. The proton spectral index is parametrized as~\citep{Werner:2016fxe}~\footnote{We assume that the proton and the electron spectral indexes are the same: $k_p = k_e$. This result is motivated by particle-in-cell simulations of magnetic reconnection with $\sigma \gg 1$~\citep{Petropoulou:2019bse}, albeit it has yet to be proven for $\sigma \approx 1$~\citep{Guo:2014via}}:
\begin{equation}
k_p \approx 1.9+\frac{0.7}{\sqrt{\langle \sigma_j \rangle }} \ . \label{eq:kp}
\end{equation}
The fraction of dissipated energy stored in accelerated protons is~\citep{Werner:2016fxe}
\begin{equation}
\varepsilon_p = 1- \frac{1}{4} \left( 1 + \sqrt{\frac{\langle \sigma_j \rangle}{10 + \langle \sigma_j \rangle}} \right) \  , \label{eq:epsp}
\end{equation}
where $\sigma_j$ is shown in Fig.~\ref{fig:magrec}.
Finally, following Ref.~\citep{Pitik:2021xhb}, we assume that protons are accelerated with a minimum energy 
\begin{equation}
E^\prime_{p, \rm{min}} = m_p c^2 \max \left[ 1, \langle \sigma_j  \rangle \frac{\varepsilon_p}{2} \frac{k_p -2}{k_p -1} \right] \ .
\end{equation}

\subsection{Neutrino fluence from magnetic reconnection}
\begin{figure}[t]
\centering
\includegraphics[width=0.47\textwidth]{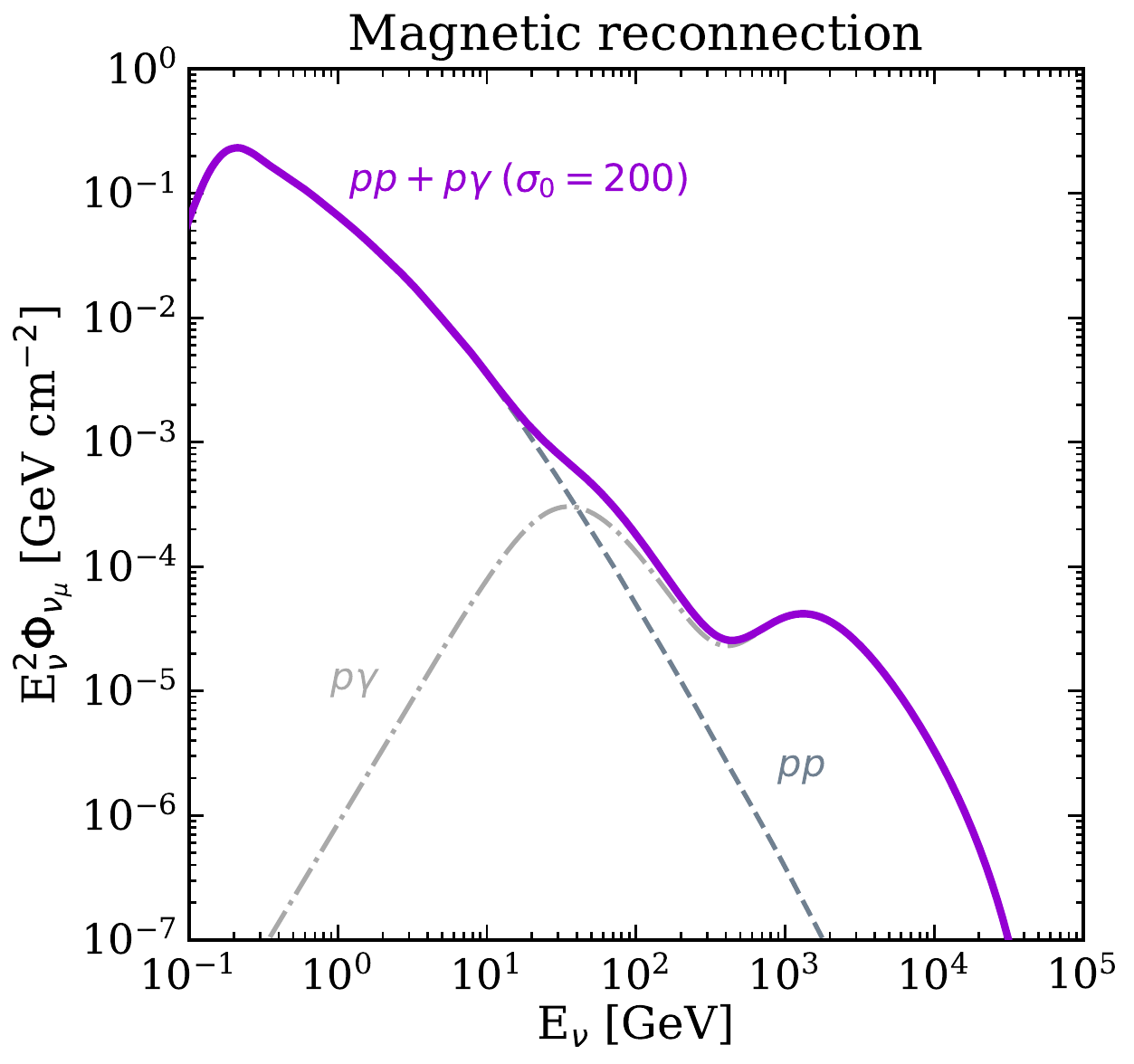}
\caption{Muon neutrino fluence on Earth for a collapsar jet at $z=2$ resulting from magnetic reconnection for our benchmark jet with $\sigma_0=200$. 
The spectral shape is determined by $p p$ interactions (dark gray dashed line) below $E_\nu \simeq 50$~GeV, and $p \gamma$ interactions (light gray dot-dashed line)  for $E_\nu \gtrsim 50$~GeV. 
The bump at $\simeq 5 \times 10^3$~GeV  is due to  kaon decay, the spectrum has a cutoff at $E_\nu \simeq 3 \times 10^4$~GeV. Magnetic reconnection is not efficient for the jet with $\sigma_0=15$ and therefore no neutrinos are produced.
}
\label{fig:nuMag}
\end{figure}
Figure~\ref{fig:nuMag}  shows the muon neutrino fluence originating from magnetic reconnection for our jet with  $\sigma_0=200$ (no neutrino production due to magnetic reconnection occurs for $\sigma_0=15$).
The neutrino distribution is determined by $p \gamma$ interactions for  $E_\nu \gtrsim 50$~GeV and  $pp$ interactions for $E_\nu \lesssim 50$~GeV. The bump in the high-energy tail of the energy distribution comes from  kaon decay, as expected due to the large magnetic fields and baryon densities along the jet, see Figs.~\ref{fig:profiles} and \ref{fig:magrec}.

The large density in the jet substantially limits the proton maximum energy, as discussed in Appendix~\ref{app:B}. Hence, the neutrino signal extends up to $E_\nu \simeq 3 \times 10^4$~GeV. 
We note that the proton spectral index in Eq.~\ref{eq:kp} becomes shallower as the radius increases, and  the corresponding proton number density decreases, causing a quick drop in the neutrino flux as the energy increases. 

\subsection{Internal sub-shocks}
\begin{figure*}[t]
\centering
\includegraphics[width=0.45\textwidth]{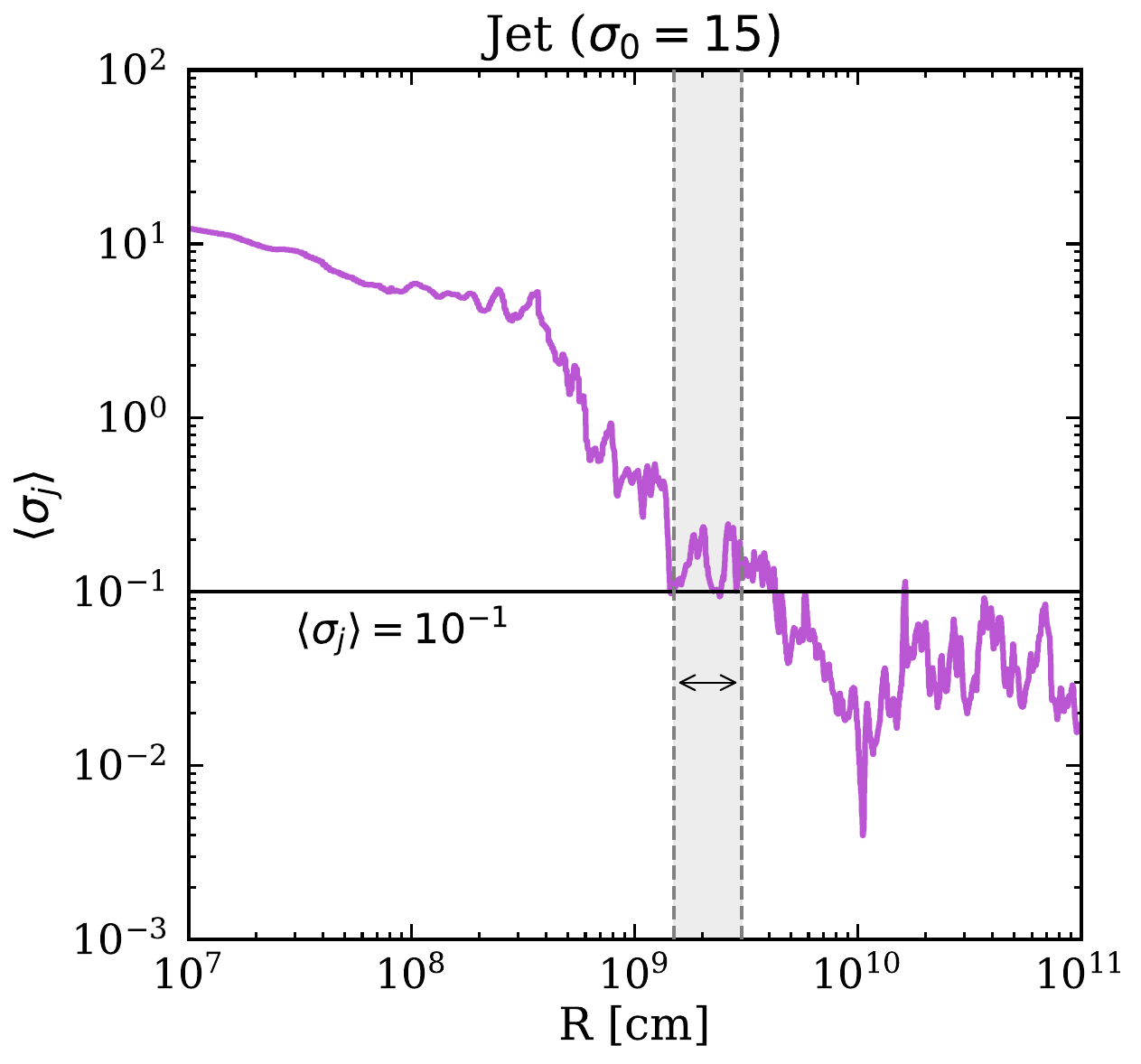}
\includegraphics[width=0.45\textwidth]{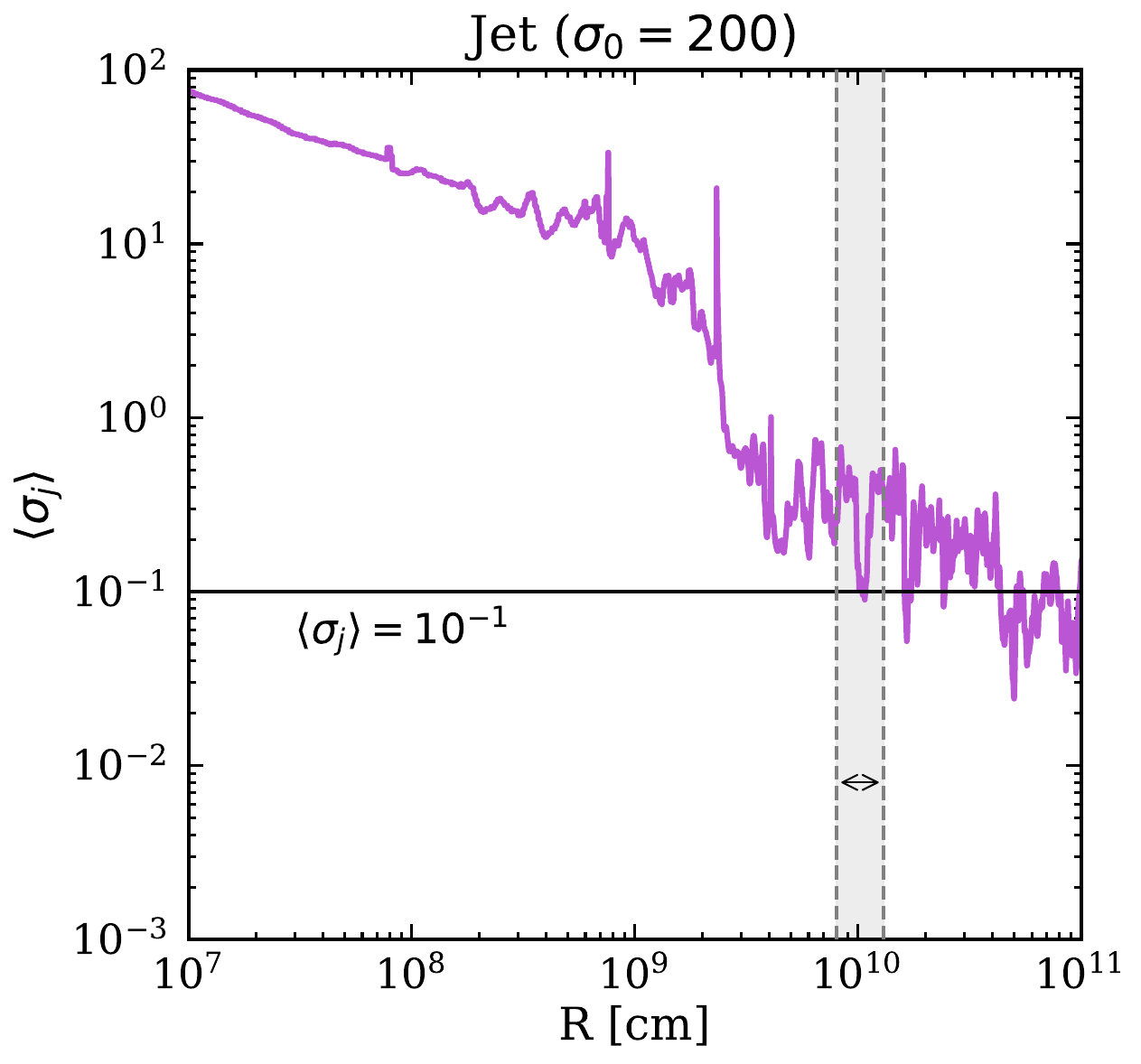}
\includegraphics[width=0.45\textwidth]{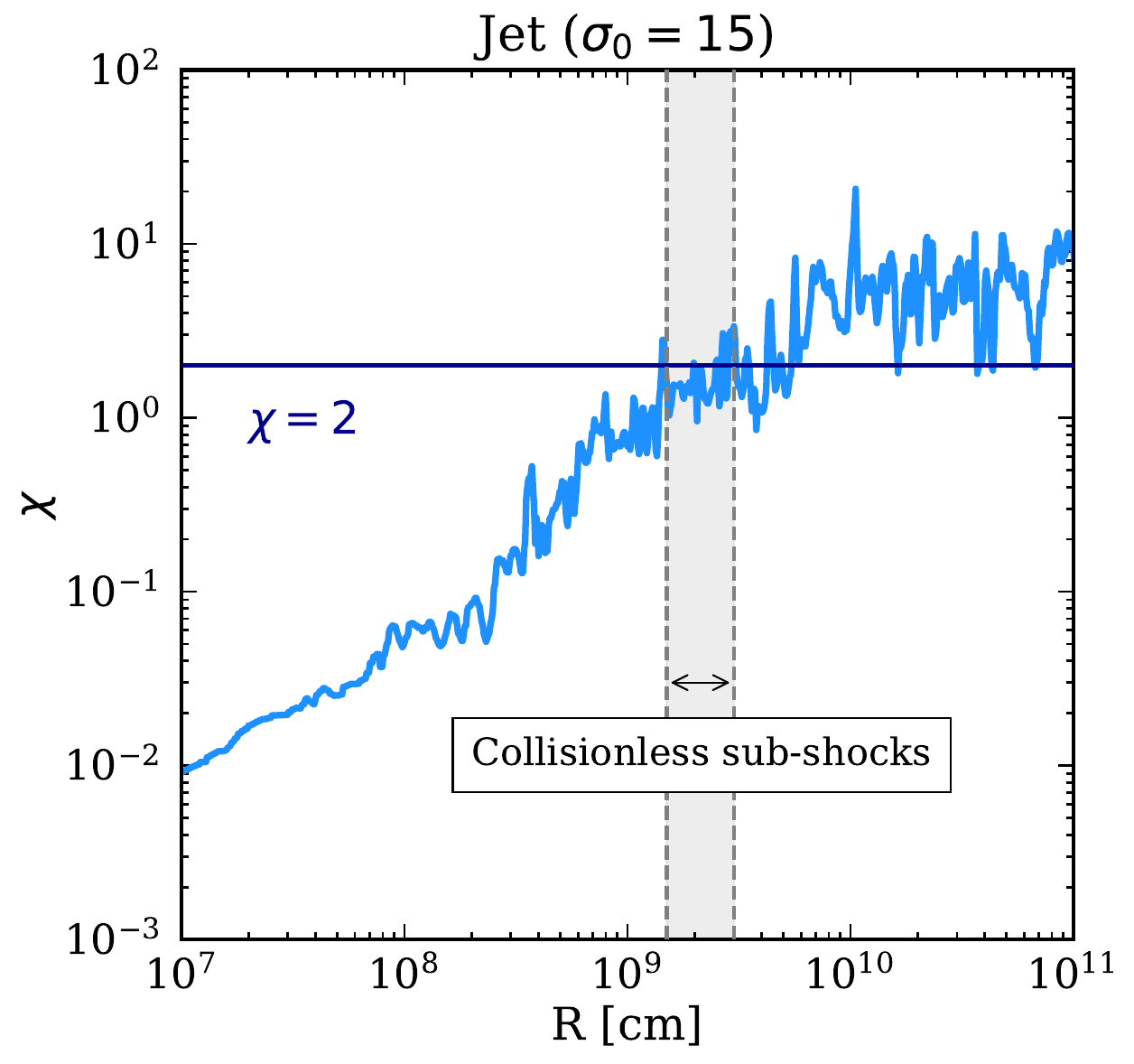}
\includegraphics[width=0.45\textwidth]{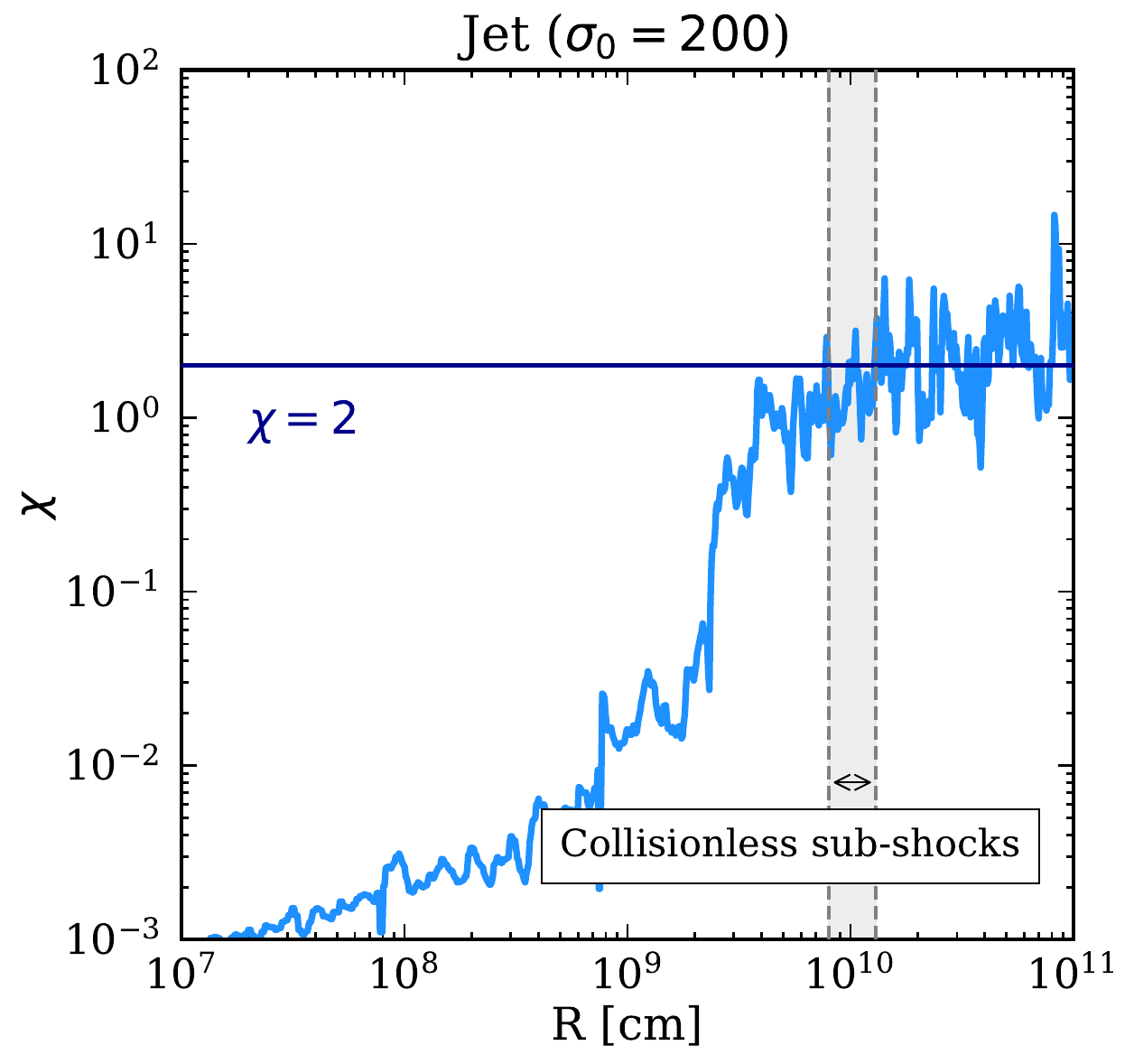}
\caption{Radial profiles of $\langle \sigma_j \rangle$ (top panels) and of the parameter $\chi$ (Eq.~\ref{eq:subshocks}, bottom panels) for $\sigma_0=15$ (left) and $\sigma_0=200$ (right). The black horizontal line in the top panels marks the condition $\langle \sigma_j \rangle =10^{-1}$, whereas the  blue horizontal line in the bottom panels marks the condition $\chi \lesssim 2$,  for which strong collisionless sub-shocks can occur within radiation mediated shocks. Collisionless sub-shocks can take place for  $ 1.5 \times 10^9 \; \rm{cm} \lesssim  R \lesssim 3 \times 10^9$~cm when $\sigma_0=15$ and $8 \times 10^9 \; \rm{cm} \lesssim R \lesssim 1.2 \times 10^{10}$~cm for $\sigma_0=200$; each of these radial regions is highlighted by  a shaded gray band.}
\label{fig:chi}
\end{figure*}
As mentioned in Sec.~\ref{sec:simulation}, the central engine powering the relativistic outflow exhibits intermittency on time scales $10$~ms~$\lesssim t_v \lesssim$~$100$~ms. It follows that the jet is not homogeneous, but it is made up of several shells of plasma moving with different velocities. These shells can collide at the internal shock  radius~\citep{Rees:1994nw}:
\begin{equation}
R_{\rm{IS}}= \frac{2 \langle \Gamma_j  \rangle^2 t_v c}{1+z} \ .
\end{equation}

Internal shocks can efficiently accelerate particles only if they are collisionless, namely when they are mediated by collective plasma instabilities, rather than  collisions~\citep{Levinson:2019usn}. Collisionless shocks can form within regions of the outflow that are optically thin (i.e.~where the Thompson optical depth is $\tau \lesssim 1$). Figure~\ref{fig:opticalDepth} shows that the outflow is highly optically thick  for $R \lesssim 4 \times 10^{11}$~cm. Hence,  even though the CO variability allows for the formation of internal shocks at $R_{\rm{IS}} \lesssim R_\star$,  {it is unlikely that these shocks are collisionless.}

If the jet is mildly magnetized ($\langle \sigma_j \rangle \gtrsim 10^{-1}$), however,  Ref.~\citep{Beloborodov:2016jmz} showed that collisionless sub-shocks may form within radiation mediated shocks when the following condition is fulfilled:
\begin{equation}
\chi \equiv \frac{p^\prime_{\rm{th}}}{p^\prime_{\rm{mag}}} \lesssim 2 \  .
\label{eq:subshocks}
\end{equation} 
In Eq.~\ref{eq:subshocks}, $p^\prime_{\rm{mag}}=\langle B_j^{\prime 2} \rangle /8 \pi$ is the comoving magnetic pressure, with $\langle B^\prime_j \rangle$ being the comoving magnetic field shown in Fig.~\ref{fig:TB}; $p^{\prime}_{\rm{th}}= (\hat{\gamma}-1) e^\prime_{\rm{th}}$ is the thermal pressure, related to the comoving internal energy of the outflow $e^\prime_{\rm{th}}= a \langle T^{\prime 4}_j \rangle$;   $\hat{\gamma}=4/3$ is the adiabatic index for an ideal polytropic fluid and $\langle T^{\prime}_j \rangle$ is the comoving temperature displayed in Fig.~\ref{fig:TB}. The radial profiles of $B^\prime_j$ and $T^\prime_j$ are displayed in Fig.~\ref{fig:TB}.  

Figure~\ref{fig:chi} shows the radial profiles of $\langle \sigma_j \rangle $ and $\chi$ (Eq.~\ref{eq:subshocks}); the horizontal lines mark the radii for which $\langle \sigma_j \rangle =10^{-1}$ and $\chi=2$, respectively. By combining the information in the top and bottom panels of Fig.~\ref{fig:chi}, we deduce that collisionless sub-shocks may occur within radiation mediated shocks for  $ 1.5 \times 10^9 \; \rm{cm} \lesssim  R \lesssim 3 \times 10^9$~cm when $\sigma_0=15$ and $8 \times 10^9 \; \rm{cm} \lesssim R \lesssim 1.2 \times 10^{10}$~cm for $\sigma_0=200$. 

As for protons accelerated at collisionless sub-shocks, we assume $k_p=2$, which is appropriate for mildly relativistic shocks~\citep{Sironi:2013ri}. The minimum energy of shock accelerated protons is $E^\prime_{p, \rm{min}}= m_p c^2 \simeq 1$~GeV.

 For  mildly relativistic sub-shocks, we rely on particle-in-cell simulations of collisionless shocks in electron-ion plasma and fix $\varepsilon_p=0.1$~\citep{Crumley:2018kvf}. We also assume  constant dissipation efficiency, $\varepsilon_d = 0.2$ for  mildly magnetized and mildly relativistic shocks~\citep{2010MNRAS.401..525M, Komissarov:2012hj}. Note that since the region where collisionless sub-shocks occur is rather small, we  rely on a  one-zone model~\citep{2010MNRAS.401..525M,2012MNRAS.422..326K}, even though $\varepsilon_d$ may depend on the details of the collision~\citep{Daigne:1998xc, Kobayashi:1997jk}. Hence,  we fix the sub-shock radius $R_{\rm{SS}} = 2.5 \times 10^9$~cm for $\sigma_0=15$ and $R_{\rm{SS}} =10^{10}$~cm for $\sigma_0=200$, where $ 0.1 \lesssim \langle \sigma_j \rangle \lesssim 1$ (see Fig.~\ref{fig:profiles}).

\subsection{Neutrino fluence from collisionless sub-shocks}
\begin{figure}[t]
\centering
\includegraphics[scale=0.4]{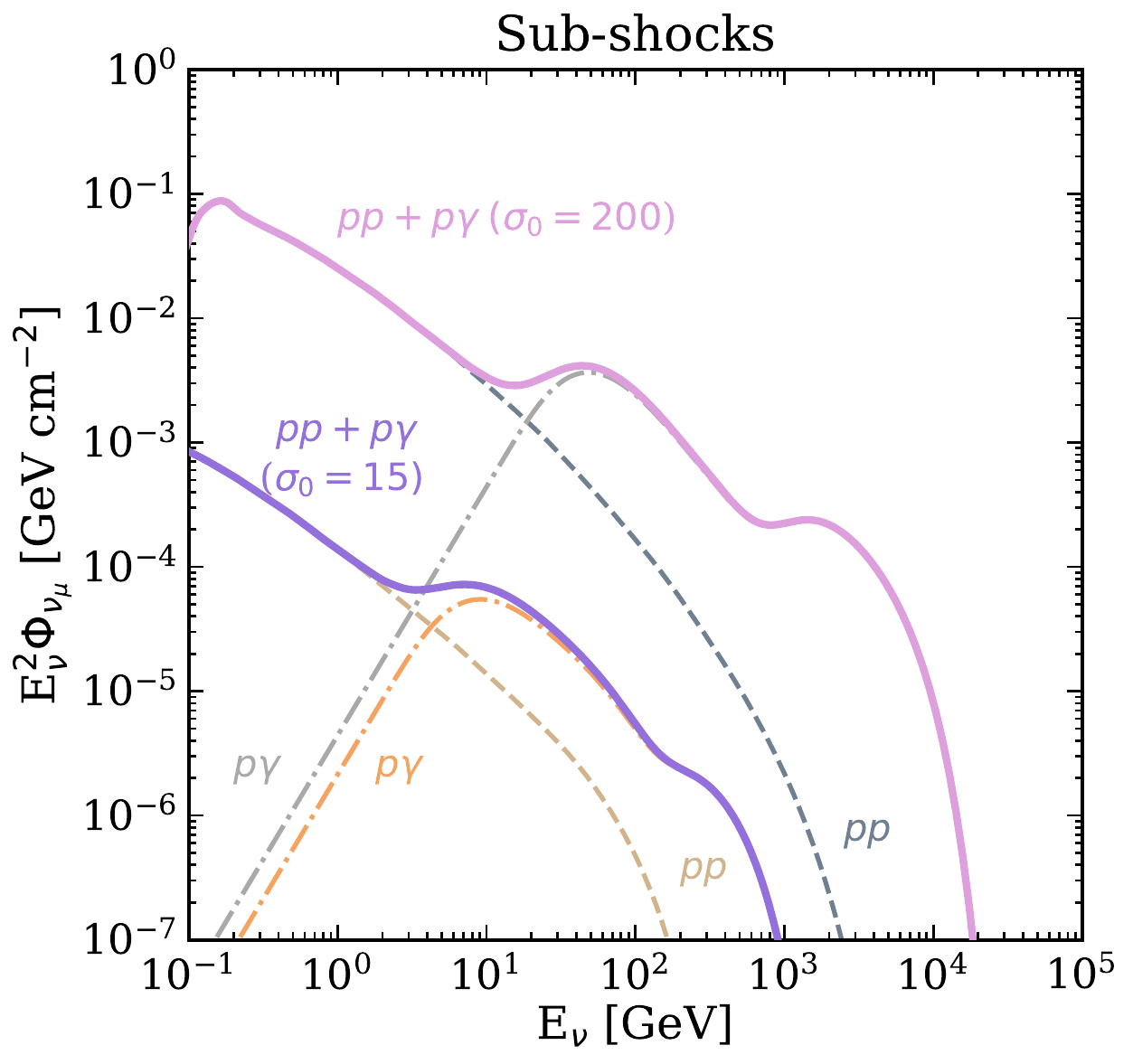}
\caption{Muon neutrino fluence on Earth for a collapsar jet at $z=2$ originating from collisionless sub-shocks for our benchmark jets with $\sigma_0=15$ (solid purple line) and $\sigma_0=200$ (solid orchid line).
For $\sigma_0=15$, $pp$ interactions (sand dashed line) dominate for $E_\nu \lesssim 1$~GeV, while $p \gamma$ (orange dot-dashed line) interactions shape the spectrum  for $E_\nu \gtrsim 10$~GeV. For $\sigma_0=200$, $pp$ interactions (dark gray dashed line) dominate for $E_\nu \lesssim 10$~GeV, while $p \gamma$ interactions (light gray dot-dashed line) are important for $E_\nu \gtrsim 10^2$~GeV. 
In both cases, the transition from the $pp$-dominated regime to the $p \gamma$-dominated one produces a bump in the neutrino spectral distribution. The second bump in the energy spectrum is due to  kaon decay. The neutrino signal is limited to energies $E_\nu \lesssim 10^3$~GeV ($E_\nu \lesssim 2 \times 10^4$)~GeV for $\sigma_0=15$ 
($\sigma_0=200$).}
\label{fig:nuSh}
\end{figure}

Figure~\ref{fig:nuSh} displays  the muon neutrino fluence from collisionless sub-shocks for our jets with $\sigma_0=15$ and $\sigma_0=200$. 
For $\sigma_0=15$, $pp$ interactions dominate the signal below $E_\nu \lesssim 1$~GeV, while $p \gamma$ interactions shape the spectrum  for $E_\nu \gtrsim 10$~GeV. For $\sigma_0=200$, $pp$ interactions are important  for $E_\nu \lesssim 10$~GeV, while $p \gamma$ interactions dominate above $E_\nu \gtrsim 10^2$~GeV. 
The first bump in the neutrino energy distribution is due to the  transition from the $pp$-dominated regime to the $p\gamma$-dominated one, while the second  bump (for $E_\nu \gtrsim 10^2$~GeV) is due to kaon decay. 
The  neutrino energy distribution has a cutoff at $E_\nu \lesssim 10^3$~GeV ($E_\nu \lesssim 2 \times 10^4$~GeV) for $\sigma_0=15$ ($\sigma_0=200$). 

The differences between the two initial magnetizations can be understood as follows. The neutrino production rate in the comoving frame obtained for $\sigma_0=200$ (see  Eqs.~\ref{eq:rate_neutrini}, \ref{eq:rate_mupp} and  \ref{eq:rate_epp})  is comparable to the one for $\sigma_0=15$. Nevertheless, the volume of the interaction region $V^\prime$ (Eq.~\ref{eq:neutrino_flux}) for $\sigma_0=200$  is larger than the one for $\sigma_0=15$, resulting in a larger fluence in the former case. Furthermore, the neutrino signal is boosted to higher energies for $\sigma_0=200$, due to the larger values of $\langle \Gamma_j \rangle$ reached in the jet; see Fig.~\ref{fig:profiles}.

\section{Outer subphotospheric particle acceleration sites} 
\label{sec:fatejets}
\begin{figure*}[t]
\centering
\includegraphics[width=0.9\textwidth]{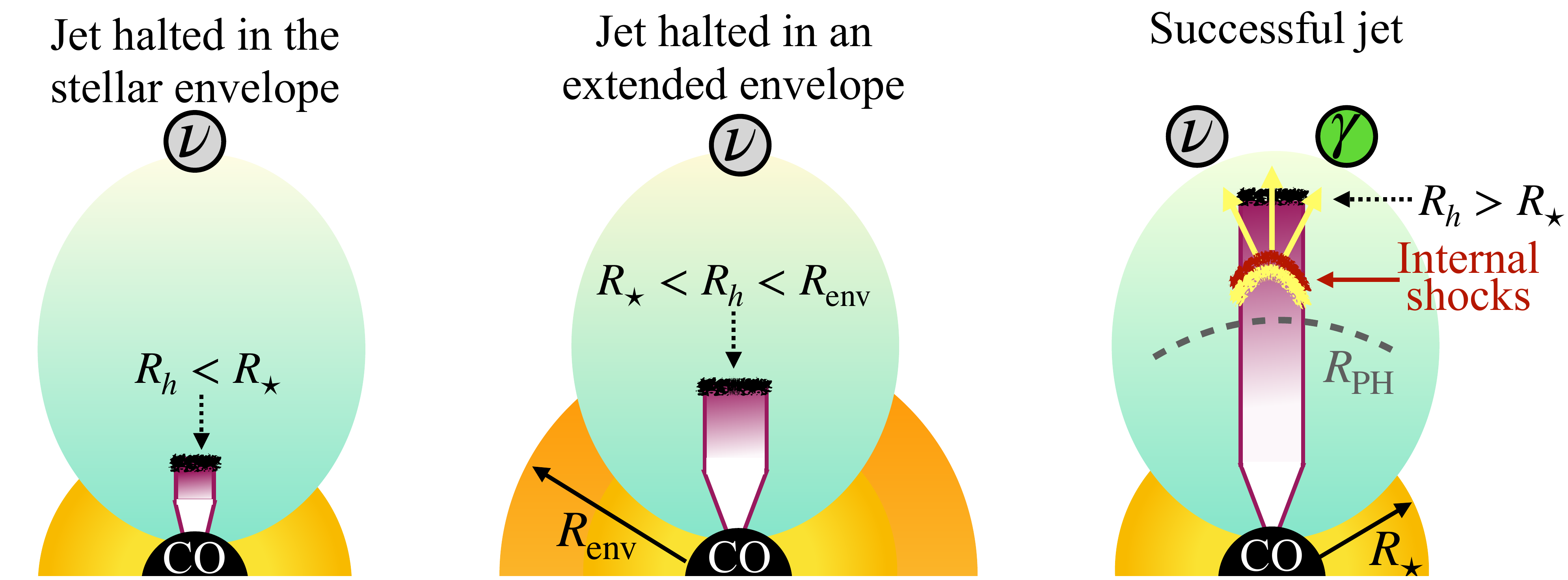}
\caption{Sketch of the fates of collapsar jets.  \textit{Left panel}:  The jet is halted in the stellar core ($R_h < R_\star$) at the end of the jet lifetime. The only particle acceleration sites are the ones displayed in Fig.~\ref{fig:sketch}. \textit{Middle panel}: The jet is halted in an extended outer envelope (dark-orange region) engulfing the star ($R_\star<R_h< R_{\rm{env}}$). If the jet is magnetized, the  acceleration sites are the ones drawn in Fig.~\ref{fig:sketch}.  \textit{Right panel}: The jet is successful and  breaks out from the stellar envelope (orange). The jet head  is above the stellar radius ($R_h > R_\star$) when the CO activity stops. 
The jet reaches the photosphere, where further energy dissipation and particle acceleration may take place.   In all three scenarios, the cocoon (aqua)  breaks out from the star and the extended envelope, if any. The case of a jet breaking out from the extended envelope is not considered, since this  is not supported by observations~\citep{Levan:2016jav}.}
\label{fig:jetfate}
\end{figure*}
 {In the context of subphotospheric particle acceleration, it is relevant to discuss unsuccessful jets,} which are smothered in the stellar envelope or within an extended envelope. A cartoon displaying the possible  {jet fate} is shown in Fig.~\ref{fig:jetfate}. We discuss the conditions that must be fulfilled in order to produce an unsuccessful jet and the relevant particle acceleration sites.  {Note that both in the case of successful and unsuccessful jets, the high-pressure cocoon breaks out from the star and  the extended envelope, if any. However, particle acceleration is not efficient at these sites; we refer the interested reader to Appendix~\ref{app:C} for an overview of the acceleration sites  in the cocoon.}

\subsection{Motivation}
A short-lived engine can generate a jet that does not break out from the stellar core and it is halted (see left panel of Fig.~\ref{fig:jetfate}). 
Another instance for which jets could be unsuccessful occurs when the progenitor star has not shed apart the hydrogen layer completely and retains an extended massive envelope engulfing its core, as sketched in the middle panel of Fig.~\ref{fig:jetfate}. This could happen, for example, for partially stripped supernovae~\citep{Gilkis:2021uht,Nakar:2015tma,Sobacchi:2017wcq}. If this is the case, even when the jet  breaks out from the stellar core, it fails to pierce through the external extended envelope. 
This scenario is of particular interest, since some GRBs or LFBOTs may harbor relativistic jets, which do not   break  out successfully. 

For example, it has been proposed that low- and high-luminosity GRBs share the same explosion mechanism, with the difference that low-luminosity GRB progenitors retain an extended low-mass envelope~\citep{Margutti:2014gha, Nakar:2015tma}. The envelope smothers the jet, which drives a mildly relativistic shock leading to a low-luminosity GRB. Choked jets could be harbored  within LFBOTs as well~\citep{Drout:2014dma, Arcavi:2015zie,Tanaka:2016ncv, DES:2018whm,Ho:2021fyb}. The lack of gamma-ray associations with known LFBOTs~\citep{Bietenholz:2019ptf} and the observation of broad hydrogen lines in some of their spectra~\citep{Perley:2018oky, Margutti:2018rri, Coppejans:2020nxp} may hint towards a jet smothered in the extended hydrogen envelope~\cite{Gottlieb:2022old}. The jet inflates the cocoon responsible for the observed electromagnetic emission in the optical and radio bands. The existence of a jet in LFBOTs would also explain the asymmetry observed in the outflow~\citep{Margutti:2018rri}. We refer the reader to Refs.~\citep{Fang:2020bkm, Guarini:2022uyp} for a discussion on the acceleration sites. 

Jets which manage to pierce through the extended envelope are not supported by observations: successful jets should produce GRBs, whose progenitors do not exhibit any hydrogen line in their spectra~\citep{Levan:2016jav}. Hence, this case is not of interest to our discussion.

Our benchmark simulations focus on   jets breaking out  from the stellar core, with no extended envelope  engulfing the progenitor star. Hence, the jet freely propagates up to its photosphere.
To date, numerical simulations tracking the   dynamics of magnetized jets that break out in an extended stellar envelope are lacking. However,  the outflow dynamics mimics the one of hydrodynamic jets above $R_\star$~\citep{Gottlieb:2022tkb}. Even though  numerical simulations would be required, we rely on previous work on hydrodynamic jets to investigate the propagation  {of our benchmark jets} in a massive envelope. Since the jet lifetime is not constrained by the simulation, we intend to explore the allowed parameter space and compute the value of  $\tilde{t}_j$ required for halting  {jets resembling the simulated ones} in the extended envelope. The goal of this section is to expand on the results of Sec.~\ref{sec:acceleration} to  unsuccessful jets.

\subsection{Conditions for halting the jet} 
We assume that the star has a  core of radius $R_\star$ and an envelope extending up to $R_{\rm{env}}$.
The stellar core is described by the following radial density profile~\citep{Matzner:1998mg, Harrison:2017jvs, Gottlieb:2022tkb}:
\begin{equation}
\rho_\star(R)= \rho_0 R^{-2} \left( 1- \frac{R}{R_\star} \right)^3 \ , 
\label{eq:init_dens}
\end{equation}
where the normalization constant is fixed by the stellar mass, namely $\rho_0= M_\star/ \left[ \int_0^{R_\star} dR^\prime 4 \pi R^{\prime^2} \rho_\star(R^\prime) \right]$. As for $M_\star$ and $R_\star$, we adopt the same values used in the simulation and listed in Sec.~\ref{sec:simulation}.

The radial density profile of the extended envelope is assumed to be~\citep{Nakar:2015tma}:
\begin{equation}
\rho_{\rm{env}}(R)= \rho_{0, \rm{env}} R^{-2} \label{eq:env_dens} \ ,
\end{equation}
where  $M_{\rm{env}}$ is its mass and $\rho_{0, \rm{env}}= M_{\rm{env}}/\left[\int_{R_\star}^{R_{\rm{env}}} dR 4 \pi R^2 \rho_{\rm{env}}(R) \right]$.
Inspired by partially stripped supernovae, we fix $R_{\rm{env}}=10^{13}$~cm and we consider two representative cases for the envelope mass: $M_{\rm{env}}=0.1 M_\odot$ and $M_{\rm{env}}=5 M_\odot$~\citep{Margutti:2014gha, Sobacchi:2017wcq, Gilkis:2021uht, Nakar:2015tma, Meszaros:2001vr}. 
Overall, the density profile of the star is parametrized as 
\begin{equation}
\rho(R)= \max \left[ \rho_\star(R), \rho_{\rm{env}}(R) \right] \ .
\label{eq:totalDens}
\end{equation}

The propagation of a hydrodynamic relativistic jet in dense media has been modeled analytically~\citep{2011ApJ...740..100B} and semi-analytically~\citep{Harrison:2017jvs}. In both cases, the jet dynamics is completely determined once its luminosity $\tilde{L}_j$, duration $\tilde{t}_j$, initial opening angle $\theta_j$, and the density profile of the medium $\rho(R)$ are fixed. Hence, in order to infer whether the jet is successful or not, we  follow the temporal evolution of its head  $R_h$. 

We stress that we  rely on hydrodynamic jets, generally different from the magnetically dominated jets considered so far. However, since the simulated jets become weakly magnetized above $R_\star$, this is a fair approximation.
The jet dynamics  is obtained by relying on the semi-analytical model presented in Ref.~\citep{Harrison:2017jvs} (we refer the interested reader to Ref.~\citep{Harrison:2017jvs} for details on the calculation). The model allows to calculate, at each time, the position of the jet head $R_h$, its proper velocity $\beta_h \Gamma_h$, and the breakout time $\tilde{t}_{\rm{BO, \star \rm{(env)}}}$ from $R_\star$ ($R_{\rm{env}}$). 

Since the jet head is relativistic, while propagating through the stellar envelope (Eq.~\ref{eq:totalDens}), the time over which the engine has to power the jet in order to allow for its breakout from the star (envelope) is
\begin{equation}
\tilde{t}_j= \tilde{t}_{\rm{BO}, \star \rm{(env)}} - \frac{R_{\star\rm{(env)}}}{c} \ .
\end{equation}
For a given pair $(\tilde{L}_j, \tilde{t}_j)$, when $\tilde{t}_j < \tilde{t}_{\rm{BO}, \star}- R_{\star}/c$ the jet is halted in the stellar core. If, instead, $  \tilde{t}_{\rm{BO}, \star}- R_{\star}/c < \tilde{t}_j < \tilde{t}_{\rm{BO, env}} - R_{\rm{env}}/c $, the jet breaks out from the stellar core, but it is halted in the envelope.

\begin{figure*}[t]
\hspace{-0.5cm}
\includegraphics[width=0.5\textwidth]{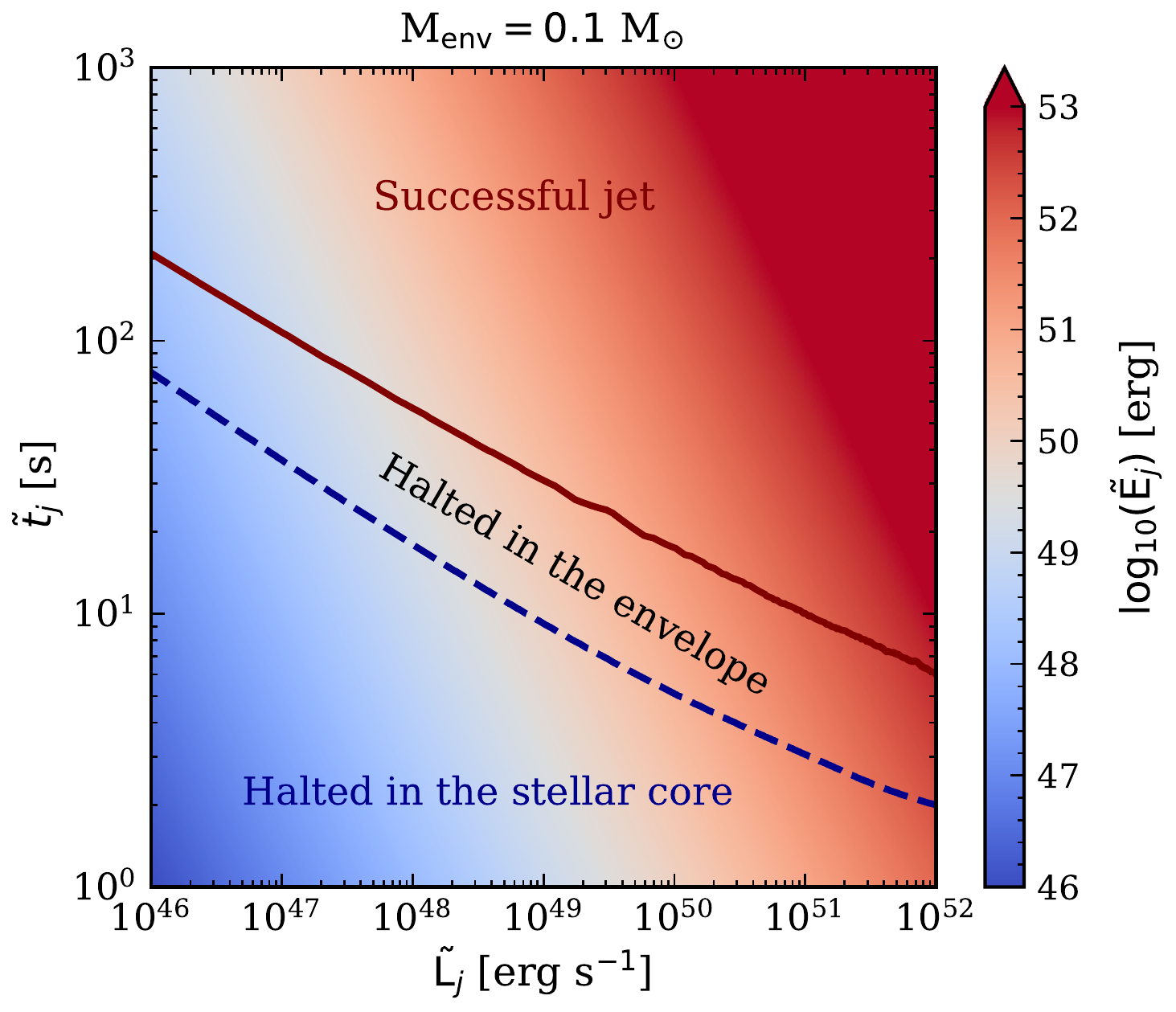}
\includegraphics[width=0.5\textwidth]{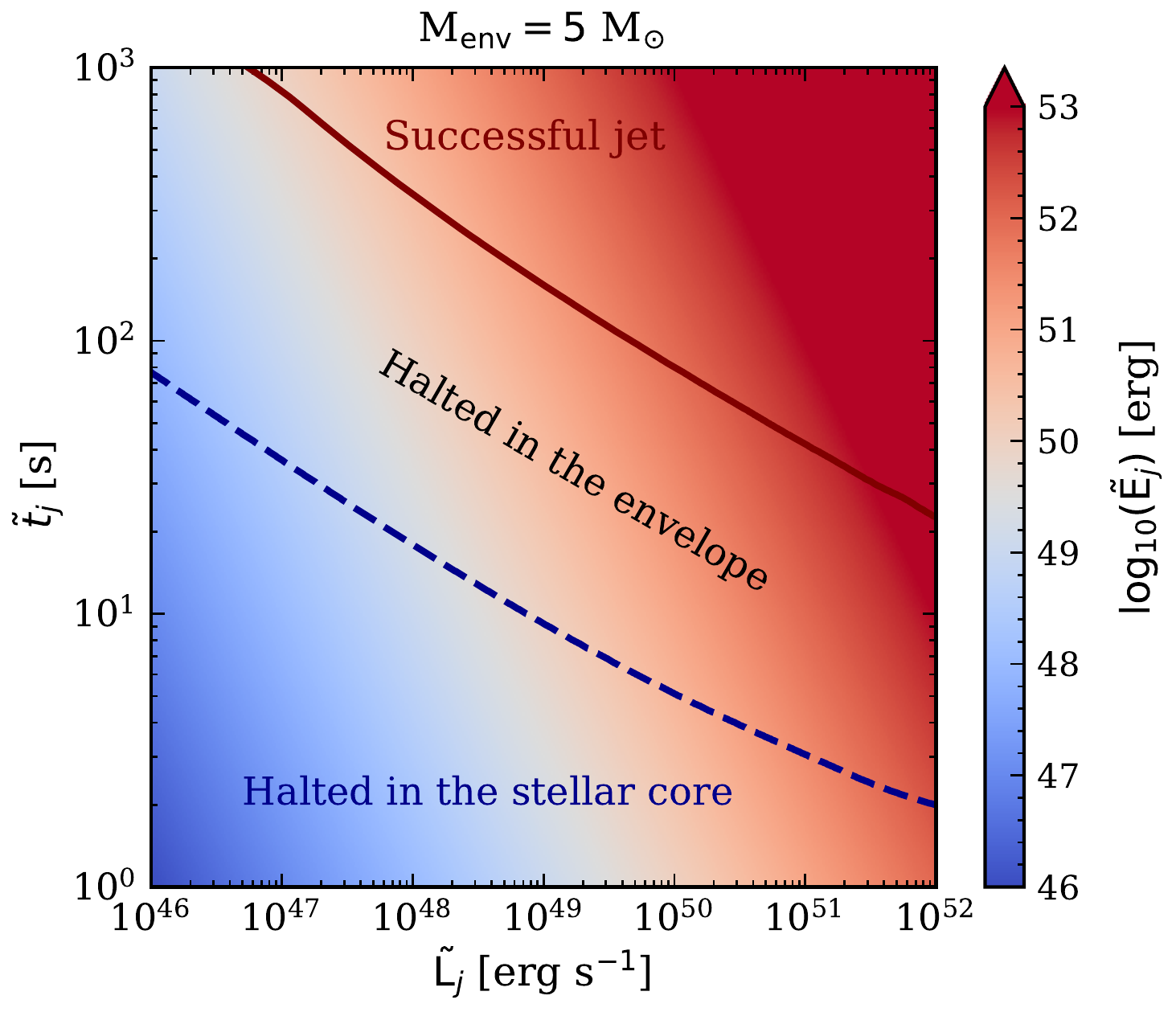}
\caption{Contour plot of the energy injected in the jet by the central engine ($\tilde{E}_j= \tilde{L}_j \tilde{t}_j$) in the plane spanned by the luminosity and engine lifetime. The jet propagates in an envelope with radius $R_{\rm{env}} = 10^{13}$~cm and mass  $M_{\rm{env}} = 0.1 M_\odot$ (left panel) or $M_{\rm{env}}=5 M_\odot$ (right panel). The region of the parameter space below the dashed blue line corresponds to jets halted within the stellar core, for which $R_h < R_\star$ at the end of the jet lifetime. The region above the solid brown line corresponds to successful jets, namely jets that breakout both from the stellar core and the extended envelope for which $ R_h > R_{\rm{env}}$. Between the blue and brown isocontours,  the pairs   $(\tilde{L}_j, \tilde{t}_j)$ lead to jets which breakout from the star, but are halted in the extended envelope, i.e.~$R_\star < R_h < R_{\rm{env}}$ at the end of the jet lifetime.}
\label{fig:contour}
\end{figure*}
Figure~\ref{fig:contour} shows the parameter space of the pairs $(\tilde{L}_j, \tilde{t}_j)$ corresponding to the same energy $\tilde{E}_j$ injected in the jet. The mass of the envelope is assumed to be $M_{\rm{env}}=0.1 M_\odot$ (left panel) and $M_{\rm{env}} = 5 M_\odot$ (right panel). The region below the dashed blue line corresponds to jets halted in the stellar core, i.e. $R_h< R_\star$ at the end of the jet lifetime. This region is not of interest for the reference simulations of  Ref.~\citep{Gottlieb:2022tkb}, since the jets are very energetic and likely to break out from $R_\star$ in any case.

The area between the dashed blue and solid brown lines in Fig.~\ref{fig:contour}  corresponds to jets halted in an extended envelope. In this case, at the end of the jet lifetime, $R_\star < R_h < R_{\rm{env}}$.  {Our simulated jets}, with total luminosity $\tilde{L}_j \simeq 5 \times 10^{51}$~erg s$^{-1}$ (at the time when the snapshots in Fig.~\ref{fig:profiles} are taken), break out from the star for $\tilde{t}_j \gtrsim 2$~s. The result is consistent with the simulations, since   the central engine is still active and powering the outflow at $2$~s.  {Our benchmark  jets} may be halted in the extended envelope if $\tilde{t}_j \lesssim 6$~s ($\tilde{t}_j \lesssim 25$~s), for $M_{\rm{env}}=0.1 M_\odot$ ($M_{\rm{env}} = 5 M_\odot$), and  we would not observe any jet-powered gamma-ray bursts.

Finally, the region above the brown line in Fig.~\ref{fig:contour} corresponds to jets able to drill out from the star, for which $R_h > R_{\rm{env}}$. As expected,  massive envelopes require long living engines in order to produce successful jets. Furthermore, for a fixed engine duration, jets  {less powerful than our simulated ones} are halted within the extended envelope more easily. 

\subsection{Neutrino production in unsuccessful jets}
From Fig~\ref{fig:contour}, we deduce that  jets can be unsuccessful only for some  $(\tilde{L}_j, \tilde{t}_j)$ pairs. 
Particle acceleration in unsuccessful jets has been discussed in the literature, both at the collimation shock~\citep{Murase:2013ffa} and at internal shocks occurring either in the outflow or at the jet head~\citep{Murase:2013ffa, Murase:2013hh,He:2018lwb,Fasano:2021bwq, Guarini:2022uyp,Tamborra:2015fzv}. These works rely on the criterion outlined in Ref.~\cite{Murase:2013ffa} for the formation of collisionless shocks and they all assume hydrodynamic jets. 

GRB like jets are expected to undergo  intense mixing  due to interactions with the cocoon~\cite{Gottlieb:2020raq}. Hence, the criterion proposed in Ref.~\citep{Murase:2013ffa}, which is given for idealized jets, has been shown to do not  be satisfied  {in regions of the jet still embedded in the stellar core ($R \lesssim R_\star$)} in numerical simulations, since the mixing slows down the jet and increases its baryon density~\citep{Gottlieb:2021avb,Gottlieb:2020raq}. Indeed, we find that the optical depth of the outflow is substantially larger than the one obtained from analytical estimations, see Fig.~\ref{fig:opticalDepth}. We conclude that particle acceleration at internal shocks occurring deep in the stellar core or at the collimation shock is  {disfavored}, contrary to what concluded  in Ref.~\citep{Murase:2013ffa} (see also the discussion in Ref.~\cite{Gottlieb:2021pzr}).

 {The picture above could change in the presence of a massive envelope surrounding the star, investigated in Refs.~\citep{He:2018lwb, Guarini:2022uyp, Senno:2015tsn}. Nevertheless, if the jet is magnetized, the results of Refs.~\citep{He:2018lwb, Guarini:2022uyp, Senno:2015tsn} may no longer hold, being the underlying jet dynamics  different and highly non-linear. The extended envelope would affect the jet dynamics above $R_\star$} and it may increase the effect of the jet-cocoon mixing,  {which would be relevant up to radii larger than the ones reached in the simulation. As a result, the outflow may become optically thin at $R_{\rm{PH}} \gg 10^{12}$~cm, possibly even above $R_{\rm{env}}$. Even though particle acceleration at internal shocks approaching the jet head seems  unlikely in magnetized jets, dedicated numerical simulations are desirable. Therefore, in the following, we limit our discussion to radii  below the edge of the envelope ($R \lesssim \mathcal{O}(10^{12})$~cm).}

This scenario could be relevant, for example, for neutrino production in LFBOTs~\citep{Gottlieb:2022old, Guarini:2022uyp}. In the case of a magnetized unsuccessful jet, the discussion on particle acceleration in Sec.~\ref{sec:acceleration} should apply.  
Since the energy of LFBOTs is expected to be smaller than the one obtained for the  jet simulations presented in Ref.~\citep{Gottlieb:2021pzr}, the overall normalization of the neutrino fluences  in Figs.~\ref{fig:nuMag} and \ref{fig:nuSh} should be affected. Yet, we expect the neutrino fluence produced from a magnetized unsuccessful jet to be limited to energies $E_\nu \lesssim 10^5$~GeV. Intriguingly, this signal would be very different from the one predicted for a hydrodynamic~\footnote{We stress that  we intend  to highlight the mechanism responsible for the jet launching through the wording ``hydrodynamic jet.''  Even though our magnetized jets resemble hydrodynamic ones after the breakout from the star, their evolution is different at the initial phase of the jet lifetime.} choked jet, which instead peaks at $E_\nu \simeq 10^5$~GeV~\citep{Guarini:2022uyp}. Hence, neutrinos could contribute not only to disentangle  the mechanism powering LFBOT sources---as suggested in Ref.~\citep{Guarini:2022uyp}---but also to discern the nature of unsuccessful jets.  {The signal calculated in Sec.~\ref{sec:acceleration} is typical of magnetized jets, while it is not expected from hydrodynamic jets, which are optically thick below $R_\star$ and do not have magnetization to sustain nor sub-shocks or magnetic reconnection~\citep{Matsumoto:2020lsw, Gottlieb:2020mmk,Gottlieb:2020ifs, Gottlieb:2020raq}.} 

Another outer particle acceleration site for hydrodynamic jets (or jets which mimic hydrodynamic ones) may be the shock which develops at the interface between the cocoon and the envelope. This shock becomes collisionless at the shock-breakout radius $R_{\rm{BO, env}}$ defined as
\begin{equation}
    \tau_{\rm{env}}(R_{\rm{BO, env}})= \int_{R_{\rm{BO, env}}}^{R_{\rm{env}}} dR \rho(R) k_{\rm{es} } = \frac{c}{v_{\rm{sh, env}}} \ . 
    \label{eq:shBO}
\end{equation}
where $\rho(R)$ is given by Eq.~\ref{eq:totalDens}, $v_{\rm{sh, env}}$ is the speed of the shock and $k_{\rm{es}}$ is the electron scattering opacity. Here we adopt $k_{\rm{es}}=0.34$, assuming solar abundances~\citep{Pan:2013nfa}. Since the cocoon fastest component moves with mildly relativistic velocities ($\langle \Gamma_c \rangle \lesssim 2$), it enters the envelope with a mildly relativistic shock, i.e.~$ {c}/ {v_{\rm{sh, env}}} \simeq 1$. 
 {Our goal is to assess whether there is a part of the parameter space for which $R_{\rm{BO, sh}} \ll R_{\rm{env}}$. This case would resemble the propagation of a mildly-relativistic shock in the circumstellar medium, see e.g.~Ref.~\cite{2013ApJ...769L...6K}, but acceleration of particles would start deep in the envelope rather than at its edge and would occur over a wide range of radii inside the star. Efficient particle acceleration when $R_{\rm{BO, env}} \simeq R_{\rm{env}}$ is possible~(see, e.g., Refs.~\cite{Nakar:2015tma, Gottlieb:2021pzr}), but we do not further investigate this case  since it is beyond the main focus of this work.}
\begin{figure}
\centering
\includegraphics[width=0.5\textwidth]{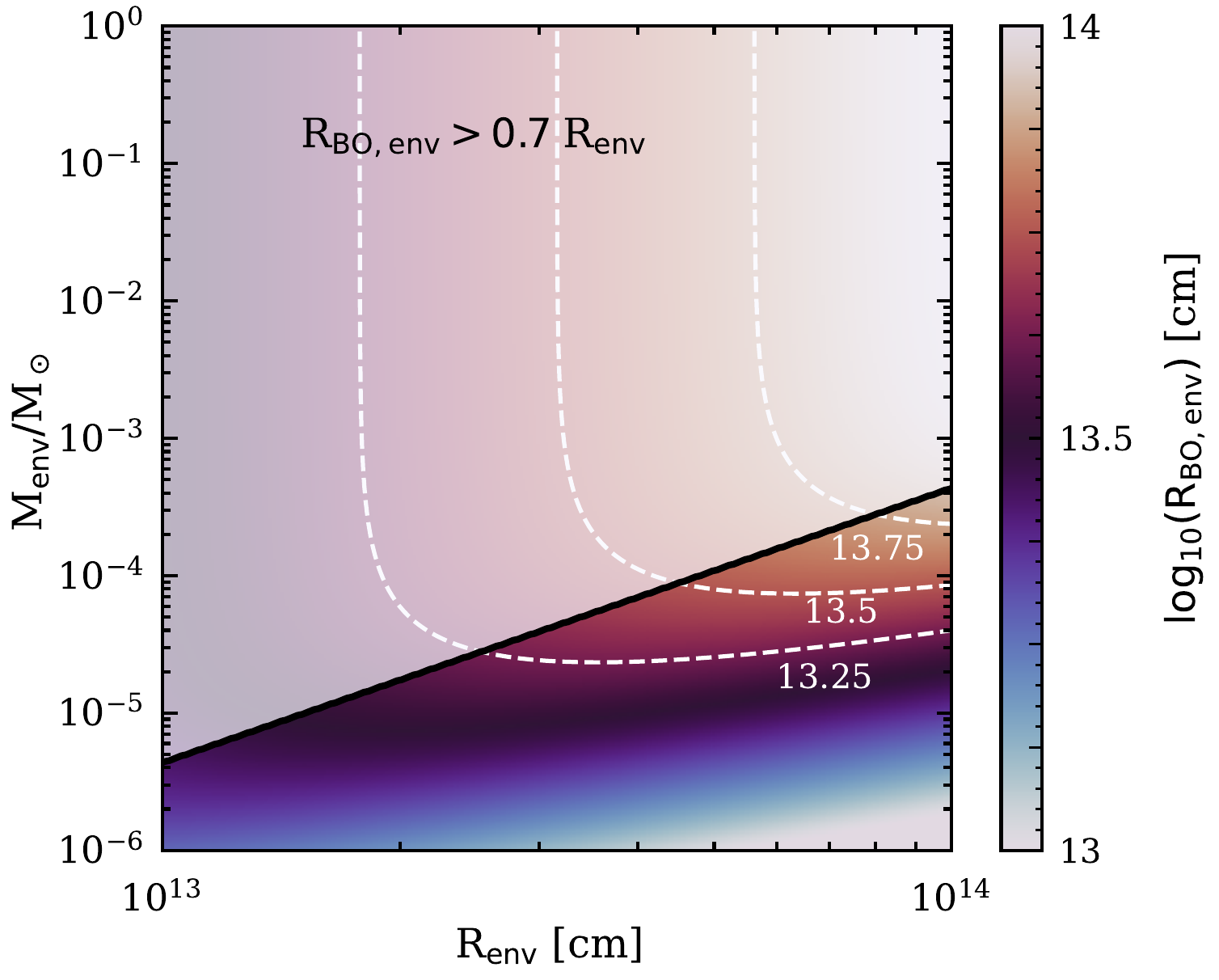}
\caption{Contour plot of the shock-breakout radius at the envelope $R_{\rm{BO, env}}$ (Eq.~\ref{eq:shBO}) in the plane spanned by the radius $R_{\rm{env}}$ of the envelope and its $M_{\rm{env}}$. The dashed white lines display representative values of $\log_{10} \left( R_{\rm{BO, env}} \right)$ to guide the eye. The shadowed region corresponds to pairs $(R_{\rm{env}}, M_{\rm{env}})$ for which the envelope becomes optically thin close to its edge, for $R_{\rm{BO, env}} > 0.7 \; R_{\rm{env}}$. The shock-breakout occurs deep in the envelope only for the right bottom corner of the parameter space, for which $M_{\rm{env}}< 5 \times 10^{-4} M_\odot$ and $R_{\rm{env}} \gg 10^{13}$~cm; these parameters are quite unusual and would require fine tuning of the initial conditions of the jet for halting it within the envelope. We conclude that in most cases the shock break out occurs very close to the edge of the envelope.}
\label{fig:breakoutEnv}
\end{figure}

Figure~\ref{fig:breakoutEnv} shows the parameter space of the  $(R_{\rm{env}}, M_{\rm{env}})$ pairs having the same shock-breakout radius $R_{\rm{BO, env}}$, as defined in Eq.~\ref{eq:shBO}. For most of the envelope masses $M_{\rm{env}}$ and radii $R_{\rm{env}}$, the shock-breakout occurs very close to the edge of the envelope, in particular at  $R \gtrsim 0.7 R_{\rm{env}}$. The breakout could occur at smaller radii only for envelopes with large extension ($R_{\rm{env}} \gg \times 10^{13}$~cm) and small masses $(M_{\rm{env}} \lesssim 5 \times 10^{-4} M_\odot)$, as visible from the right bottom corner of the parameter space in Fig.~\ref{fig:breakoutEnv}. Such envelopes are not dense enough to halt relativistic jets and are   poorly motivated theoretically~\cite{Margutti:2014gha, Sobacchi:2017wcq, Gilkis:2021uht, Nakar:2015tma, Meszaros:2001vr}. 

Combining the results from Figs.~\ref{fig:contour} and~\ref{fig:breakoutEnv}, we deduce that small envelope masses  require fine tuning of the jet lifetime and luminosity to  {simultaneously} smother the jet  {and allow for neutrino production in the range $R_\star \lesssim R \lesssim R_{\rm{env}}$}. Therefore, particle acceleration at the shock between the cocoon of unsuccessful jets and the envelope is either inhibited or it occurs in a very narrow range of radii, making it a subleading process for neutrino production  {in the region $R \ll R_{\rm{env}}$.} 

We conclude that, if a magnetized jet is halted in the extended envelope, neutrino production is possible at the sites discussed in Sec.~\ref{sec:acceleration}. For instance,  if the simulated jets were to breakout from the stellar core in an envelope with $R_{\rm{env}} = 10^{13}$~cm and $M_{\rm{env}}= 5 M_\odot$, for the fixed lifetime ${t}_j = 10$~s,  the  neutrino fluence from magnetic reconnection processes and collisionless sub-shocks would be the same as the one displayed in Figs.~\ref{fig:nuMag} and~\ref{fig:nuSh}, respectively, with the  results being sensitive to the initial magnetization of the jet. As for jets which are hydrodynamically launched and choked in the extended envelope, neutrino production may occur at the sites discussed in Refs.~\citep{He:2018lwb, Fasano:2021bwq, Guarini:2022uyp}.  {It is still to be proven whether further particle acceleration can occur in magnetized unsuccessful jets at the same sites, namely at $R_{\rm{IS}} \simeq R_{h} \lesssim R_{\rm{env}}$.}

If the jet head is halted in the extended envelope at the position $R_h$, the neutrino signal produced at the acceleration sites discussed in Sec.~\ref{sec:acceleration} can be attenuated because of neutrino propagation in matter between $R_h$ and $R_{\rm{env}}$. The attenuation factor for the  neutrino fluence  scales approximately as $ f_{\rm{att}} \simeq \exp[ - \int_{R_h}^{R_{\rm{env}}} \rho(R)/(2 m_p) \sigma_{\nu}^{\rm{CC}} (E_\nu)]$, where $\rho(R)$ is given in Eq.~\ref{eq:totalDens} and $\sigma_{\nu}^{\rm{CC}}$ is the cross section for neutrino-charged current interactions which  is the dominant process in the GeV--TeV energy range of interest~\citep{Formaggio:2012cpf}. Attenuation  is relevant when $f_{\rm{att}} \ll 1$; for the density profile in Eq.~\ref{eq:totalDens}, we find that this condition is fulfilled for  $E_\nu \gtrsim 100$~TeV, i.e.~it is negligible for the scenarios investigated in this paper. Neutrino flavor conversion may also occur in choked jets~\citep{Mena:2006eq, Razzaque:2009kq, Carpio:2020app},  nevertheless for our  collapsar scenarios  the flavor composition at Earth is not substantially altered~\citep{Sahu:2010ap}. Further attenuation of the neutrino signal may be caused by the increase of the jet-cocoon mixing in the presence of a massive envelope, which cannot be analytically estimated.
Hence, the results presented in Sec.~\ref{sec:acceleration} for the subphotospheric neutrino signal expected on Earth still shall be interpreted as an upper limit for a  {magnetized} jet halted in an extended envelope.

\section{Expected subphotospheric neutrino emission}\label{sec:fluence}
By relying on the findings of Secs.~\ref{sec:acceleration} and \ref{sec:fatejets}, in this section we present the total fluence expected for subphotospheric neutrinos  produced in collapsar jets. We also compare our finding with the existing literature.  {Our results are sensitive to the underlying reference simulations. Yet they urge to move towards a more robust modelling than the one provided by analytical treatments.}
\subsection{Neutrino fluence}
\begin{figure}[t]
\centering
\includegraphics[width=0.47\textwidth]{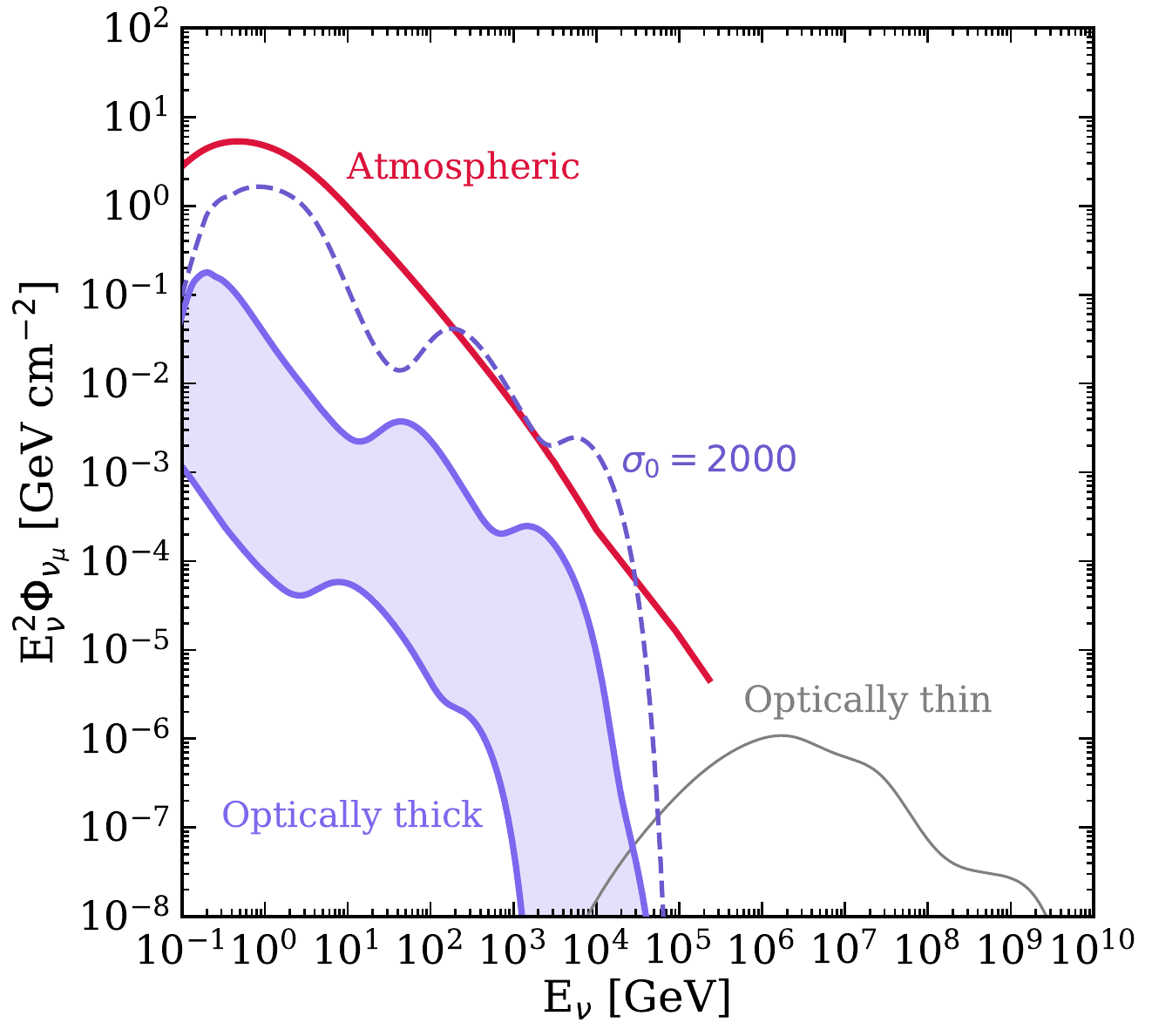}
\caption{Muon neutrino fluence on Earth for a collapsar jet at $z=2$. The purple band represents the range of variability of the subphotospheric neutrino production (optically thick region); the lower limit  corresponds to the fluence obtained for $\sigma_0=15$ (as displayed in Fig.~\ref{fig:nuSh}), while the upper limit is obtained for $\sigma_0=200$ (see Figs.~\ref{fig:nuMag} and \ref{fig:nuSh}). The purple dashed line corresponds to the neutrino fluence expected for $\sigma_0=2000$; see main text for details. For comparison, we show the benchmark muon neutrino fluence from the optically thin region (above the photosphere) of a successful collapsar jet, namely a GRB  {(see Appendix~\ref{app:D})}. The red line represents the  atmospheric background expected during the jet lifetime~\citep{Super-Kamiokande:2015qek, IceCube:2015mgt, IceCube:2014slq}. The neutrino signal  in the optically thick region of the outflow extends up to $E_\nu \simeq 4 \times 10^4$~GeV ($E_\nu \simeq 10^3$~GeV) for $\sigma_0=200$ ($\sigma_0=15$) and it lies below the atmospheric background.  For  $\sigma_0=2000$, the neutrino signal extends up to $E_\nu \lesssim 7 \times 10^4$~GeV and it is comparable in intensity to the atmospheric background.}
\label{fig:fluence}
\end{figure}
 Figure~\ref{fig:fluence} shows the total subphotospheric muon neutrino fluence, where the lower limit is set by $\sigma_0=15$ and the upper limit by $\sigma_0=200$. In the former case, only internal sub-shocks are a viable mechanism for neutrino production, since the magnetization along the jet is not large enough to sustain magnetic reconnection; see Fig.~\ref{fig:nuSh}. In the latter scenario, both  sub-shocks and magnetic reconnection contribute to shape the neutrino energy distribution from the optically thick region; see Figs.~\ref{fig:nuMag} and  \ref{fig:nuSh}. 
The neutrino fluence  has a cutoff  at $E_\nu \simeq 4 \times 10^4$~GeV ($E_\nu \simeq 10^3$~GeV) for $\sigma_0=200$ ($\sigma_0=15$). This is due to the large baryon density in the outflow, which substantially limits the maximum energy at which protons can be accelerated. 

As pointed out in Ref.~\citep{Gottlieb:2022tkb},  GRB jets may have  initial magnetization larger than the ones considered in this paper ($\sigma_0 \gtrsim 1000$) in order to reach the observed Lorentz factors of a few hundreds. Because of numerical limitations, jet simulations with such large $\sigma_0$ are not yet available. Nevertheless,  we extrapolate the radial profiles of the jet characteristic quantities ($\langle \rho^\prime_j \rangle$, $\langle \sigma_j \rangle$, $\langle \Gamma_j \rangle$)  for a relativistic jet with  $\sigma_0 = 2000$ by assuming a constant  scaling ratio on the basis of the simulations with $\sigma_0= 15$ and  $\sigma_0=200$ (see Fig.~\ref{fig:profiles}), while the temperature is kept unchanged. The corresponding  neutrino fluence increases up to one order of magnitude compared to the one obtained for $\sigma_0=200$, as shown in Fig.~\ref{fig:fluence} (dashed purple line). Yet, the  larger baryon density and magnetic field in the jet are such that  the neutrino spectrum extends up to energies $\lesssim 7 \times 10^{4}$~GeV.
While this result should be interpreted as  an order of magnitude computation and may change if it were to be obtained by relying on self-consistent jet simulations, it  provides a good insight on what to expect.

For comparison,  the   neutrino fluence produced above the photosphere (optically thin region) in the case of a successful jet is also shown in Fig.\ref{fig:fluence}; see also Appendix~\ref{app:D}. We compute this fluence by assuming that  the target photon energy distribution is shaped by a dissipative photosphere and internal shocks occur above the photosphere, as discussed in Appendix~\ref{app:D}. The photospheric efficiency of the jet is $\epsilon_{\rm{PH}} \simeq 0.1$, which is the fraction of the jet isotropic energy energy emerging from the photosphere. The radiative efficiency at the photosphere is obtained by solving the hydrodynamic equations for the fireball model, within the assumption that the jet is almost hydrodynamic; see e.g. Ref.~\citep{Gottlieb:2019aae}. Our benchmark simulations hint that $\epsilon_{\rm{PH}} \gtrsim 10 \%$ could be reached for jets with $\sigma_0 \gtrsim 1000$. All other jet parameters follow the ones adopted in Ref.~\citep{Pitik:2021xhb}, chosen  to match GRB observations (see Ref.~\citep{Pitik:2021xhb} and references therein). We can see that the neutrino fluence from the optically thin region has a lower overall normalization, but it  extends up to $E_\nu \simeq 10^9$~GeV. We stress that this result is only shown to favor a direct comparison between the subphotospheric neutrino signal and the one produced above the jet photosphere, if the jet is successful.

In Fig.~\ref{fig:fluence}, we also show the expected fluence of atmospheric muon neutrinos during the jet lifetime~\citep{Super-Kamiokande:2015qek, IceCube:2015mgt, IceCube:2014slq}. Our neutrino fluence from the optically thick region of the outflow lies below the atmospheric background both for $\sigma_0=200$ and $\sigma_0=15$, while it becomes  comparable to the atmospheric one for a jet launched with $\sigma_0=2000$.

\subsection{Comparison with  existing literature}
Our findings are in contrast with existing literature. In fact, under the assumption of  collisionless internal shocks taking place in parts of the jet deeply embedded in the stellar envelope, Refs.~\citep{Razzaque:2004yv, Murase:2013ffa,Ando:2005xi,Tamborra:2015fzv} conclude that  TeV--PeV neutrinos could be produced. The main difference with our work is that the aforementioned papers overlooked the role of jet-cocoon mixing, underestimating the optical depth of the outflow; we find that shocks in the innermost parts of the jet are  {likely} radiation mediated when the role of mixing is consistently accounted for in the jet dynamics~\citep{Gottlieb:2022tkb}.  {As mentioned in Sec.~\ref{sec:simulation}, low baryon densities may be allowed if the jet accelerates at small radii to large Lorentz factors. This might favor acceleration of particles through internal shocks~\citep{Murase:2013ffa}. Nevertheless, such large Lorentz factors seem to be disfavored from state-of-the-art numerical simulations of collapsar jets.}

Our results are in  agreement with Ref.~\citep{Gottlieb:2021pzr}, which investigated the neutrino production at internal sub-shocks in the optically thick region of short GRBs, by relying on the outputs of  numerical simulations artificially  launching the jet. Yet, the self-consistent jet launching of our benchmark jet simulations~\citep{Gottlieb:2022tkb} affects the jet fate.

Intriguingly, subphotospheric production of  neutrinos in the same energy range displayed in Fig.~\ref{fig:fluence} can occur if collisional heating is considered as the mechanism responsible for energy dissipation in collapsar jets~\citep{Bartos:2013hf, Zegarelli:2021vuf}. In this scenario, neutrinos are produced through neutrino-proton interactions along the outflow.

\section{Detection prospects}\label{sec:results}
The subphotospheric neutrino fluence shown in Fig.~\ref{fig:fluence} spans an energy range below $\mathcal{O}(100)$~TeV, where the  IceCube Neutrino Observatory is most sensitive to astrophysical neutrinos. Hence, contrarily to the conclusions drawn in Refs.~\citep{Murase:2013ffa, Denton:2018tdj,Tamborra:2015fzv,Ando:2005xi,Razzaque:2003uv}  {for hydrodynamic jets}, unsuccessful  {magnetized} jets cannot contribute to the diffuse  neutrino flux detected by the IceCube Neutrino Observatory~\citep{IceCube:2013low,IceCube:2016umi}. These conclusions might change if the jet should be halted in an extended envelope and neutrino production should take place close to the jet head, at $R \lesssim 10^{13}$~cm~\citep{Senno:2015tsn, Fasano:2021bwq, He:2018lwb}.  {As extensively discussed in Sec.~\ref{sec:fatejets}, we expect this scenario to be  unlikely for magnetized jets.}

The  detection of subphotospheric neutrinos is hampered by the  atmospheric neutrino flux, as discussed in Sec.~\ref{sec:fluence}; however, we  investigate whether   astrophysical neutrinos could be discriminated from the atmospheric background by exploiting the directionality of the incoming astrophysical neutrinos. To this purpose, we rely on  the Hyper-Kamiokande neutrino detector~\citep{Hyper-Kamiokande:2018ofw} and IceCube DeepCore, designed to detect neutrinos with energy as low as $E_\nu \simeq 10$~GeV~\citep{IceCube:2011ucd}. 

As for Hyper-Kamiokande, the event directionality can be reconstructed by relying on the  elastic scattering of neutrinos on electrons: $\nu_\alpha + e^- \rightarrow \nu_\alpha + e^{-} \; (\bar{\nu}_\alpha + e^- \rightarrow \bar{\nu}_\alpha + e^{-})$~\footnote{Note that in this case  we need to distinguish between neutrinos and antineutrinos, since the respective cross-sections are different.}. The dominant contribution to the $\nu_\alpha e^{-}$ elastic scattering channel comes from the electron flavor, while the contribution from  muon or tau flavors is subleading 
(see e.g.~Refs.~\citep{Formaggio:2012cpf,Valera:2022ylt} for a review). Hence, we only consider $\alpha= e$ at Hyper-Kamiokande. 
The total number of subphotospheric neutrino events is~\citep{Abe:2011ts}
\begin{eqnarray}
\label{eq:hk}
N_{\nu_e+ \bar{\nu}_e}(z) &=& \epsilon N_e \int d E_\nu  [ \sigma_{\nu_e + e^{-}}(E_\nu) \Phi_{\nu_e}(E_\nu, z)  \\ 
&+& \sigma_{\bar{\nu}_e + e^{-}}(E_\nu)  \Phi_{\bar{\nu}_e}(E_\nu, z) ] \ , \nonumber
\end{eqnarray}
where $\sigma_{\nu + e^{-}}$ [$\sigma_{\bar{\nu} + e^{-}}$] is the cross-section for the neutrino (antineutrino)--electron elastic scattering~\citep{Formaggio:2012cpf} and $\epsilon$ is the detector efficiency, which we optimistically  assume to be $1$. The total number of electron targets is $N_e = 1.13 \times 10^{34}$ for a water Cherenkov detector with a fiducial volume of $0.188$~Mton~\citep{Hyper-Kamiokande:2022smq}. The number of atmospheric neutrino events is calculated through Eq.~\ref{eq:hk}, by using the neutrino atmospheric flux in Refs.~\citep{Super-Kamiokande:2015qek, IceCube:2015mgt, IceCube:2014slq}. 

For  IceCube-DeepCore, the total number of subphotospheric neutrino events is 
\begin{equation}
N_{\nu_\mu+ \bar{\nu}_\mu} = \int_{10 \; \rm{GeV}}^{100 \; \rm{GeV}} dE_\nu A_{\rm{eff}}(E_\nu) \Phi_{\nu_\mu+ \bar{\nu}_\mu} (E_\nu, z) \ ,
\end{equation}
where $A_{\rm{eff}}$ is the energy-dependent effective area of the detector~\citep{IceCube:2011ucd}. The rate of atmospheric neutrinos in the $10$--$100$~GeV range is obtained from Ref.~\citep{Wiebusch:2009jf}. 

\begin{figure*}[t]
\centering
\includegraphics[width=0.47\textwidth]{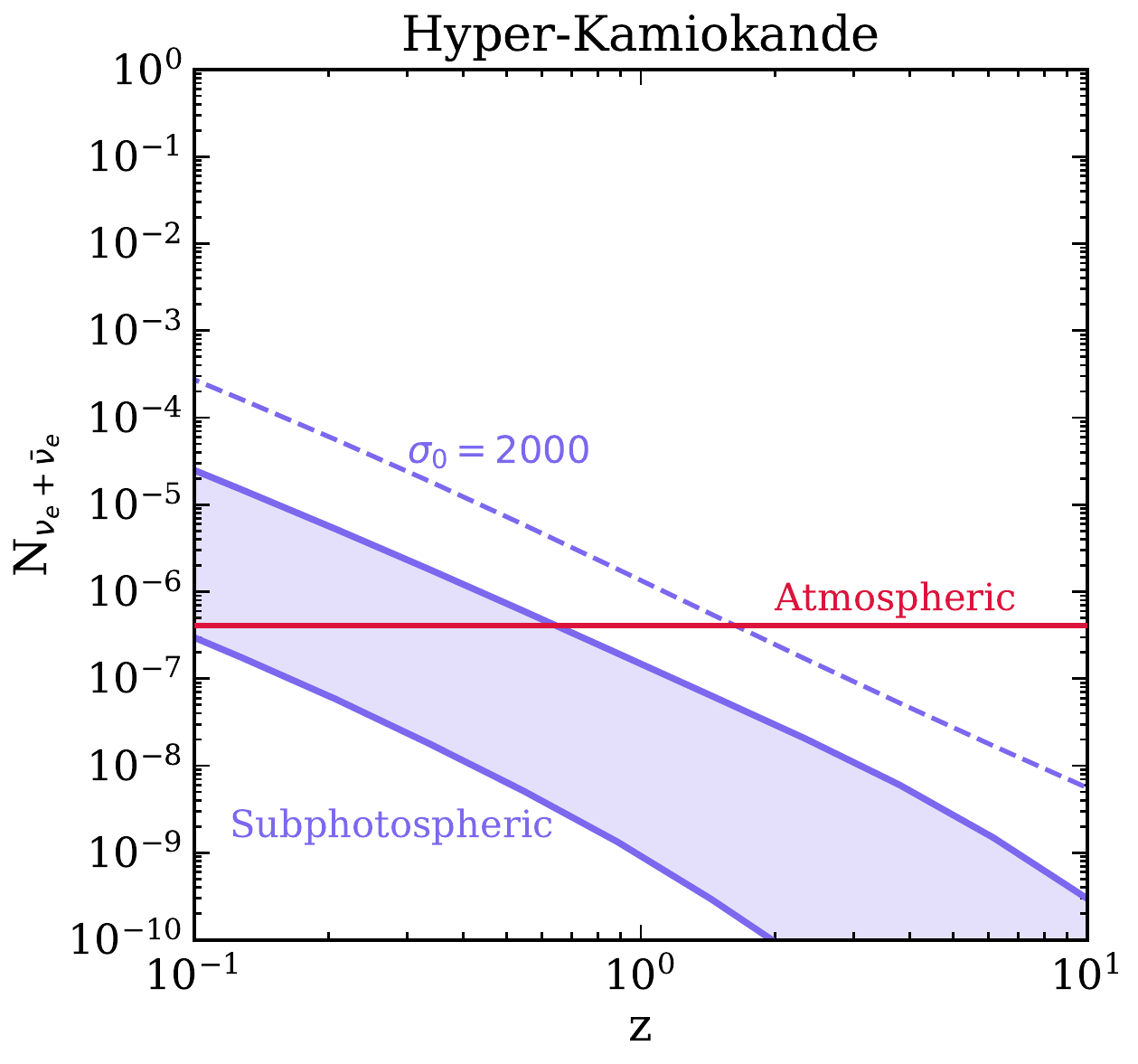}
\includegraphics[width=0.47\textwidth]{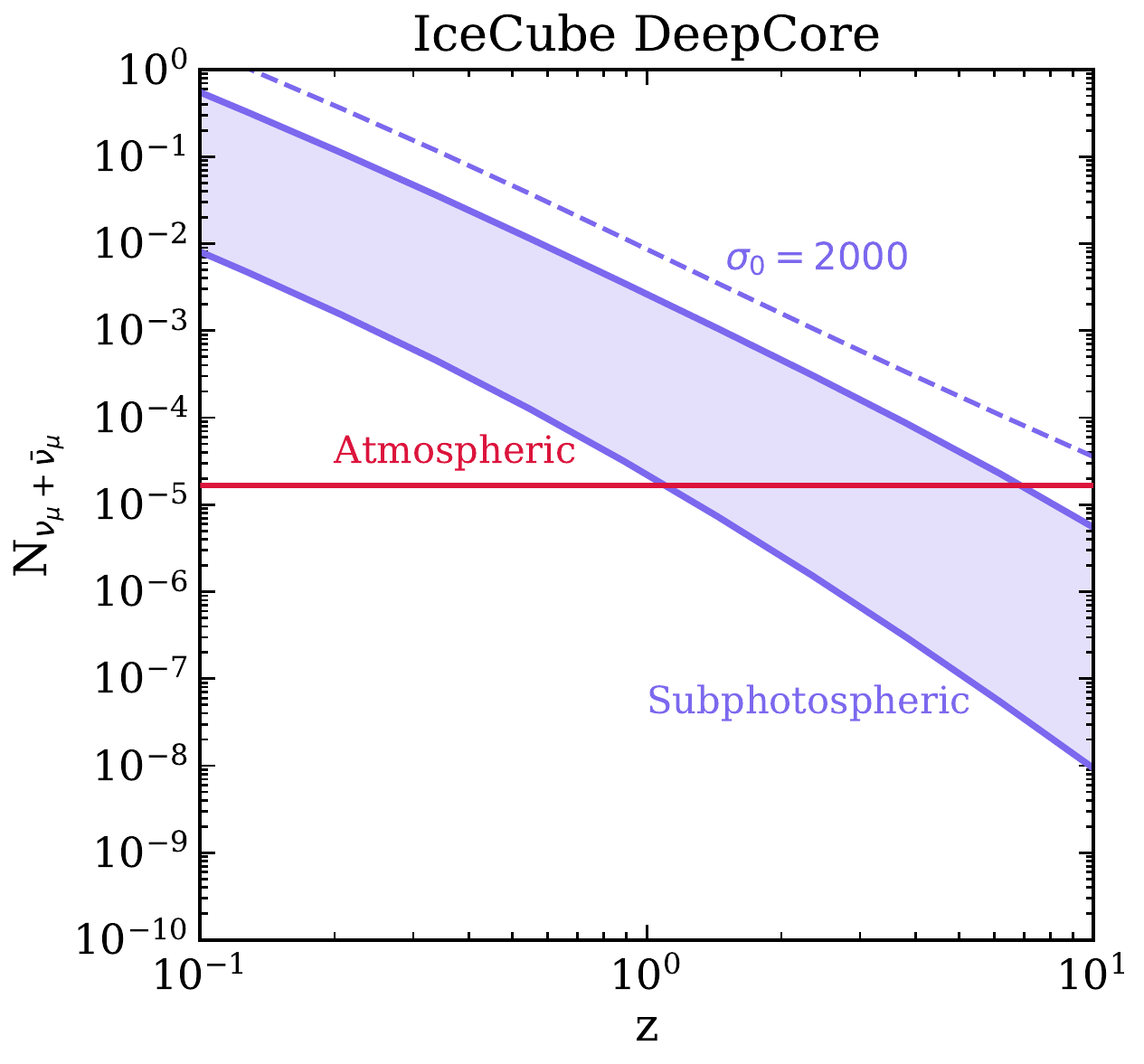}
\caption{Number of subphotospheric neutrino events  (purple band) expected at Hyper-Kamiokande (left panel) and IceCube DeepCore (right panel). The upper and lower solid lines of each band correspond to $\sigma_0=200$ and $\sigma_0=15$, respectively. We also show the expected number of neutrinos extrapolated for a jet with  $\sigma_0=2000$ (dashed purple line). For comparison,  the background of atmospheric neutrino events is plotted (solid red line). The number of suphotospheric neutrino events is larger than the  atmospheric ones in Hyper-Kamiokande, if the source is placed at $z\lesssim 0.8$ ($z \lesssim 0.1$) for a jet with $\sigma_0=200$ ($\sigma_0=15$). While for  IceCube DeepCore, this should happen  for a jet located at  $z \lesssim 7$. As for the initial magnetization $\sigma_0=2000$, the number of suphotospheric neutrino events is larger than the  atmospheric ones  for $z \simeq 2$ for Hyper-Kamiokande and for $z \gtrsim 10$ for IceCubeDeep Core. }
\label{fig:detection}
\end{figure*}
Figure~\ref{fig:detection} shows the total number of subphotospheric neutrino events expected  at Hyper-Kamiokande (on the left) and IceCube DeepCore (on the right) as a  function of the source redshift, for our benchmark jets. For comparison,  the number of atmospheric neutrino events is also plotted in Fig.~\ref{fig:detection}. We can see that the number of events  expected at Hyper-Kamiokande is significantly lower than the one observable at IceCube DeepCore, due to the smaller  
cross-section. 

The number of events from subphotospheric neutrinos  would be larger than the atmospheric neutrino number of events  at Hyper-Kamiokande for a jet at $z \lesssim 0.8$ ($z \lesssim 0.1$) for $\sigma_0=200$ ($\sigma_0=15$). On the other hand, in principle, the astrophysical signal may be  larger than the atmospheric one for sources at  $z \lesssim 7$ at IceCube DeepCore. 
As for the jet with initial magnetization $\sigma_0=2000$, the astrophysical signal becomes comparable to the atmospheric one at $z \simeq 2$ ($z \gtrsim 10$) in Hyper-Kamiokande (IceCube DeepCore).

The detection of $N_{\nu_\alpha + \bar\nu_\alpha} \gtrsim 1$  might be possible if the magnetized collapsar jet is located within $z \lesssim 0.1$ for all $\sigma_0$'s considered in this work. These findings  
are in agreement with Ref.~\citep{Zegarelli:2021vuf}, that investigated the detection of neutrinos in a similar energy range for collisionally heated GRBs. 
While the detection horizon is limited, the existence of bright GRBs at fairly moderated redshift is not ruled out---see, e.g.,  GRB 221009A which occurred at $z \simeq 0.151$~\cite{2022GCN.32648....1D}. {These detection prospects   may further  improve with the upcoming IceCube Upgrade~\citep{Ishihara:2019aao, IceCube:2019pna}, expected to lead to a more accurate event reconstruction in the energy region where IceCube DeepCore is sensitive.} {Moreover, as also pointed out in Ref.~\citep{Zegarelli:2021vuf}, stacking searches of collapsar jets could enhance the detection chances of subphotospheric neutrinos;  dedicated forecast work on  stacking searches is left as future task, as it would require a large set of GRMHD simulations.}

\section{Conclusions} \label{sec:conclusions}
A realistic modelling of relativistic jets and their related particle production is not only relevant for interpreting electromagnetic observations of a growing number of astrophysical transients, but also for investigating the associated high-energy neutrino production. 
While analytical treatments hold in the optically thin region of the outflow, they are no longer adequate to describe the evolution of the jet in the optically thick regime. This is due to the large jet--cocoon mixing revealed in a  range of numerical simulations, both for hydrodynamic and magnetized jets.

In this work, we explore neutrino production in the optically thick region of relativistic  jets by relying on the numerical simulations carried out in Ref.~\citep{Gottlieb:2022tkb}, with  initial magnetization  $\sigma_0=15$ and $\sigma_0=200$. As the jet propagates through the star, it efficiently converts magnetic  into kinetic energy. While the formation of collisionless shocks  {seems to be disfavored} due to the large optical depth, the mild magnetization $\langle \sigma_j \rangle \lesssim 0.1$ reached at  $R \simeq 10^9$--$10^{10}$~cm could sustain the formation of collisionless sub-shocks for both $\sigma_0$. The intrinsic magnetization of the jet may also trigger magnetic reconnection events, especially for jets with $\sigma_0=200$. Hence, both sub-shocks and magnetic reconnection are viable particle acceleration mechanisms.

Our findings reveal that the subphotospheric neutrino signal spans an energy range with  $E_\nu \lesssim 10^{4}$~GeV ($10^{5}$~GeV) for $\sigma_0=15$ ($\sigma_0=200$). This  result also holds for larger initial magnetizations of the jet, e.g. $\sigma_0=2000$, for which we extrapolate the dynamics from the two simulated jets with lower $\sigma_0$. The maximum neutrino energy is limited due to the large baryon density and high magnetic field, which causes the cooling of secondary mesons and it is consistent with the results of Ref.~\citep{Gottlieb:2021pzr}. These findings are in contrast with the ones previously  reported in Refs.~\citep{Razzaque:2004yv,Murase:2013ffa,Tamborra:2015fzv,Ando:2005xi}, where TeV---PeV neutrinos were produced in the  star under the assumption of  collisionless internal shocks, which we show  {are unlikely} because of the large  optical depth of the outflow determined by  the jet-cocoon mixing. 

If the jet is halted in an extended envelope engulfing the progenitor star, the same conclusions concerning neutrino production hold,  {if the jet is magnetized.} 
We find that no particle acceleration can occur at the shock developing at the cocoon front as it propagates in the extended envelope, unless the properties of the envelope and the jet are fine-tuned. Extreme conditions, which are not physically motivated, are required to simultaneously halt the jet and allow for particle acceleration. 

Because of their low energies  {and based on our benchmark simulated jets}, subphotospheric neutrinos  {from magnetized jets unlikely} contribute to the  high-energy diffuse neutrino flux observed by the IceCube Neutrino Observatory, contrarily to what suggested in the literature~\citep{Murase:2013ffa}. 
Yet, we investigate the detection perspectives in the upcoming  water Chereknov detector Hyper-Kamiokande and IceCube DeepCore. The subphotospheric signal could be  discriminated by  the atmospheric background by exploiting the directional information of the astrophysical neutrinos, with the expected number of neutrino events being  larger than the atmospheric one for a jet located at $z \lesssim 0.8$ ($z \lesssim 0.1$) for $\sigma_0=200$ ($\sigma_0=15$) in Hyper-Kamiokande and $z \lesssim 7$ ($z \lesssim 1$) for $\sigma_0=200$ ($\sigma_0=15$) in IceCube DeepCore.

Our results might not hold if a hydrodynamic jet is launched and halted in an extended envelope. In this case, particle acceleration at internal shocks approaching the jet head cannot be ruled out, albeit numerical simulations of this scenario are lacking. This might be the case for choked jets accompanying some Type-II supernovae~\citep{He:2018lwb} and LFBOTs~\citep{Guarini:2022uyp}. 

In conclusion, our work highlights the importance of an advanced modeling of  particle production and acceleration in collapsar jets, which takes into account the jet dynamics and related non-linearities. As shown in this work, such modeling may largely affect previous conclusions on the subphotospheric neutrino detection prospects.\\

\noindent {\bf Note added.---} While this project was in its final stages of completion, we became aware of work in progress by Carpio et al.~\cite{CarpioTalk,Bhattacharya:2022btx}, which focuses on  high-energy neutrino emission from magnetized jets propagating in different stellar progenitors. Reference~\cite{CarpioTalk,Bhattacharya:2022btx} relies on an analytic model  with  magnetization at the base of the jet growing as a function of time. This is intrinsically different from our work, which is based on post-processing of realistic 3D GRMHD collapsar jet simulations. Reference~\cite{CarpioTalk,Bhattacharya:2022btx} also overlooks the effects of jet-cocoon mixing and it considers neutrino production in uncollimated jets in collapsars, while our benchmark jets are naturally collimated by the cocoon. Uncollimated jets imply jet energies that are orders of magnitude higher that those observed among GRBs and are thus not supported by observations. 

\begin{acknowledgments}
We thank Kohta Murase and Annika Rudolph for comments on the manuscript.
In Copenhagen, this project has received funding from the  Villum Foundation (Project No.~37358), the Carlsberg Foundation (CF18-0183), and the Deutsche Forschungsgemeinschaft through Sonderforschungsbereich
SFB~1258 ``Neutrinos and Dark Matter in Astro- and Particle Physics'' (NDM). OG is supported by a CIERA Postdoctoral Fellowship and  aknowledges support by Fermi Cycle 14 Guest Investigator program 80NSSC22K0031. 
\end{acknowledgments}

\appendix
\section{Photon thermalization}\label{app:A}
Electrons are assumed to be accelerated to a power-law distribution $N(\gamma^{\prime}_e) \propto \gamma^{\prime -k_e}_e$, where $k_e$ is the electron spectral index. 
Both at collisionless sub-shocks and at magnetic reconnection sites, they are expected to cool through the emission of synchrotron radiation~\citep{2009ApJ...707L..92S,Beniamini:2017fqh,Gill:2020oon}. 

The synchrotron spectrum is defined in terms of three characteristic electron Lorentz factors:  the minimum, the cooling and the self-absorption Lorentz factors ($\gamma^\prime_{e, \rm{min}}$, $\gamma^\prime_{e, \rm{cool}}$, and $\gamma^\prime_{e, \rm{abs}}$), respectively.
These are defined as~\citep{Sari:1999iz, Kobayashi:2003zk, 2011MNRAS.415.1663T, zhang_2018, Thompson:2006fp}:
\begin{eqnarray}
 \gamma^\prime_{e, \rm{min}} &=& \varepsilon_d \varepsilon_e \frac{m_p}{m_e} \frac{k_e-2}{k_e-1} \ , \label{eq:gammamin} \\ 
\gamma^\prime_{e, \rm{cool}} &=& \frac{6 \pi m_e c}{\sigma_T B^{\prime 2} t^{\prime}_{\rm{dyn}}} \, \label{eq:gammacool} \\
\gamma^\prime_{e, \rm{abs}} &=& \left( \frac{\varepsilon_{\rm{abs}}}{\varepsilon_B \alpha} \right)^{1/7} \left(  \frac{B^\prime}{B_Q}  \right)^{-1/7} \ ,
 \label{eq:gammabs}
\end{eqnarray}
where $\varepsilon_e$ and $\varepsilon_B$ are the fractions of the dissipated energy that is stored in accelerated electrons and into magnetic field, respecitvely.  With $\varepsilon_{\rm{abs}}$ we denote the fraction of energy that goes into accelerated electrons radiating at $\gamma^\prime_{e, \rm{abs}}$; $\sigma_T$ is the Thompson cross-section, $\alpha= 1/137$ is the fine-structure constant, $m_e$ the electron mass and $B_Q= 4.41\times 10^{13}$~G. The dynamical time scale of the acceleration process is $t^{\prime}_{\rm{dyn}} = R/(2 c \langle \Gamma_j \rangle)$, where $R$ is the radius at which the process takes place.

Motivated by the results of particle in cell simulations, as for mildly magnetized sub-shocks, we assume $\varepsilon_e= 5 \times 10^{-4}$, $\varepsilon_B=0.1$~\citep{Crumley:2018kvf} and $k_e=2.5$~\citep{Sironi:2013ri}. In the case of magnetic reconnection, $k_e$ is given by Eq.~\ref{eq:kp} and $\varepsilon_e \approx 1- \varepsilon_p$, with $\varepsilon_p $ given by Eq.~\ref{eq:epsp}. Finally, following Ref.~\citep{Thompson:2006fp}, we assume $\varepsilon_{\rm{abs}}= 0.1 \varepsilon_e$.
With this choice of parameters, we get that electrons are always in the fast-cooling regime, namely $\gamma^\prime_{e, \rm{min}} \gg \gamma^\prime_{e, \rm{cool}} $, both for internal sub-shocks and magnetic reconnection. 

The characteristic Lorentz factors in Eqs.~\ref{eq:gammamin}, \ref{eq:gammacool}, and \ref{eq:gammabs} result in three break energies in the photon spectrum, given by
\begin{equation}
E^\prime_{\gamma} (\gamma_e^\prime)= \frac{3}{2} \frac{\hbar e}{m_e c} \gamma_e^\prime B^\prime \, .
\end{equation}
In particular, the self-absorption frequency $\nu_{\gamma, \rm{abs}}^\prime = E^\prime_{\gamma, \rm{abs}}/ h$ gives an estimation of the time over which the synchrotron spectrum becomes self-absorbed and relaxes to a black-body: $t^{\prime}_{\gamma, \rm{abs}} = \nu_{\rm{abs}}^{-1}$. 

The main goal of this paper  is to compute the neutrino production when the jet is  optically thick. Hence, we need to check whether the synchrotron photons thermalize before undergoing $p \gamma $ interactions, whose cooling time can be approximated by
\begin{equation}
t^\prime_{p \gamma} \simeq (n^\prime_{\gamma, \rm{synch}} \sigma_{p \gamma} f_{p \gamma} c)^{-1} \ .
\end{equation}
Here, $\sigma_{p \gamma} \simeq 10^{-28}$~cm$^2$ and $f_{p \gamma} \approx 0.2$ are the cross-section and the multiplicity of $p \gamma$ interactions, respectively~\citep{Razzaque:2005bh}; $n^\prime_{\gamma, \rm{synch}}$ is the number of synchrotron photons  defined as in Eq.~6 of Ref.~\citep{Thompson:2006fp}, with the appropriate energy density of the outflow. The latter is obtained from our benchmark simulations.

As an example,  we get $t^{\prime}_{\gamma, \rm{abs}} \simeq 2.2 \times 10^{-19}$~s at sub-shocks for the jet with $\sigma_0=15$. The photo-hadronic cooling time at the same position is $t^\prime_{p \gamma} \simeq 5 \times 10^{-5}$~s, namely  self-absorption  is much faster than $p \gamma $ interactions.  Similar results hold also for $\sigma_0=200$, both for internal sub-shocks and magnetic reconnection processes. Hence, we can safely assume a black-body spectrum in the optically thick region of the outflow.

\section{Proton and meson cooling rates} \label{app:B}
\begin{figure*}[t]
\centering
\includegraphics[width=0.45\textwidth]{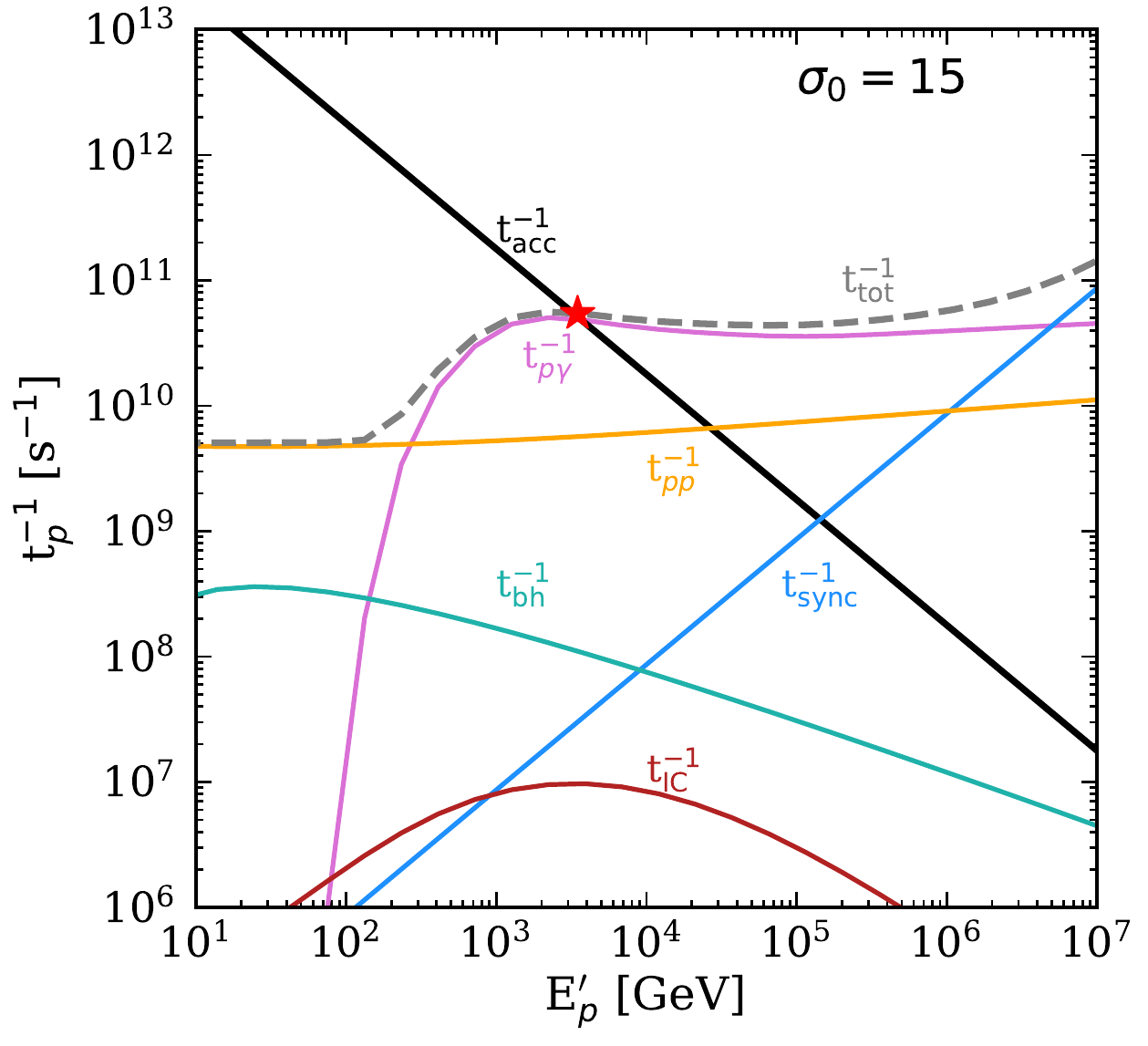}
\includegraphics[width=0.45\textwidth]{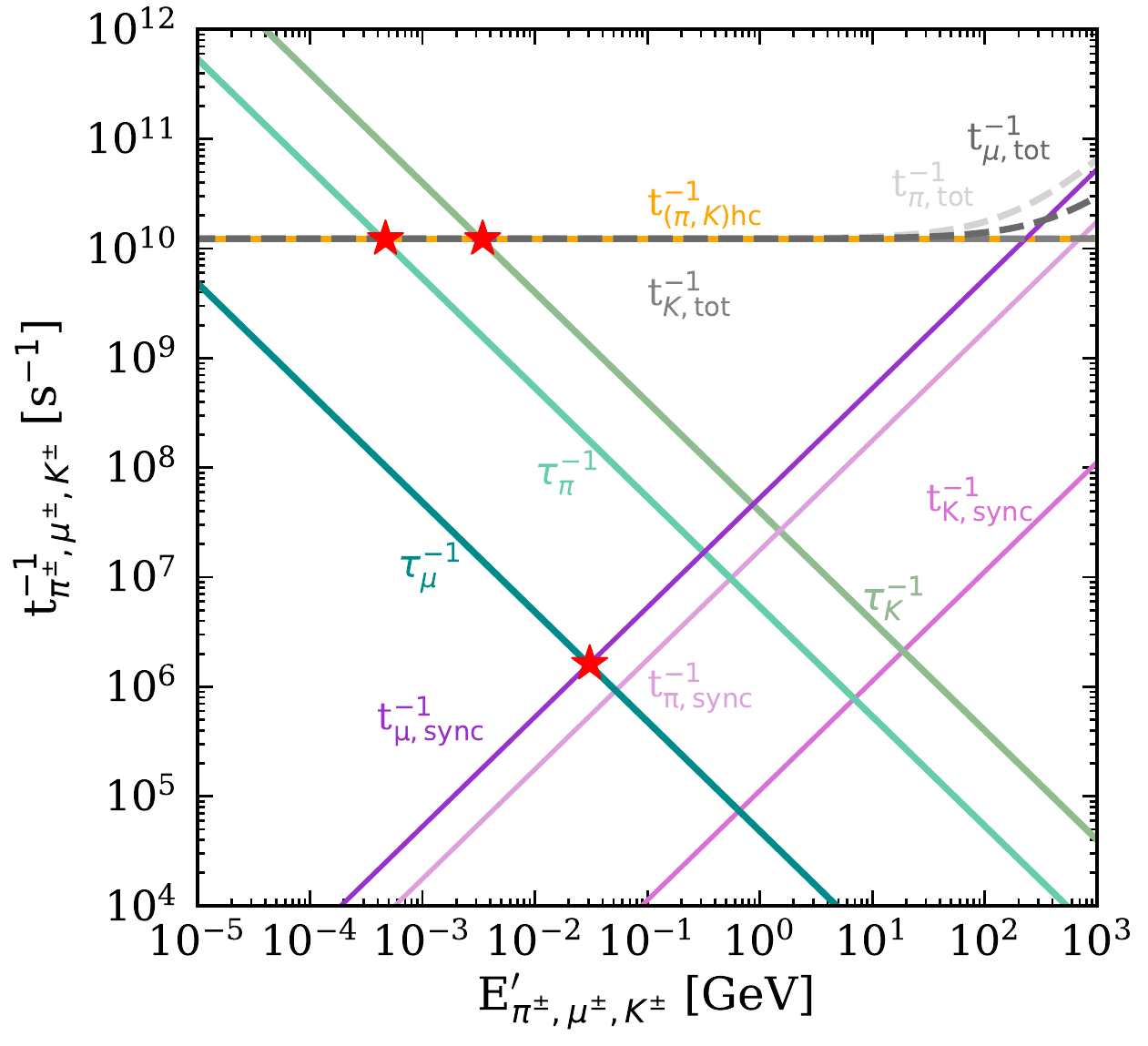}
\caption{\textit{Left panel}: Comoving cooling rates of protons for our benchmark jet with $\sigma_0=15$, calculated for internal sub-shocks at $R_{\rm{SS}}=2.5 \times 10^9$~cm. The red star marks the maximum energy at which protons can be accelerated. Protons mainly cool through $p \gamma$ interactions, while $pp$ interactions become important for $E^{\prime}_p \lesssim 10^2$~GeV. \textit{Right panel}: Same as the left panel, but for secondary mesons. Kaons and pions mainly cool through hadronic processes, while muons undergo strong synchrotron losses. Similar results hold for $\sigma_0=200$. }
\label{fig:cooling}
\end{figure*}
The comoving acceleration rate of protons is
\begin{equation}
    t^{\prime -1}_{p, \rm{acc}} = \frac{c e B^\prime}{\xi E^\prime_p} \ ,
    \label{eq:acceleration}
\end{equation}
where $B^\prime$ is the magnetic field in the acceleration region and it is shown in Fig.~\ref{fig:TB}, $e= \sqrt{\hbar \alpha c}$ is the electric charge, with $\hbar$ being the reduced Planck constant, and $\alpha$ is the fine structure constant. $\xi$ corresponds to the number of gyroradii required for accelerating protons; following Ref.~\cite{Gao:2012ay}, we assume $\xi=10$. 

Accelerated protons undergo several energy loss procceses, parametrized through the total cooling rate: 
\begin{equation}
    t^{\prime -1}_{p, \rm{cool}}= t^{\prime -1}_{p, \rm{ad}}+ t^{\prime -1}_{p, p \gamma} + t^{\prime -1}_{p, pp} + t^{\prime -1}_{p, \rm{BH}} + t^{\prime -1}_{p, \rm{IC}}+ t^{\prime -1}_{p, \rm{sync}} \; ,
\end{equation}
where $t^{\prime -1}_{p, \rm{ad}}$, $t^{\prime -1}_{p, p\gamma}$, $t^{\prime -1}_{p, pp}$, $t^{\prime -1}_{\rm{BH}}$, $t^{\prime -1}_{p, \rm{IC}}$, and $t^{\prime -1}_{p, \rm{sync}}$ are the adiabatic, photo-hadronic ($p \gamma$), hadronic ($pp$), Bethe-Heitler (BH, $p \gamma \rightarrow p e^+ e^-$), inverse Compton (IC) and 
 synchrotron cooling rates, defined as~\citep{dermer_book, Gao:2012ay, Razzaque:2005bh}:
\begin{eqnarray}
 t^{\prime -1}_{\rm{ad}} &=& \frac{2 c \Gamma}{R}\ , \\
 t^{\prime -1}_{p \gamma} &=& \frac{c}{2 \gamma^{\prime 2}_p} \int_{E_{\rm{th}}}^\infty dE^\prime_\gamma \frac{n^\prime_{\gamma} (E^\prime_\gamma)}{E^{\prime 2}_\gamma}  \\  \nonumber & & \; \times \int_{E_{\rm{th}}}^{2 \gamma^\prime_p E^\prime_\gamma} dE_r E_r \sigma_{p \gamma}(E_r) K_{p \gamma}(E_r)\ ,  \\
 t^{\prime -1}_{{pp}} &=& c n^\prime_{p, j} \sigma_{pp} K_{pp}\ ,  \\
 t^{\prime -1}_{p, \rm{BH}} &=& \frac{7 m_e \alpha \sigma_T c}{9 \sqrt{2} \pi m_p \gamma^{\prime 2}_p} \int_{\gamma_p^{\prime -1}}^{\frac{E^\prime_{\gamma, \rm{max}}}{m_e c^2}} d\epsilon^\prime \frac{n^\prime_{\gamma} (\epsilon^\prime)}{\epsilon^{^\prime 2}} \\  \nonumber &  & \; \times  \biggl\{ (2 \gamma^\prime_p \epsilon^\prime)^{3/2} \biggl[\ln(\gamma^\prime_p \epsilon^\prime) -\frac{2}{3} \biggr] +  \frac{2^{5/2}}{3} \biggr\}\ ,  \\
 t^{\prime -1}_{p, \rm{IC}} &=& \frac{3 (m_e c^2)^2 \sigma_T c}{16 \gamma_p^{\prime 2}( \gamma^\prime_p-1) \beta^\prime_p} \int_{E^\prime_{\gamma, \rm{min}}}^{E^\prime_{\gamma, \rm{max}}} \frac{dE^\prime_\gamma}{E_\gamma^{^\prime 2}} \\ &\times& F(E^\prime_\gamma, \gamma^\prime_p) n^\prime_{\gamma}(E^\prime_\gamma)\ , \nonumber \\
 t^{\prime -1}_{p, \rm{sync}} &=& \frac{4 \sigma_T m_e^2 E^\prime_p B^{\prime 2}}{3 m_p^4 c^3 8 \pi} \ .
\end{eqnarray}
Here, $\gamma^\prime_p = E^\prime_p/m_p c^2 $ is the proton Lorentz factor, $E_{\rm{th}} = 0.150$~GeV is the photo-pion production energy threshold, and $\epsilon^\prime= E^\prime_\gamma/m_e c^2$. The comoving proton density $n^\prime_{p, j}$ is given by Eq.~\ref{eq:protonStatic}. The energy dependent cross-sections $\sigma_{p \gamma}$ and $\sigma_{pp}$ are provided by Ref.~\cite{ParticleDataGroup:2020ssz}. The inelasticity for $p \gamma$ interactions is taken from Ref.~\citep{dermer_book}: 
\begin{equation}
K_{p\gamma}(E_r) = 
\begin{system}
0.2 \; \; \; \; \; \;  \; \; \;  E_{\rm{th}} < E_r < 1~\rm{GeV}\\
0.6 \; \; \; \; \; \;  \; \; \;  E_r > 1~\rm{GeV}\ ,
\end{system} \
\end{equation}
where $E_r = \gamma^\prime_p E^\prime_\gamma (1 - \beta^\prime_p \cos\theta^\prime)$ is the relative energy between a photon with energy $E^\prime_\gamma$ and a proton with Lorentz factor $\gamma^\prime_p$, moving in the comoving frame of the interaction region along the directions defined by the  angle $\theta^\prime$.
The inelasticity for $pp$ interactions is $K_{pp}=0.5$. Finally, the function $F(E^\prime_\gamma, \gamma^\prime_p)$ is defined as in Ref.~\cite{1965PhRv..137.1306J}, replacing $m_e \rightarrow m_p$.

As an example,  Fig.~\ref{fig:cooling} shows the proton cooling rates for the optically thick region of our jet with $\sigma_0=15$, at the internal sub-shock radius $R_{\rm{SS}}= 2.5 \times 10^9$~cm. Protons mainly cool through $p \gamma$ interactions, while $pp$ interactions become relevant for $E^\prime_p \lesssim 100$~GeV. Synchrotron losses are important for $E^\prime_p \gtrsim 10^6$~GeV. Similar results hold for $\sigma_0=200$, both for collisionless sub-shocks and magnetic reconnection events.
 
Before decaying, mesons undergo several cooling processes as well. In particular, they suffer adiabatic, synchrotron and hadronic losses, the latter affecting only pions and kaons with the cross-section $\sigma_{\rm{h}}= 5 \times 10^{-26} $~cm$^2$~\citep{ParticleDataGroup:2020ssz}. Their cooling rates are defined as for protons, with the replacement $m_p \rightarrow m_{\pi, K, \mu}$. The meson cooling times are shown in the right panel of Fig~\ref{fig:cooling} for internal sub-shocks ($\sigma_0=15$). Pions and kaons substantially suffer hadronic losses, while muons mainly cool through synchrotron radiation.

\section{  {Particle acceleration in the cocoon and at the interface between the cocoon and counter-cocoon}} \label{app:C}
\begin{figure*}[t]
    \centering
    \includegraphics[width=0.7\textwidth]{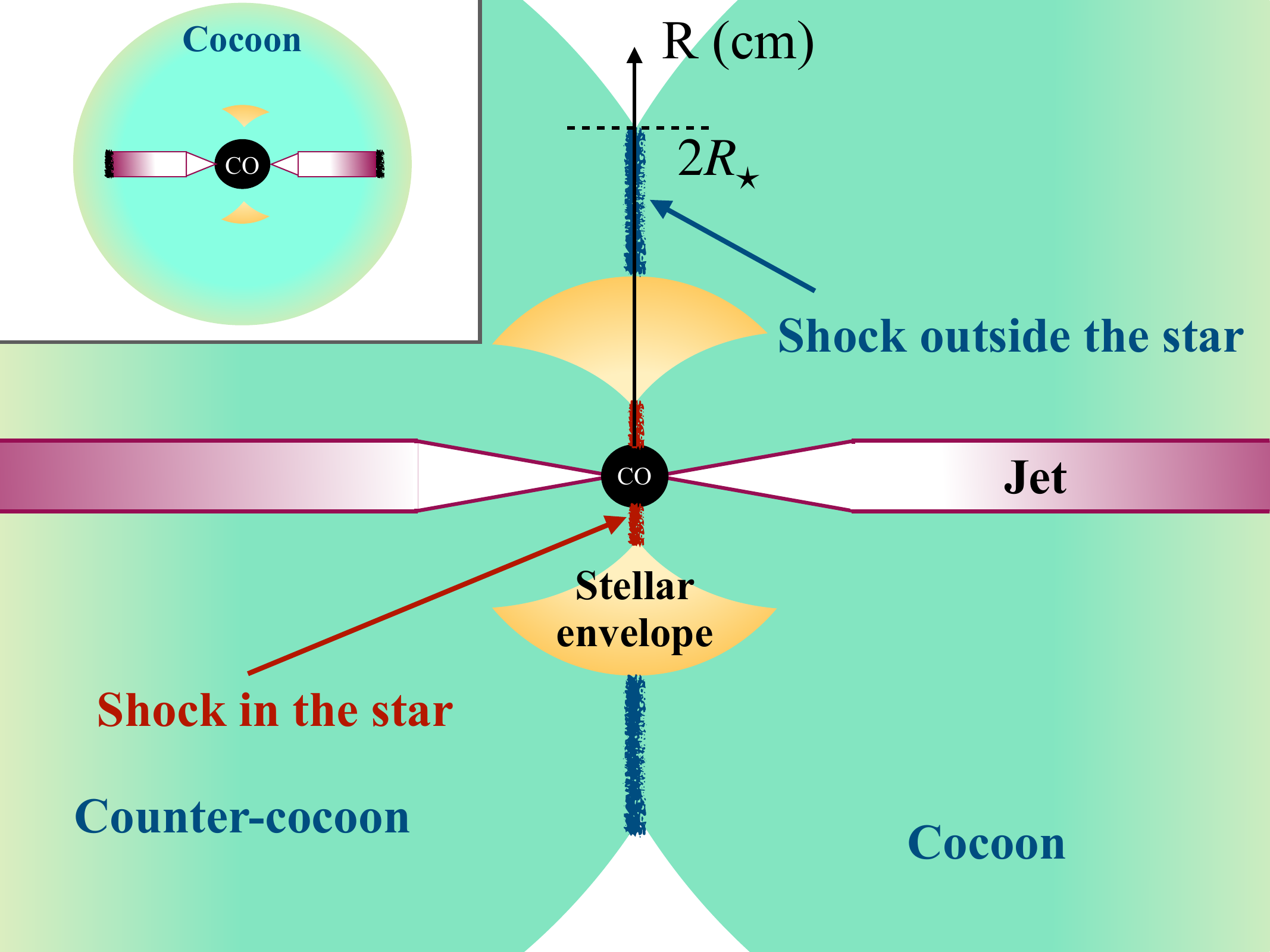}
    \caption{Sketch of the interaction between the cocoon and the counter-cocoon, after the cocoon (aqua) breaks out from the star  (orange), expands and engulfs the progenitor star. The inset plot on the top left corner displays a late-time snapshot, when the cocoon engulfs the star. The cocoon and its counterpart interact inside the star (red line) and outside  (blue line) at $R \simeq 2 R_\star$. Interactions take place in the form of shocks. The large optical depth and very low magnetization of these regions of the outflow inhibit the formation of collissionless shocks and particle acceleration. }
    \label{fig:cocoon}
\end{figure*}
 {While propagating inside the star, the jet inflates a high-pressure region called cocoon. After breaking out, the cocoon expands and engulfs the whole star whether the jet is successful or not. In some cases the fastest component moves with mildly relativistic velocities~\citep{Nakar:2016cih, Gottlieb:2021ebc}. 
Particle acceleration at internal shocks  in the cocoon was assumed to take place in Ref.~\cite{Xiao:2017blv}.  However, as shown in Fig.~\ref{fig:profiles}, the innermost regions of the cocoon are non-relativistic with $ \langle \Gamma_c \rangle \simeq 1$, preventing the formation of strong mildly relativistic shocks. 
Furthermore, we can see from Fig.~\ref{fig:opticalDepth} that the optical depth of the cocoon is extremely large up to $R \simeq 4 \times 10^{11}$~cm; collisionless internal shocks are unlikely to take place. Following  Sec.~\ref{sec:magnetic},  we deduce from Fig.~\ref{fig:profiles} that the cocoon average magnetization is not large enough to trigger magnetic reconnection. Hence, we conclude that particle acceleration in the cocoon is disfavored, contrary to what pointed out in Ref.~\cite{Xiao:2017blv}.}

 {The simulations of Ref.~\cite{Gottlieb:2022tkb} also show interactions between the cocoon and the counter-cocoon both inside and outside the stellar progenitor, as sketched in Fig.~\ref{fig:cocoon}. The interaction outside the star takes place in the form of a shock, occurring at $R \simeq 2 R_\star= 8 \times 10^{10}$~cm. Nevertheless, the outflow is optically thick at this radius (see Fig.~\ref{fig:opticalDepth}) and the corresponding magnetization is $\sigma \simeq 10^{-2} $ (see Fig.~\ref{fig:profiles}). Hence, there is no mechanism able to efficiently accelerate particles at the shock between the cocoon and counter-cocoon shock.}

\section{ {Successful jets}} \label{app:D}
 {The jet is successful when  its energy and  Lorentz factor are such that the jet  drills out of the stellar envelope, eventually reaching the photosphere,  as sketched in the right panel of Fig.~\ref{fig:jetfate}. In this Appendix, we briefly discuss this case for completeness and in order to help the reader to compare the related neutrino emission to the subphotospheric one; the latter being the main focus of this work.} 

 {In our benchmark simulations, the jet breaks out from the star  for both $\sigma_0=15$ and $\sigma_0=200$;  the photospheric radius is  $R_{\rm{PH}} \gtrsim 10^{12}$~cm. The position of the photospheric radius is independent on the jet magnetization, since the jet mimics a hydrodynamic one upon its breakout from the star.}

 {The scenario of successful jets is of particular interest for GRBs. Once the jet reaches the photosphere, it produces the gamma-ray and neutrino bursts eventually observable on Earth. The gamma-ray signal corresponds to the GRB prompt emission, whose origin is still under debate~\citep{2022Galax..10...38B, Zhang:2014qta}. The findings of  Ref.~\citep{Gottlieb:2022tkb} hint towards a hybrid composition of the jet, since both internal shocks and magnetic reconnection may contribute to energy dissipation. }

 {Because of the strong energy dissipation occurring below the photosphere, the prompt signal originates from a non-thermal spectrum at $R_{\rm{PH}}$. The spectral peak and the low-energy part below it are determined by quasi-thermal Comptonization of photons by electrons accelerated to mildly relativistic velocities in the regions of the outflow with $1 \lesssim \tau \lesssim 100$~\citep{Giannios:2006jb, Thompson:2013yna}. Further dissipation may occur above the photosphere, for example through internal shocks. 
GRBs with a dissipative photosphere plus internal shocks has been widely discussed in the literature, see e.g.~Refs.~\citep{1994MNRAS.270..480T, Giannios:2006mx, 2011MNRAS.415.1663T, 2013ApJ...764..143V, Gill:2014fwa, Rees:2004gt, 2013ApJ...764..157B,Pitik:2021xhb}.  A summary of the  neutrino signal from GRBs for various  mechanisms proposed to  model the prompt emission is provided in Ref.~\citep{Pitik:2021xhb}.}

\newpage
\bibliography{main.bib}

\end{document}